\documentclass[11pt]{article}

\usepackage{amsthm,amsmath,amssymb,bm}
\usepackage{graphicx,color,epsfig}
\usepackage[margin=1in]{geometry}

\newcommand{\C}{\mathbb{C}}
\newcommand{\N}{\mathbb{N}}
\newcommand{\R}{\mathbb{R}}

\newcommand{\Z}{\mathbb{Z}}

\newtheorem{Thm}{Theorem}[section]
\newtheorem{Claim}[Thm]{Claim}
\newtheorem{Pro}[Thm]{Proposition}
\newtheorem{Lem}[Thm]{Lemma}
\newtheorem{Cor}[Thm]{Corollary}
\newtheorem{Rem}[Thm]{Remark}

\begin{document}

\title{A simple model for the population dynamics in OTC wholesale fresh product markets}
\author{Ali Ellouze and Bastien Fernandez}
\date{}
\maketitle

\begin{center}
Laboratoire de Probabilit\'es, Statistique et Mod\'elisation\\
CNRS - Univ. Paris Cit\'e -  Sorbonne Univ.\\
Paris, France\\
ellouze@lpsm.paris and fernandez@lpsm.paris
\end{center}

\begin{abstract}
Given the combined evidences of bounded rationality, limited information and short-term optimization, over-the-counter (OTC) fresh product markets provide a perfect instance where to develop a behavioural approach to the analysis of micro-economic systems. Aiming at characterizing via a rigorous mathematical analysis, the main features of the spontaneous organization and functioning of such markets, we introduce and we study a stylized dynamical model for the time evolution of buyers populations and prices/attractiveness at each wholesaler. The dynamics is governed by immediate reactions of the actors to changes in basic indicators. Buyers are influenced by some degree of loyalty to their regular suppliers. Yet, at times, they also prospect for potential better offers. On the other hand, sellers primarily aim at maximising their profit. Yet, they can be also prone to improving their competitiveness in case of clientele deficit.

Our results reveal that, in spite of being governed by simple and immediate rules, the competition between sellers self-regulates in time, as it constrains to bounded ranges the dispersion of both prices and clientele volumes, does similarly for the mean clientele volume, and it generates oscillatory behaviours that prevent any seller to dominate permanently its competitors (and to be dominated forever). Long-term behaviours are also investigated, with focus on asymptotic convergence to an equilibrium, as can be expected for a standard functioning mode. In particular, in the simplest case of $2$ competing sellers, a normal-form-like analysis proves that such convergence holds, 
provided that the buyer's loyalty is sufficiently high or the sellers' reactivity is sufficiently low. In other words, this result identifies and proves those characteristics of the system that are responsible for long term stability and asymptotic damping of the oscillations. 

\end{abstract}

\leftline{\small\today.}


\section{Introduction}
\subsection{OTC wholesale fresh product markets and their analysis}
In spite of efforts to structure wholesale fresh product markets as centralised entities via auction sales and, more recently, digital platforms and electronic marketplaces, many such professional markets across the world, including the largest ones which provide daily foodstuff supply to large megalopolis (e.g.\ Central de Abastos in Mexico City, Paris Rungis Market, Mercamadrid), continue to operate based on over-the-counter (OTC) transactions that involve unregulated interactions between buyers and wholesalers. For markets of such perishable goods, the persistence and stability of this functioning mode, in which strong competition and limited public information prevail, are remarkable. This contrasts with the contemporary trend that aims to formalise and to improve these marketplaces in order to make information widely accessible. 

Indeed, much as in traditional bazaars and souks in Eastern Asia and North Africa \cite{G78}, in these markets, the sellers/wholesalers are grouped by product categories\footnote{It may also be the case that the market consists only of goods of a unique category, for instance a fish market.}, sitting next to each other in the same halls, with their products often displayed in an imposing way.  The prices are usually not posted (or they merely serve as a starting point for negotiation) and the final prices are discreet and they often remain private.\footnote{In some cases, some overall public information about prices is available. For instance, at Paris Rungis Market, the authorities daily post the mean prices of a selection of goods, obtained from declarations of wholesalers in a sample, collected the previous day. Likewise, Central de Abastos's website provides price ranges for certain products.}
 In addition, while standards and certificates exist for most goods, their quality and availability fluctuate depending on the wholesalers's provisions, which are not only seasonal but also depend on other random external factors such as weather conditions and sanitary issues. 

While OTC fresh product markets, whether they are restricted to professionals or are open to the general public, are extremely common, studies dedicated to these markets in the economic literature are scarce beyond 
price data analysis in fish markets \cite{G06,HVS18,WA22}\footnote{Some professional fish markets function based on auction sales \cite{GGKP11} or adopt auction sales and OTC in parallel. In the latter case, it has been show that the agents regularly alternate between the two modes, suggesting that the OTC market remains attractive even in presence of auctions \cite{HVS18,MTV12}. In addition, \cite{WA22} demonstrates through extensive statistics about fish markets in France that prices differences between the two modes are in fact very small.} and related modelling in the game theoretic framework \cite{GH11,HK95}. 

In addition, some behavioural models aiming at reproducing the main features of the buyers' dynamics in OTC (fish) markets have been developed and investigated. In particular, a model has been introduced in \cite{WKH00} for the time evolution of the buyers' preference towards sellers, that includes dependence on sellers' attractiveness. When envisaged at the population scale, the model is sufficiently simple to be amenable to a rigorous mathematical analysis \cite{EF25b}. Yet, it  shows interesting parameter-dependent bifurcations and non-trivial asymptotic functioning modes. However, the attractiveness are assumed to be constant in time and the model does not include any (clientele-dependent) feedback from the merchants. 

This limitation has been addressed in \cite{EF24} where we have included some bi-directional interactions between buyers and sellers in the population dynamics, and in particular a negative feedback from the sellers onto the buyers. Such negative retro-action generates oscillatory behaviours and the main results of that paper identify conditions on the ingredients of the model for the stationary points to be locally/globally asymptotically stable, ensuring convergence to an equilibrium in the long term. Yet, some of the features of the stationary points, such arbitrarily large ratios of the price at competing merchants or the potential permanence of intrinsically repulsive sellers, have limited economic relevance.      

At the individuals' scale, agent-based models have been implemented, based on the major features of buyer–seller interactions, including feedback and bargaining. Numerical simulations of their dynamics have revealed proved capacity to adequately reproduce salient characteristics of the market, such as persistent price dispersion and high loyalty \cite{KV00}. The numerics also have provided quantitative evaluations of the various bargaining strategies \cite{CCB09,MR08}. However, the proposed high level of detail in these models prevents one to proceed to a mathematically rigorous analysis of the dynamics and of the resulting asymptotic behaviours.

\subsection{Main characteristics of the markets}
In this paper, we mathematically investigate a behavioural model for the population dynamics in OTC wholesale fresh product markets, that includes clientele-dependent sellers feedback. As a complement to previous works in the field, to \cite{EF24} and \cite{WKH00} in particular,  the aim is to determine the long-term consequences on prices and clientele volumes, of (daily) repeated basic buyer-seller interactions, namely the immediate clientele variations due to prices changes and, conversely, the immediate price updates that result from changes in clientele volumes. 

The model has been elaborated based on the main empirical features of such markets as they have been reported in the literature, and summarized in this section. 

\paragraph{Buyers' features.}
The buyers are themselves professionals, mostly retailers and restaurateurs. Most of them are loyal, regular customers who have long term (informal) agreements with one/some wholesalers that they use as referent \cite{B-R03,C10}. Yet, there is real competition among wholesalers because at times, the buyers seek for new information, also via their professional networks, in order to check that the agreements they have settled remain competitive. This competition is further enhanced by another type of buyers, namely the "nomads" \cite{VE11}, who do not have stable ties with sellers and who are constantly trying to find the best deal by comparing offers.
 
In addition, buyers' profiles and expectations are variable and heterogeneous \cite{G06}. Depending of their own retail clientele, some buyers focus almost exclusively on prices, while for others, product quality equally matters. 

\paragraph{Sellers' features.} 
When elaborating their offers, sellers take into account their stocks (quantity and quality) and those available at their competitors. They also evaluate the demand, in particular depending on the actual showing up of their regular buyers, lowering the prices at days of scarce clientele and {\sl vice-versa} \cite{C10}. Less evidenced in such markets, although plausible given that stocks and demand are hardly foreseeable, is some of form of cooperation between the wholesalers, or at least some type of communication, in an attempt to reduce the effects of the fluctuations of the demand \cite{VC09}.

\paragraph{Transaction and prices characteristics.}
As mentioned before, prices are not posted and the final prices are not disclosed \cite{G06}. Sellers may charge different prices to different buyers and there are evidences of significant variations of prices paid for the same good \cite{VE11}. The final price usually results from pairwise negotiations and intensive bargaining between the parties. However, in some cases such as fish markets, there is little negotiation and the prices are more of the "take-it-or-leave-it" type \cite{K01}. At Paris Rungis Market, the price the wholesalers pay their providers can be determined {\sl a posteriori}, once the corresponding stock have been sold (post-transaction prices) \cite{C10}. This rule gives the wholesaler additional room to negotiate prices with the clients. More generally, negotiations originate from product heterogeneities, quality variations and also some kind of information asymmetry, namely the fact that the wholesalers know the expectations of their buyers' clients and the retail prices, while the buyers have little knowledge about the wholesalers constraints, especially those due to unforeseeable external factors such as weather conditions or socio-economic/political context. Depending on buyers' needs, the deals can be influenced by various aspects such as interpersonal relationships, accompanying service, delivery, and product follow-up. The combination of so many different factors may be the cause of the observed price dispersion.  

\subsection{Theoretical considerations and modelling}
Bounded rationality, as evidenced by the existence of implicit contracts between buyers and their regular suppliers, combined with limited horizon for optimization due to rapid perishability and to marked unpredictability of prices and product quality, suggest that, in OTC wholesale fresh product markets, a behavioural approach to modelling \cite{D16,H13} based on direct interactions may be best appropriate. 

Besides, the inherent lack of public information about (effective) prices and, more generally, about any element of a seller's attractiveness, advocates that a proper economic appreciation of the market vitality and business should also incorporate volumes of trade, or at least, quantifiers of clientele.

Accordingly, our modelling process aims at investigating in the simplest setting, how basic actions and reactions of the various actors in the market, as they result from direct perceptions, can impact the distribution of both clientele volumes and prices among the competing sellers. It has been based on the following considerations:

\begin{itemize}
\item As it is standard in population dynamics in mathematical biology, we assume, as a first approximation,  that the clientele at each seller is sufficiently large so that it can be considered as an homogeneous collection of agents described by a fraction $x\in [0,1]$ of the total buyers' population. 
    
\item The mentioned  loyalty-induced bounded rationality is taken into account as follows. We assume that the agreements between sellers and buyers imply that, independently of the offer on a given day, a fraction $\alpha x$ of the previous's day clientele systematically returns to the same wholesaler. Only the remaining population $(1-\alpha)x$ is sensitive to current offers. Assuming that behaviours are homogeneous in time, $\alpha$ can be regarded as the fraction of the times at which any buyer returns to their previous merchant(s). 
    
\item Of note, the nomad buyers' population may be associated with a vanishing loyalty parameter ($\alpha=0$). Without clear evidence of direct interactions between loyal and nomad buyers and of their combined impact on prices, we assume that the two populations are fully decoupled and that they can be studied separately, by considering different values of $\alpha$.
    
\item Importantly, buyers can purchase distinct goods of the same category at distinct merchants the same day. Also possible but less likely, some buyers may not purchase any good at all if they are not satisfied with any of the offers. In any case, we do not impose any conservation of the total mass of the effective buyers, {\sl viz.}\  we do not require that ${\displaystyle\sum_{i=1}^N} x_i=1$ where $x_i$ denotes the fraction of buyers that purchases at seller $i\in \{1,\cdots,N\}$. Instead, this number can a priori take any value in $[0,N]$ and it can vary in time.
    
 \item We assume that sellers charge the same price to all their clients, independently of previous visits and of the volume of goods purchased. In other words, loyalty and volumes of trade do not impact (individual) prices. Any future effort to incorporate more realistic features in the model should probably address this assumption in the first place.
    
\item The dynamics is self-contained, namely the time evolution of the variables (cliente volumes and prices) does not involve any time-dependent external input such as production/maintenance costs or any contextual/seasonal effect.
    
\item No {\sl a priori} bounds are imposed on prices and clientele volumes. Instead, a sanity check of the model in the series of results below shows that the dynamics spontaneously regulate these quantities and constraint them to a limited range. 
\end{itemize}

The rest of the paper is organized as follows. The next section provides a formal definition of the model, together with the necessary technical assumptions on its constituents that are required in the proofs. Basic properties of the dynamics and preliminary features are also highlighted. Section \ref{S-MAINR} contains the main mathematical results, which cover several facets of the behaviours, from emerging bounds on the various variables, to a criterion for convergence to equilibrium, to granting of perpetual oscillatory behaviours and finally, for $N=2$, to a claim of asymptotic local stability of stationary points. For the sake of the presentation, all results are grouped together and their proofs are postponed to the following section. A final discussion is given in the last section.

\section{Definitions and preliminary properties}
\subsection{Definitions}\label{S-DEF}
\paragraph{Formal definition of the dynamical system under consideration.} 
The model is a discrete time dynamical system whose variables are quantifiers of clientele volumes and wholesaler prices. In order to provide a formal definition, we consider that the market is composed by $N\in \Z^+\setminus\{1\}=\{2,3,4,\cdots\}$ competing sellers of a single product category. The market is open at repetitive and regular events, typically every working day. The events are labelled by the discrete variable $t\in\N=\{0,1,2,\cdots\}$ and we assume for simplicity that the variables are updated once per event (NB: For a finer temporal modelling, one may instead envisage several updates per event). The state of the market on day $t$ is given by the $2N$-dimensional vector 
\[
(\mathbf x^t,\mathbf p^t)\in M_N=[0,1]^N\times (\R_\ast^+)^N,
\]
where $\mathbf x^t=(x^t_1,\cdots ,x^t_N)$ and $\mathbf p^t=(p^t_1,\cdots ,p^t_N)$ are the variables, which are interpreted as follows.  
\begin{itemize}
\item The {\bf volume of clientele} of merchant $i\in [1,N]=\{1,\cdots ,N\}$ at day $t$ is given by the fraction $x_i^t\in [0,1]$ of the total buyer population present in the market that day. 
As mentioned before, because of potential diversified purchases, the cumulated clientele volume ${\displaystyle\sum_{i=1}^N}x_i^t$ needs not be normalized (and constant). Instead, the {\bf mean clientele volume} 
\[
\langle \mathbf x^t\rangle=\frac1{N}{\displaystyle\sum_{i=1}^N}x_i^t,
\]
can be {\sl a priori} any number in $[0,1]$. Yet, we shall prove that this quantity is naturally constrained by the dynamics and it cannot reach the boundaries of that interval. 
\item Following {\sl in-situ} observations reported above, the quantity $p_i^t\in\R_\ast^+$, which is called {\bf price} for simplicity, intends to aggregate all contributions to the overall attractiveness of seller $i$'s offer on day $t$, namely the price, the product quality, the willingness to negotiation, the accompanying services, etc. 
\end{itemize}
The rules for the time evolution of the variable $(\mathbf x^t,\mathbf p^t)$ are based on the following principles, inspired by the discussion in the previous section.
\begin{itemize}
\item In addition to a certain degree of loyalty, buyers' behaviours are driven by offers' comparisons. Accordingly, we assume that the fraction $x_i^t$ is influenced by the ratio $\frac{p_i^t}{\langle \mathbf p^t\rangle_i^c}$ that involves the mean price $\langle \mathbf p^t\rangle_i^c$ of seller's $i$ competitors, where we use the notation
\[
\langle \mathbf p\rangle_i^c=\frac1{N-1}\left({\displaystyle \sum_{j=1}^N}p_j-p_i\right).
\]
When $p_i^t<\langle \mathbf p^t\rangle_i^c$, merchant $i$ is competitive and attracts new customers; hence $x_i^t$ should increase. On the opposite, $p_i^t>\langle \mathbf p^t\rangle_i^c$ means that seller $i$ is not competitive on day $t$ and repels prospective buyers, lowering the value of $x_i^t$.\footnote{Of note, the reason for considering the ratio $\frac{p_i^t}{\langle \mathbf p^t\rangle_i^c}$  instead of $\frac{p_i^t}{\langle \mathbf p^t\rangle}$ where $\langle \mathbf p\rangle=\frac1{N}{\displaystyle \sum_{j=1}^N}p_j$, is to avoid unintentional bias in the dynamics. Indeed, while both ratios can be made arbitrarily small, close to 0, we have ${\displaystyle \lim_{\langle \mathbf p\rangle_i^c\to 0}}\frac{p_i}{\langle \mathbf p\rangle_i^c}= +\infty$ whereas $\frac{p_i}{\langle \mathbf p\rangle}\leq N$ for all $\mathbf p\in (\R_\ast^+)^N$. A consequence on the dynamics of this asymmetry in the model with ratios $\frac{p_i^t}{\langle \mathbf p^t\rangle}$  is presented in Appendix \ref{A-ALTERNATIVE}.}
\item On the other side, the sellers adjust their prices according to clientele volumes, increasing them when they have more customers than the average of their competitors, and reducing them otherwise.
\end{itemize}
Altogether, the rules for time evolution are formally given by the following set of iterations, whose details are explained immediately below
\begin{equation}
\left\{\begin{array}{l}
x_i^{t+1}=f_\alpha\left(\frac{p_i^{t+1}}{\langle \mathbf p^{t+1}\rangle_i^c},x_i^{t}\right)\\
p_i^{t+1}=p_i^{t}\left(1+g(x_i^{t}-\langle \mathbf x^t\rangle_i^c)\right)
\end{array}\right.\ \text{for}\ i\in [1,N],
\label{DEFDYNAM}
\end{equation}
where $\alpha\in [0,1)$, the parametrized maps $f_\alpha$ are defined by
\[
f_\alpha(\rho,x)=\alpha x+(1-\alpha) f(\rho,x),\ \forall \rho\in\R_\ast^+,x\in [0,1]
\]
and where the maps $f$ and $g$ will be specified below. 

The iteration \eqref{DEFDYNAM} can be regarded as the repeated action of some {\bf multidimensional map} $F$ acting in the $2N$-dimensional phase space $M_N=[0,1]^N\times (\R_\ast^+)^{N}$ with variables $(\mathbf x,\mathbf p)$. In order words, equation \eqref{DEFDYNAM} implicitly defines a map $F$ such that $(\mathbf x^{t+1},\mathbf p^{t+1})=F(\mathbf x^t,\mathbf p^t)$ for all $t\in\N$.\footnote{The conditions to be given on $f$ and $g$ imply that  $F:M_N\to M_N$, as claimed in the next section.} In this context, we recall that the {\bf orbit} that starts at $t=0$ from the initial condition $(\mathbf x^0,\mathbf p^0)$ is the sequence $\{(\mathbf x^t,\mathbf p^t)\}_{t\in\N}$ defined by $(\mathbf x^t,\mathbf p^t)=F^t(\mathbf x^0,\mathbf p^0)$.

\paragraph{Interpretation of the iteration rules.}
The expression in the first row of \eqref{DEFDYNAM} considers that, as indicated previously, a fraction $\alpha$ of the buyers' population is fully loyal and systematically returns to the sellers they purchased at the previous day (term $\alpha x_i^t$). The loyalty parameter $\alpha$ is fixed once for all and remains unchanged throughout the paper. For the sake of notation, {\bf we will not mention any explicit dependence on this parameter $\alpha$}. 

The remaining $1-\alpha$ fraction of buyers is assumed to be volatile and sensitive to prices. The term $(1-\alpha) f\left(\frac{p_i^{t+1}}{\langle \mathbf p^{t+1}\rangle_i^c},x_i^t\right)$ stipulates that the volume of such buyers who purchase at merchant $i$ at $t+1$ not only depends on the price ratio $\frac{p_i^{t+1}}{\langle \mathbf p^{t+1}\rangle_i^c}$ on that day, but also on the fraction $x_i^t$ at previous day, expressing some contagion process driven by mimetic behaviours. 

The expression in the second row of \eqref{DEFDYNAM} considers that sellers simply update their prices depending on the difference of the volumes between their own clientele $x_i^t$ and the mean clientele $\langle \mathbf x^{t}\rangle_i^c$ at their competitors. Accordingly, the function $g$ involved in the multiplicative coefficient should be positive when this difference is positive, and negative (but not smaller than $-1$) when the difference is negative. 

These seller reactions regulate the dynamics through negative feedback. Indeed, the first row of \eqref{DEFDYNAM} and the assumptions on $f_\rho$ imply that the clientele's volume decreases at every wholesaler whose price is above the average of their competitors' prices. Accordingly, it is likely to eventually fall below the clientele's volumes of the competitors. However, the second row implies that, from the day this happens, these wholesalers react by reducing their price, possibly below the average of competitors' price, which then triggers an increase of clientele. In short terms and anticipating a statement below, dominant and dominated sellers, as well as the volumes of their clientele, are prone to alternate in time. Interestingly, such oscillatory behaviours driven by negative feedback is reminiscent of the principles governing homeostasis in biological systems, see e.g.\ \cite{KST07,PMO95} for theoretical results in that context. 

Finally we mention that, by assuming that sellers' influence operates through price ratios, the system \eqref{DEFDYNAM} has been designed to address the main limitations of the plain relevance of the model in \cite{EF24} (where this influence is based on an elementary absolute scale), especially the fact that the ratios of prices at competitive sellers can be arbitrary large in stationary states. In particular, Proposition \ref{UNIFORM-BOUNDS-PRICESRAT}  states that all prices ratios must be eventually uniformly bounded in every orbit. Besides, notice that oscillatory behaviours in \cite{EF24} eventually stop when approaching a fixed point, and this property largely facilitates the stability analysis. Instead, as we shall see below (Proposition \ref{ALTERNATIONS}), oscillations perdure forever in \eqref{DEFDYNAM}, which makes the proofs of asymptotic stability, and more generally, any control of the dynamics, much more challenging.  

\paragraph{Assumptions on the maps $f$.} 
In order to prove results on the dynamics of the system \eqref{DEFDYNAM}, a number of technical assumptions are imposed on the map $f$ that are inspired by the considerations above about sellers' attractiveness. The basic assumptions that are required for results on the global dynamics in phase space, for every $N$, are the following ones, which specify the dependence on $x$ and $\rho$. 
\begin{figure}[ht]
\begin{center}
\includegraphics*[width=90mm]{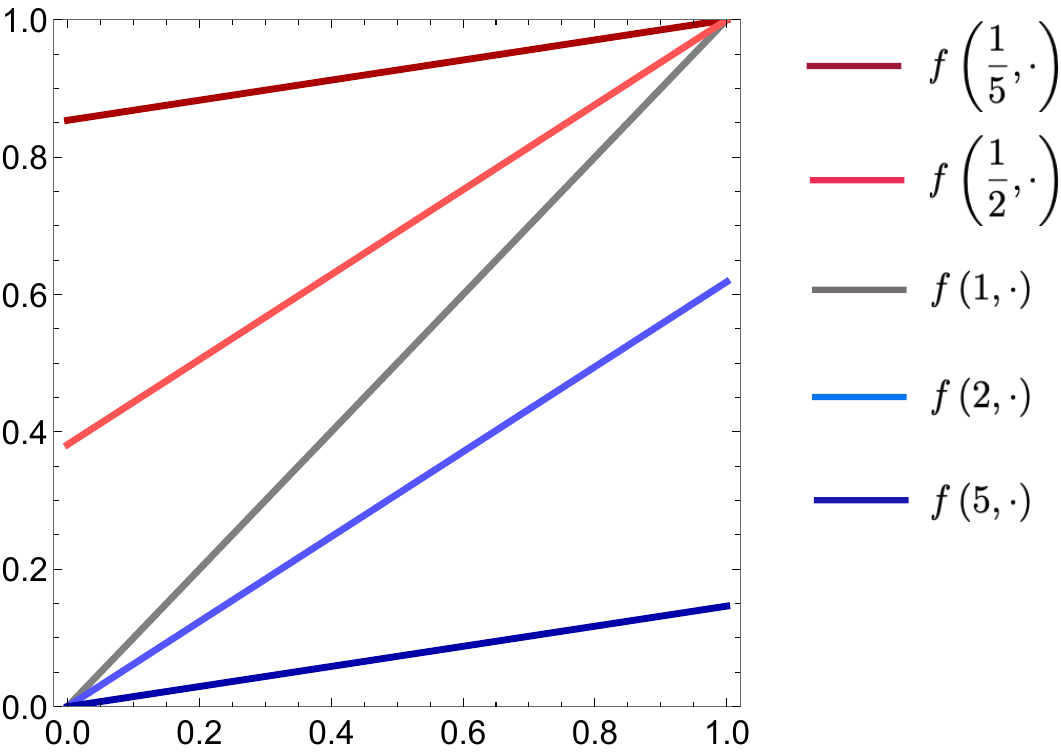}
\end{center}
\caption{Graphs of $f(\rho,\cdot)$ for the piecewise affine example \eqref{LINEARMAP} with $c(\rho)=e^{-|\ln \rho|}$ (so that $c(\rho)=\frac1{\rho}$ for $\rho\geq 1$ and $c(\rho)=\rho$ for $\rho<1$) and $\rho\in \{\frac14,\frac12,1,2,4\}$. These plots illustrate the assumptions (Hf1-2), and in particular, the monotonicity in $x$ and $\rho$, that the graph lies above the diagonal, excepted at $x=1$, when $\rho<1$, and it lies below the diagonal, excepted at $x=0$, when $\rho>1$, and that we have $f'_x(\rho,\cdot)<1$ for all $\rho\neq 1$. The symmetry in (Hf3) is also evident.}
\label{GRAPHF-AFFINE}
\end{figure}
\begin{itemize}
\item[(Hf1)] ({\em Dependence on $x$.}) $f(\rho,\cdot)$ is continuously differentiable and increasing on $[0,1]$ for every $\rho\in\R^+_\ast$ and we have 
\begin{itemize}
\item[$\bullet$] $f(\rho,0)=0$ and $f(\rho,1)<1$, for every  $\rho> 1$,
\item[$\bullet$] $f(\rho,0) >0$ and $f(\rho,1)=1$,  for every  $\rho< 1$,
\item[$\bullet$] ${\displaystyle\sup_{x\in [0,1]}}f'_x(\rho,x)<1$ for every  $\rho\neq 1$.
\end{itemize}
\item[(Hf2)] ({\em Dependence on $\rho$.}) 
\begin{itemize}
\item[$\bullet$] $f(\cdot,x)$ is continuous and decreasing on $\R^+_\ast$, for every  $x\in (0,1)$, 
\item[$\bullet$] $f(\cdot,0)$ is continuous and decreasing on $[1,+\infty)$,
\item[$\bullet$] $f(\cdot,1)$ is continuous and decreasing on $(0,1]$,
\item[$\bullet$] $f(1,\cdot)=\mathrm{Id}$, 
\item[$\bullet$] $f(0^+,x)=1$ and $f(+\infty,x)=0$, for every  $x\in [0,1]$.
\end{itemize}
\end{itemize}

Further conditions are imposed in addition to (Hf1-2), that depend on $N$. In the case $N=2$, the following symmetry is required.  
\begin{itemize}
\item[(Hf3)] ({\em Symmetry.}) $f(\frac1{\rho},x)=1-f(\rho,1-x)$ for all  $(\rho,x)\in \R^+_\ast\times [0,1]$,
\end{itemize}
That the buyers' instant behaviours obey such prices-clientele volumes symmetry is a simplifying assumption that has important consequences on the dynamics of clientele volumes. However, buyers behaviours need not be symmetric, and for completeness, we present in Appendix \ref{A-NONSYM} those features of clientele volumes that prevail when (Hf3) fails. 

For example,  the following simple affine map of the variable $x$
\begin{equation}
f(\rho,x)=\left\{\begin{array}{ccl}
c(\rho)x&\text{if}&\rho\geq 1\\
1-c(\rho) (1-x)&\text{if}&\rho\leq 1
\end{array}\right.
\label{LINEARMAP}
\end{equation}
satisfies the assumptions (Hf1-3), provided that the function $c$ has the following properties
\begin{itemize}
\item $c(\frac1{\rho})=c(\rho)$ for all $\rho\in \R_\ast^+$,
\item $c(\rho)\in (0,1)$ for all $\rho>1$ and $c(1)=1$,
\item $c$ is continuous and decreasing on $[1,+\infty)$ and ${\displaystyle\lim_{\rho\to +\infty}}c(\rho)=0$.
\end{itemize}
An illustration for $c(\rho)=e^{-|\ln \rho|}$ is given on Fig.\ \ref{GRAPHF-AFFINE}.

Besides, for the local stability result for $N=2$ (Theorem \ref{STABILITY}), other conditions on $f$ will be required. An example that satisfies all conditions at once will be given after that statement.
 
In the case $N>2$, the map $f$ will be an extension of the map in \eqref{LINEARMAP} with $c(\rho)=e^{-|\ln \rho|}$. More precisely, it is given by the following expression
\begin{equation}
f(\rho,x)=\left\{\begin{array}{ccl}
\frac{x}{\rho}+f_\mathrm{dev}(\rho,x)&\text{if}&\rho\geq 1\\
1-\rho(1-x)-f_\mathrm{dev}(\rho,1-x)&\text{if}&\rho\leq 1
\end{array}\right.\ \forall x\in [0,1].
\label{SPEFAM}
\end{equation}
where $f_\mathrm{dev}(\rho,\cdot)\geq 0$ is differentiable for all $\rho\in \R_\ast^+$ and 
\[
f_\mathrm{dev}(\rho,x)\leq \left\{\begin{array}{ccl}
\left(1-\frac{1}{\rho}\right)x&\text{if}&\rho\geq 1\\
(1-\rho)x&\text{if}&\rho\leq 1
\end{array}\right.\ \forall x\in [0,1],
\]
which in particular, implies that $f_\mathrm{dev}(\rho,0)=0$ for all $\rho\in \R_\ast^+$ and $f_\mathrm{dev}(1,\cdot)=0$. The condition $f_\mathrm{dev}(\rho,\cdot)\geq 0$ implies a quantitative control of the attraction rate to 0 when $\rho>1$ (which is at most $\frac{1}{\rho}$), and to 1 when $\rho<1$. In other words, in absence of sellers' feedback, buyers' fractions cannot approach too fast (depending on $\rho$) the fixed points 0 and 1. 

\paragraph{Assumptions on the map $g$.} 
Combining the considerations above about sellers' reactions with technical assumptions that will be employed in the proofs, we require that the map $g$ (which is defined on $[-1,1]$) be differentiable, increasing and such that 
\[
g(x)>-1,\ \forall x\in [-1,1]\quad\text{and}\quad g(0)=0.
\]
When focus is made on orbits for which $\mathbf x^t\in (0,1)^N$ for all $t\in\N$ (which, up to an initial transient period, is the case of every orbit except for the two fixed points with either $x_i=0$ for all $i$ or $x_i=1$ for all $i$), it suffices to assume the following condition instead, in particular, in order to control the relative prices $\frac{p_i^t}{p_N^t}$
\begin{itemize}
\item[(Hg1)] $g$ is differentiable and  increasing on $(-1,1]$, $g(x)>-1,\ \forall x\in (-1,1]$ and $g(0)=0$.
\end{itemize}
While the previous assumption does not hold for the elementary example $g(x)=a x$ with arbitrary $a\in (0,1]$, this weaker condition holds for this example.

Moreover, the following assumptions will be employed in order to ensure either a uniform control of these ratios and a control of the price amplitudes themselves.
\begin{itemize}
\item[(Hg2)] Given $N\geq 2$, we have ${\displaystyle\sum_{i=1}^N}g(x_i-\langle x\rangle_i^c)\leq 0$ for all $x\in (0,1)^N$.
\item[(Hg3)] $S_g={\displaystyle\sup_{(x,y)\in (-1,1)^2}}\frac{1+g(x)}{1+g(y)}<+\infty$.
\end{itemize}
For instance, the map $g$ defined by $g(x)=ax+bx^2$, where $a\in (0,1)$ and $b\in (\max\{a-1,-\frac{a}2\},0]$, satisfies {\rm (Hg1-3)}. 
  
\subsection{Preliminary properties}
The goal of the analysis presented below is to investigate the behaviours of the orbits of the dynamical system generated by \eqref{DEFDYNAM}, and in particular the long term behaviours as $t\to +\infty$. In this preliminary section,  we provide a number of general features that set the context of the analysis to follow. 

\paragraph{Well-defined dynamics.} 
The convex combination in the first equation of \eqref{DEFDYNAM}, together with the property $f(\rho,[0,1])\subset [0,1]$, and the fact that $g(x)> -1$ for all $x\in [-1,1]$ imply that for every initial condition $(\mathbf x^0,\mathbf p^0)\in M_N$, the subsequent orbit $\{(\mathbf x^t,\mathbf p^t)\}_{t\in\N}$ is well defined, and we have $(\mathbf x^t,\mathbf p^t)\in M_N$ for all $t\in\N$. 

Moreover, the monotonicity of the maps $f(\rho,\cdot)$ implies that we have $f(\rho,(0,1))\subset (0,1)$ which, together with the condition (Hg1), suffices to ensure that if $\mathbf x^0\in (0,1)^N$, then $\{(\mathbf x^t,\mathbf p^t)\}_{t\in\N}$ is well defined and we have $\mathbf x^t\in (0,1)^N$ for all $t\in\N$.

In the analysis below, {\bf we only consider the orbits with initial buyer population $\mathbf x^0\in (0,1)^N$} and assume (Hg1).

Notice also that the map $F$ is invertible, {\sl ie.}\ $(\mathbf x^t,\mathbf p^t)$ in \eqref{DEFDYNAM} can be uniquely determined by $(\mathbf x^{t+1},\mathbf p^{t+1})$, and $F^{-1}(M_N)\subset M_N$, so that pre-images arbitrarily far in the past are all well defined.

\paragraph{Synchronized dynamics and fixed points.}
As a particular instance of coupled map system without spatial structure \cite{CF05}, the dynamics \eqref{DEFDYNAM} commutes with the simultaneous permutations of the $\mathbf x$- and $\mathbf p$-coordinates, {\sl viz.}\ if the sequence $\{(\mathbf x^t,\mathbf p^t)\}_{t\in\N}$ is an orbit of \eqref{DEFDYNAM}, then for every permutation $\pi$ of $\{1,\cdots ,N\}$, the sequence $\{(\pi \mathbf x^t,\pi \mathbf p^t)\}_{t\in\N}$ where we use the generic notation
\[
\pi \mathbf x=(x_{\pi(1)},\cdots,x_{\pi(N)}),
\]
is also an orbit of this system. This symmetry implies in particular that if two sellers $i$ and $j$ are at {\bf equilibrium} at some instant $t$, meaning that $x_i^t=x_j^t$ and $p_i^t=p_j^t$, then this feature will be preserved at all future instants $t'>t$. This property is the basis of the so-called partial synchronization (when groups of nodes in a network evolve in synchrony), a phenomenon that has been largely investigated in the context of coupled dynamical systems, see for instance \cite{PSN02} and also \cite{BKOVZ02} for a review on synchronization.

In particular, in the case of full synchrony, {\sl ie.}\ when all sellers are at equilibrium with respect to each other, the assumption $f(1,\cdot)=\text{Id}$ implies that $(\mathbf x^t,\mathbf p^t)$ must be a fixed point of $F$, namely that the orbit must be constant. Every couple $(\mathbf x,\mathbf p)\in M_N$ in which the coordinates of $\mathbf x$ and $\mathbf p$ are equal ({\sl ie.} $x_i=x\in [0,1]$ and $p_i=p\in \R_\ast^+$ for all $i$) is a {\bf fixed point} and these couples are the only fixed points of $F$.\footnote{Indeed, by contradiction, if a fixed point existed with ${\displaystyle\min_{i\in [1,N]}}p_i<{\displaystyle\max_{i\in [1,N]}}p_i$, then we would have 
\[
\min_{i\in [1,N]}\frac{p_i}{\langle \mathbf p\rangle_i^c}<1<\max_{i\in [1,N]}\frac{p_i}{\langle \mathbf p\rangle_i^c}
\]
implying that the existence of $i\neq j\in [1,N]$ such that $x_i=0$ and $x_j=1$. Yet, the second row in \eqref{DEFDYNAM} and the assumption (Hg1) impose that the coordinates $x_i$ of any fixed point cannot depend on $i$.}

We claimed above that the map $F$ is invertible. Its inverse $F^{-1}$ also commutes with the simultaneous permutations of the $\mathbf x$- and $\mathbf p$-coordinates. As a consequence, for every $t\in\N$, no orbit can become (partly) synchronized at instant $t$ if its initial condition is not so. For the sake of the presentation, only those orbits that are not partly synchronized will be considered in the rest of the paper.

\paragraph{Dynamics in terms of relative prices}
According to expression \eqref{DEFDYNAM}, the dynamics of the clientele volumes $x_i$ only depends on the relative prices $\frac{p_i}{p_j}$, and in particular on the ratios $\rho_i=\frac{p_i}{p_N}$. In addition, the dynamics of these ratios is self-contained. In other words, letting $\boldsymbol\rho=\{\rho_i\}_{i=1}^{N-1}$, under the change of variables $\mathbf p\mapsto (\boldsymbol\rho,p_N)$ the variables $(\mathbf x,\boldsymbol \rho)$ evolve in an autonomous way (hence independent of the variable $p_N$) whereas the iterations of $p_N$ depend on all the variables $(\mathbf x,\boldsymbol \rho,p_N)$.\footnote{The complete system acting on $(\mathbf x,\boldsymbol \rho,p_N)$ is said to be a skew-product dynamical system.} The iterations of the variables $(\mathbf x,\boldsymbol \rho)$ simply read 
\begin{equation}
\left\{\begin{array}{l}
x_i^{t+1}=f_{\alpha,\frac{\rho_i^{t+1}}{\langle \boldsymbol \rho^{t+1}\rangle_i^c}}(x_i^{t})\quad \text{for}\ i\in [1,N]\\
\rho_i^{t+1}=\rho_i^{t}\frac{1+g(x_i^{t}-\langle \mathbf x^t\rangle_i^c)}{1+g(x_N^{t}-\langle \mathbf x^t\rangle_N^c)}\quad \text{for}\ i\in [1,N-1]
\end{array}\right..
\label{BASIS}
\end{equation}
and we shall denote $F_\mathrm{skew}$ the corresponding map acting in $[0,1]^N\times (\R_\ast^+)^{N-1}$. Focusing on the dynamics of the clientele volumes and of the relative prices, the analysis below mostly investigates the dynamics of $F_\mathrm{skew}$, whose {\bf fixed points} are every couple $(\mathbf x,\mathbf 1)$ where $x_i=x$ for any $x\in [0,1]$ and $\mathbf 1$ stands for the vector with components $\rho_i=1$ for $i\in [1,N-1]$). Yet, as mentioned above, some considerations will also be made about the behaviours of the $p_i^t$ themselves, and especially about their location in a closed interval inside $\R_\ast^+$. 

\section{Main results}\label{S-MAINR}
\begin{figure}[ht]
\begin{center}
\includegraphics*[width=140mm]{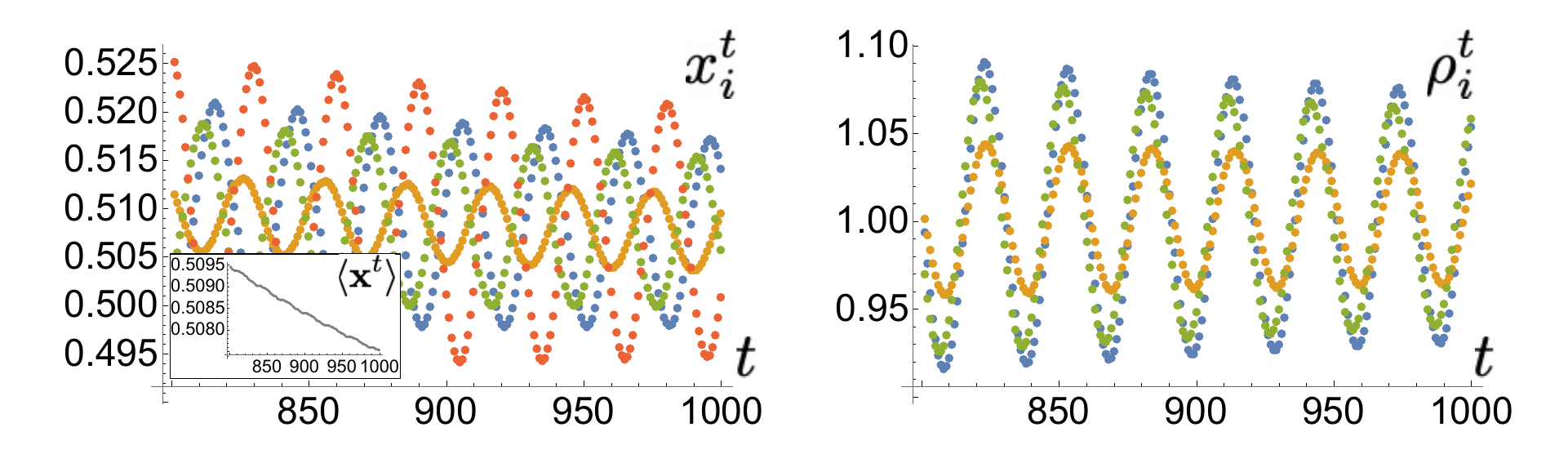}
\end{center}
\caption{Illustration for $N=4$ of the various results on clientele fractions and prices, their asymptotic behaviours and their perpetual crossings. Plots of the times series of clientele fractions $\{x_i^t\}_{i=1}^4$ (left),  mean value $\langle \mathbf x^t\rangle$ (inset) and prices ratios $\{\rho_i^t\}_{i=1}^3$ (right) associated with an orbit of the system \eqref{DEFDYNAM} with $\alpha=0.9$, $f$ as in Fig.\ \ref{GRAPHF-AFFINE} and $g(x)=\frac{x}2$. The initial condition has been chosen at random. The picture illustrates in particular the perpetual crossings claimed in Proposition \ref{ALTERNATIONS} and it suggests that the long-term convergence to equilibrium of Proposition \ref{REALIST-PRICES} holds in this case. The time series in the inset indicates that monotonous decay of the mean value $\langle \mathbf x^t\rangle$ prevails below the value $N-\frac1{N-1}$ given in Proposition \ref{CONSTRAINTS-X}.}
\label{ILLUSTR-ALTERN}
\end{figure}

This section collects the main results about the dynamics and orbits' behaviours. While most results holds for arbitrary $N\geq 2$, as can be expected, more thorough conclusions have been obtained in the simplest case $N=2$. An overall illustration by means of times series of the various coordinates, is given in Fig.\ \ref{ILLUSTR-ALTERN}. Moreover, a specific illustration of the fixed point asymptotic stability in the case $N=2$ (Theorem \ref{STABILITY}) will be given in Fig.\ \ref{ILLUSTR-TRAJEC} below.

\subsection{Emerging bounds on prices and their ratios}\label{S-CONSTR}
By itself, the definition of the phase space $M_N$ does not constrain the amplitudes of the prices $p_i^t$ and their ratios $\rho_i^t$. Hence, {\sl a priori}, these quantities could become arbitrarily large or arbitrarily small as $t\to +\infty$. However, as far as economic relevance is concerned, it would be problematic that the prices and their ratios asymptotically diverge. 
Accordingly, a first sanity check of the model is to make sure that, in every orbit, prices ratios remain constrained to some bounded interval away from 0. The first result of this paper claims that this is indeed the case provided that $f$ and $g$ satisfy the primary $N$-independent conditions listed above. 
\begin{Pro}
Let $N\geq 2$ be arbitrary and assume that $f$ satisfies the assumptions {\rm (Hf1-2)} and $g$ satisfies {\rm (Hg1)}. Then, for every orbit $\{(\mathbf x^t,\boldsymbol\rho^t)\}_{t\in\N}$ of \eqref{BASIS}, there exists $M\in\R^+_\ast$, $M>1$, such that 
\[
\max_{i\in [1,N-1]}\max\left\{\rho_i^t,\frac1{\rho_i^t}\right\}\leq M,\ t\in\N.
\]
\label{BOUNDED-RHO}
\end{Pro}
\noindent
The proof is given in Section \ref{S-BOUNDED-RHO}.  Since the initial condition $(\mathbf x^0,\boldsymbol\rho^0)$ can be chosen so that ${\displaystyle \max_{i\in [1,N-1]}}\left\{\rho_i^0,\frac1{\rho_i^0}\right\}$ be arbitrarily large, the bound $M$ in this statement evidently depends on the orbit under consideration. 

However, one may wonder if, after a transient period, a uniform, orbit-independent, bound can apply for all $t$ sufficiently large. In other words, does the time evolution naturally impose universal bounds on prices ratios in the long term and does it constrain these ratios to eventually enter in an invariant interval in the interior of $\R_\ast^+$? 

The existence of such universal bounds requires an additional condition on $g$ because periodic orbits with arbitrarily large values of $\max\left\{\rho_i^t,\frac1{\rho_i^t}\right\}$ can exist when the instantaneous evolution rate of $\rho_i$, namely the ratio
\[
\frac{1+g(x_i-\langle \mathbf x\rangle_i^c)}{1+g(x_N-\langle \mathbf x\rangle_N^c)}
\]
is unbounded. Indeed, for $N=2$ and the piecewise affine map $f$ defined by \eqref{LINEARMAP} with $c(\rho)=e^{-|\ln\rho|}$ and $g(x)=x$, one can easily prove the existence of 4-periodic orbit for which 
\[
\rho_1^{4t}=\rho_1^{4t+2}=1,\quad \rho_1^{4t+1}=\rho\quad \text{and}\quad \rho_1^{4t+3}=\frac1{\rho},
\]
for every $\rho\in\R_\ast^+$ \cite{E25}. Nonetheless, when the ratio above is bounded (as the assumption (Hg3) imposes) and provided that it is sufficiently small depending on $N$ (for $N>2$), one proves that the above questions have affirmative answers and that price ratios discrepancies in the market are indeed uniformly limited at large times. This is the purpose the statement below. 

In order to get quantitative estimates, given an arbitrary $N\geq 2$, let
\[
T_N=\min\left\{t\in\Z^+\ :\ f_\alpha^{t}\left(\frac{N(N-1)}{1+(N-2)(N+1)},1\right)<f_\alpha^{t}\left(\frac{N-1}{N},0\right)\right\},
\]
where the notation $f^t$ stands for the iterations of the $x$ variable defined by the following induction
\[
f^1_\alpha(\rho,x)=f_\alpha(\rho,x)\quad \text{and}\quad f^{t+1}_\alpha(\rho,x)=f_\alpha(\rho,f^t_\alpha(\rho,x)),\ \forall (\rho,x)\in \R^+_\ast\times [0,1],t\in\Z^+.
\]
Given the properties on $f$, we must have $T_N<+\infty$.
\begin{Pro}
Assume that $f$ satisfies {\rm (Hf1-2)} and $g$ satisfies {\rm (Hg1)} and {\rm (Hg3)}.

\noindent
{\rm Case $N=2$.} For every orbit $\{(\mathbf x^t,\rho_1^t)\}_{t\in\N}$ of \eqref{BASIS}, there exists $t'\in\N$ so that 
\[
\max\left\{ \rho_1^t,\frac1{\rho_1^t}\right\}\leq 2 S_g^{T_2},\ \forall t\geq t'.
\]

\noindent
{\rm Case $N>2$.} Assume in addition that $S_g\leq \left(\frac{N+1}{N}\right)^\frac1{T_N}$. Then, for every orbit $\{(\mathbf x^t,\boldsymbol\rho^t)\}_{t\in\N}$ of \eqref{BASIS}, there exists $t'\in\N$ so that 
\[
\max_{i\in [1,N-1]}\max\left\{\rho_i^t,\frac1{\rho_i^t}\right\}\leq N+1,\ \forall t\geq t'.
\]
\label{UNIFORM-BOUNDS-PRICESRAT}
\end{Pro}
\noindent
The proof is given in Section \ref{S-UNIFORM-BOUNDS-PRICESRAT}. 

What about bounds on the prices? As before, it is legitimate to ask whether or not these quantities can become arbitrarily large (or small) as $t\to+\infty$. Notice that the dynamics of prices commutes with the multiplication by any number in $\R_\ast^+$, {\sl viz.} if $\{(\mathbf x^t,\mathbf p^t)\}_{t\in\N}$ is an orbit of \eqref{DEFDYNAM}, then for every $a\in\R_\ast^+$, the sequence $\{(\mathbf x^t,a\mathbf p^t)\}_{t\in\N}$ is also an orbit of this system. Therefore, no orbit-independent bound alike in Proposition \ref{UNIFORM-BOUNDS-PRICESRAT} can hold for the prices. One can only expect to show that the prices remain bounded and do not diverge in the long term. Proposition \ref{REALIST-PRICES} below claims that this is indeed the case provided that in addition to (Hg1), $g$ also satisfies the weak concavity assumption (Hg2). 

Furthermore, Proposition \ref{REALIST-PRICES} does not prevent the prices to asymptotically converge to 0. However, it establishes a relationship between such putative convergence and the asymptotic behaviour of the orbit itself, 
underlying the importance of the convergence to equilibrium in order to maintain realistic prices in the market. More precisely, the only case in which prices can asymptotically vanish is when the orbit does not converge to the set of stationary points. 
\begin{Pro}
Let $N\geq 2$ be arbitrary and assume that $f$ satisfies {\rm (Hf1-2)} and $g$ satisfies {\rm (Hg1-2)}.
 Then, for every orbit $\{(\mathbf x^t,\mathbf p^t)\}_{t\in\N}$ of \eqref{DEFDYNAM}, we have
\[
\sup_{t\in\N}\max_{i\in [1,N]} p_i^t<+\infty.
\]
Moreover, assume that ${\displaystyle\varlimsup_{t\to +\infty}\max_{i\in [1,N]}}p_i^t>0$. Then the distance to the set of fixed points of $F$, namely, 
\[
\max_{i,j\in [1,N]}\max\{|x_i^t-x_j^t|,|p_i^t-p_j^t|\},
\]
converges to 0 as $t\to +\infty$.  
\label{REALIST-PRICES}
\end{Pro}
\noindent
The proof is given in Section \ref{S-REALIST-PRICES}. It is unclear if, under the condition ${\displaystyle\varlimsup_{t\to +\infty}\max_{i\in [1,N]}}p_i^t>0$, the orbit itself converges to one of the fixed points, or instead, if it keeps slipping back and forth along this set. 

In addition, notice that while (Hg2) is a mild condition, it may not always hold. Yet, other bounds on prices may apply when (Hg2) fails as the next statement illustrates in the case $N=2$ (see again Section \ref{S-REALIST-PRICES} for a proof).
\begin{Rem}
Let $N=2$, assume that $f$ satisfies {\rm (Hf1-2)} and let $g(x)=ax+bx^2$ where $a\in (0,1]$ and $b\in (\frac{a^2}2,\frac{a}2)$. Then, $g$ satisfies {\rm (Hg1)} but not {\rm (Hg2)}.\footnote{Recall that, for any $a\in (0,1)$ and $b\in (\max\{a-1,-\frac{a}2\},0]$, the map $g(x)=ax+bx^2$ does satisfy (Hg1) and (Hg2).} Moreover, for every orbit $\{(\mathbf x^t,\mathbf p^t)\}_{t\in\N}$ of \eqref{DEFDYNAM}, we have
\[
\inf_{t\in\N}\min\{p_1^t,p_2^t\}>0,
\]
and if ${\displaystyle\varliminf_{t\to +\infty}}\min\{p_1^t,p_2^t\}<+\infty$, then the distance to the set of fixed points of $F$, namely, 
\[
\max\{|x_1^t-x_2^t|,|p_1^t-p_2^t|\}
\]
converges to 0 as $t\to +\infty$.
\label{REALIST-PRICES-N2}
\end{Rem}

\subsection{Emerging bounds on clientele volumes}
As for prices, one may question those potential constraints that the dynamics may generate on clientele volumes. Indeed, given that 0 (respectively 1) is a globally attracting fixed point of $f(\rho,\cdot)$ for every $\rho>1$  (resp.\ $\rho<1$), these volumes could {\sl a priori} approach arbitrarily close the boundaries of the interval $[0,1]$ in the course of time. Evidently, such behaviour would be questionable as far as economic relevance is concerned. Hence, another sanity check of the dynamics is to make sure that clientele volumes remain at moderate values inside the interval, away from 0 and 1. 

\paragraph{Emerging bounds on mean volumes.}
Prior to considering the individual fractions $x_i^t$, bounds on the mean value $\langle \mathbf x^t\rangle$, which represents an overall measure of clientele volumes, can be investigated firstly. This is the purpose of the upcoming statements below, which consider separately the cases $N=2$ and $N>2$. 

The result in the first case relies on the symmetry assumption (Hf3).\footnote{See however Appendix \ref{A-NONSYM} for results in absence of symmetry.} 
It paves the way for the asymptotic stability of the fixed point $(\frac12,\frac12,1)$ of \eqref{BASIS}, to be established later on. In the second case, the symmetry assumption does not matter but we rely instead on the explicit expression \eqref{SPEFAM}. Moreover, the control on $\langle \mathbf x^t\rangle$ is comparatively weaker in this case and, in particular, it is insufficient to address the stability of fixed points.
 \begin{Lem}
Let $N=2$ and assume that $f$ satisfies {\rm (Hf1-3)} and $g$ satisfies {\rm (Hg1)}. Then, the following properties hold for every orbit $\{(\mathbf x^t,\rho_1^t)\}_{t\in\N}$ of \eqref{BASIS}.
\begin{itemize}
\item[$\bullet$] If $\langle \mathbf x^t\rangle=\frac12$ for some $t\in\N$, then $\langle \mathbf x^{t+1}\rangle=\frac12$. 
\item[$\bullet$] If $\langle \mathbf x^t\rangle\neq \frac12$ and $\rho_1^{t+1}\neq 1$ for some $t\in\N$, then $(2\langle \mathbf x^{t+1}\rangle-1)(2\langle \mathbf x^t\rangle-1)> 0$ and $\left|\frac{2\langle \mathbf x^{t+1}\rangle-1}{2\langle \mathbf x^t\rangle-1}\right|<1$.
\end{itemize}
\label{CONSTRAINTS-X-N2}
\end{Lem}
\noindent
The proof is immediate, see beginning of Section \ref{S-CONSTRAINTS-X}.  The Lemma implies that $\langle \mathbf x^t\rangle$ (monotonically) converges as $t\to +\infty$. However,  from (Hf2), we have $f'_x(1,x)=1$ for all $x\in [0,1]$; {\sl viz.}\ the rate at which $|2\langle \mathbf x^t\rangle-1|$ decays vanishes when $\rho_1^t\to 1$. Therefore, the limit of $\langle \mathbf x^t\rangle$ depends on the asymptotic behaviour of $\{\rho_1^t\}$ with respect to 1 and needs not be equal to $\frac12$ when $\rho_1^t$ approaches 1 too fast. 

As far as economic interpretation is concerned, when it holds, convergence of $\langle \mathbf x^t\rangle$ to $\frac12$  in markets where only two sellers compete together means a kind of spontaneous normalization. That is to say, even though these features are not initially prescribed, in the long term, the system approaches a functioning mode in which all buyers do purchase regularly at every market event, and at each event, they only do it at a single seller. 

As announced, the next statement claims that the dynamics of $\langle \mathbf x^t\rangle$ can also be constrained in the case $N>2$, provided that $f$ is given by \eqref{SPEFAM} and some suitable control on the deviation $f_\mathrm{dev}$ holds.
\begin{Pro}
Let $N>2$ be arbitrary, assume that $g$ satisfies {\rm (Hg1)} and $f$ is given by \eqref{SPEFAM} with $f_\mathrm{dev}(\rho,\cdot)\geq 0$ for all $\rho\in \R_\ast^+$, and it satisfies {\rm (Hf1-2)}. Moreover, assume that 
\begin{itemize}
\item $f_\mathrm{dev}(\rho,\cdot)=0$ for all $\rho\in \R_\ast^+$, if $N\in [3,4]$,
\item ${\displaystyle\sup_{x\in (0,1)}}\frac{f_\mathrm{dev}(\rho,x)}{x}<C_N(\rho)$ for all $\rho\in (0,N-1)$ (where $C_N(\rho)$ is known explicitly) if $N\geq 5$.
\end{itemize}
Then, the following properties hold for every orbit $\{(\mathbf x^t,\boldsymbol \rho^t)\}_{t\in\N}$ of \eqref{BASIS}.
\begin{itemize}
\item[(i)] If $N\langle \mathbf x^t\rangle\geq \frac1{N-1}$ (resp.\ $\leq N-\frac1{N-1}$) for some $t\in\N$, then $N\langle \mathbf x^{t+1}\rangle\geq \frac1{N-1}$ (resp.\ $\leq N-\frac1{N-1}$). 
\item[(ii)] If $N\langle \mathbf x^t\rangle<\frac1{N-1}$ and $\boldsymbol\rho^t\neq \mathbf 1$ for some $t\in\N$, then $\langle \mathbf x^{t+1}\rangle>\langle \mathbf x^t\rangle$.
\item[(iii)] If $N\langle \mathbf x^t\rangle>N-\frac1{N-1}$ and $\boldsymbol\rho^t\neq \mathbf 1$ for some $t\in\N$, then $\langle \mathbf x^{t+1}\rangle<\langle \mathbf x^t\rangle$.
\end{itemize}
\label{CONSTRAINTS-X}
\end{Pro}
\noindent
The proof is given in Section \ref{S-CONSTRAINTS-X} and the expression of $C_N(\rho)$ is given at the end of that section. The constraints on $f$ in this statement are a consequence of the approach in the proof. Yet, we believe that the results hold in more general cases, and in particular for nonlinear maps $f$ in the case $N\in [3,4]$.\footnote{For $N\in [3,4]$, the lower bound in statement {\em (i)} and statements {\em (ii)} and {\em (iii)} actually hold for some maps of the form \eqref{SPEFAM} with $f_\mathrm{dev}(\rho,\cdot)\neq 0$, see the statements in Section \ref{S-CONSTRAINTS-X}.}

Proposition \ref{CONSTRAINTS-X} implies that if $\langle \mathbf x^t\rangle$ belongs to $[\frac1{N(N-1)},1-\frac1{N(N-1)}]$ for some $t\in\N$, then it remains in this interval at all future times. In addition, if it never enters this interval, then it must monotonically converge, moving away from the boundaries of $[0,1]$. Even though the conclusion on ${\displaystyle\lim_{t\to +\infty}} \langle \mathbf x^t\rangle$ is not as stringent as in the case $N=2$, it shows that, for any number of competing sellers in the market, a kind of (soft) spontaneous normalisation emerges in long term.

\paragraph{Emerging bounds on individual fractions.}
The properties of the mean volume $\langle \mathbf x^t\rangle$ given by Lemma \ref{CONSTRAINTS-X-N2} and Proposition \ref{CONSTRAINTS-X} constraint the individual fractions $x_i^t$. In particular, it is impossible that they all simultaneously approach the boundaries of $[0,1]$.  The next statement shows that all of them must remain away from these boundaries. In other words, no seller can ever capture an arbitrarily large fraction of the buyer population and, conversely, no seller's clientele volume can decrease below a certain threshold. 
\begin{Cor}
Assume that either $N=2$ and the conditions of Lemma \ref{CONSTRAINTS-X-N2} hold, or $N>2$ and the conditions of Proposition \ref{CONSTRAINTS-X} hold. Then, for every orbit $\{\mathbf x^t,\boldsymbol\rho^t\}_{t\in\N}$, there exists $\epsilon\in (0,1)$ such that 
\[
\epsilon\leq x_i^t\leq 1-\epsilon,\ \forall i\in [1,N-1],t\in\N.
\]
\label{BOUNDED-X}
\end{Cor}
\noindent
For the proof, see Section \ref{S-BOUNDED-X}. In addition, the existence of fixed points with $x_i=x$ for any $x\in [0,1]$ arbitrary indicates that no uniform, orbit independent, bounds on the variables $x_i^t$ can hold. One cannot {\sl a priori} exclude that, depending on the initial condition, the clientele fraction at each merchant remains arbitrarily small, or arbitrarily close to 1. However, when focus is made on the differences between these fractions, a uniform bound applies at large times, as claimed in the next statement. 
\begin{Pro}
Let $N\geq 2$ be arbitrary and assume that the conditions of Proposition \ref{UNIFORM-BOUNDS-PRICESRAT} hold. Then, there exists $\epsilon_N\in (0,1)$ such that, for every orbit $\{(\mathbf x^t,\boldsymbol\rho^t)\}_{t\in\N}$, there exists $t'\in\N$ so that 
\[
\max_{i\in [1,N-1]}|x_i^t-x_N^t|\leq \epsilon_N, \ \forall t\geq t'.
\]
\label{UNIFORM-BOUNDS}
\end{Pro}
\noindent
(Sub-optimal) Estimates of the bound $\epsilon_N$ can be deduced from the proof, which is given in Section \ref{S-UNIFORM-BOUNDS}. It is worth noticing that this statement only relies on the assumptions of Proposition \ref{UNIFORM-BOUNDS-PRICESRAT}, and not on those of Lemma \ref{CONSTRAINTS-X-N2} or Proposition \ref{CONSTRAINTS-X}. Together with Proposition \ref{UNIFORM-BOUNDS-PRICESRAT}, Proposition \ref{UNIFORM-BOUNDS} claims that if the rate at which the sellers change their prices is bounded (adequately depending on $N,\alpha$ and $f$), then while there is no {\sl a priori} blatant dissipation in this system, the dynamics actually forces the price ratios and the differences between clientele fractions, and hence the distance to the set of fixed points, to be uniformly bounded at large times, which can be interpreted as a kind of weak form of long-term homogenisation in this system.   

\subsection{Perpetual crossings of fractions and prices}
The negative feedback induced by the sellers' reactions to clientele changes is the main cause of the above bounds on prices and clientele volumes. Besides, negative feedback loops are known to cause oscillatory behaviours, as it has been identified in the literature, especially in the modelling of genetic regulation, see {\sl e.g.}\ \cite{G98,KST07,PMO95}.

It follows from the next statement that oscillations do take place here, as a consequence of the fact the seller with maximum/minimum price or clientele fraction must change infinitely often. In other words, no seller can be forever more competitive (or less competitive) than its competitors and no seller can have forever more clientele (or less clientele) than its competitors (NB: this comment does not consider those orbits in which prices or clientele volumes can become indistinguishable, even temporarily.)
\begin{Pro}
Let $N\geq 2$ be arbitrary and assume that $f$ satisfies the assumptions {\rm (Hf1-2)} and $g$ satisfies {\rm (Hg1)}. 
%
Let $\{(\mathbf x^t,\mathbf p^t)\}_{t\in\N}$ be a trajectory for which, for each $t\in\N$, all coordinates of $\mathbf x^t$ and of $\mathbf p^t$ differ. Given an arbitrary $t\in\N$, define
 \[
 \underline{i}^t_{\mathbf x}=\arg\min_{i\in [1,N]}x_i^t\quad\text{and}\quad \overline{i}^t_{\mathbf x}=\arg\max_{i\in [1,N]}x_i^t,
 \]
 and define $\underline{i}^t_{\mathbf p}$ and $\overline{i}^t_{\mathbf p}$ similarly. Then, these 4 quantities must change infinitely often. That is to say, for every $t\in\N$, there exists $t'>t$ such that 
 \[
\underline{i}^{t'}_{\mathbf x}\neq \underline{i}^t_{\mathbf x},
\]
and a similar statement holds for $\overline{i}^t_{\mathbf x},\underline{i}^t_{\mathbf p}$ and $\overline{i}^t_{\mathbf p}$.
\label{ALTERNATIONS}
 \end{Pro}
 \noindent
 The proof is given in Section \ref{S-ALTERNATIONS}. Notice that in the simple case $N=2$ of two sellers, this statement implies perpetual crossings of the variables $x_1^t$ and $x_2^t$, as well as of the price variables $p_1^t$ and $p_2^t$.
 
\subsection{Long term equilibration in the case $N=2$} 
Given the conclusion of Proposition \ref{ALTERNATIONS}, a natural subsequent question is whether or not the amplitude of the oscillations remains bounded from below or if the oscillations are asymptotically damped and the prices and the fractions equilibrate in the long term. 

In the case $N=2$, in which the variables $x_1^t$ and $x_2^t$, as well as $p_1^t$ and $p_2^t$, must cross infinitely often according to Proposition \ref{ALTERNATIONS}, this question is particularly relevant as far as economic interpretation is concerned. Indeed, if one can prove that the difference $x_1^t-x_2^t$ asymptotically vanishes and the ratio $\rho_1^t$ converges to 1, 
then, together with the convergence of $\langle \mathbf x^t\rangle$, this would imply that the whole system converges to a fixed point, {\sl viz.}\  spontaneous convergence to a steady state takes place in the long term.  

Accordingly and for simplicity, assuming the symmetry (Hf3) -- which ensures that $f(1,\frac12)=\frac12$ and hence that $(\frac12,\frac12,1)$ is a fixed point of $F_\mathrm{skew}$ -- focus is made here on the behaviour of $(x_1^t-x_2^t,\rho_1^t)$ when close enough to $(0,1)$ and when $\langle \mathbf x^t\rangle$ is close enough to $\frac12$. 
In order to set the context, assume for now that $x\in (0,1)$ is a fixed point of $f(1,\cdot)$ (which is assumed to be differentiable) and we have $f'_x(1,x)=1$ (NB: Both assumptions evidently hold for all $x\in (0,1)$ when $f(1,\cdot)=\mathrm{Id}$). Then, direct computations show that the Jacobian matrix of $F_\mathrm{skew}$ at $(x,x,1)$ reads
%
\[
\left(\begin{array}{ccc}
1+2(1-\alpha)f'_\rho(1,x)g'(0)&-2(1-\alpha)f'_\rho(1,x)g'(0)&(1-\alpha)f'_\rho(1,x)\\
-2(1-\alpha)f'_\rho(1,x)g'(0)&1+2(1-\alpha)f'_\rho(1,x)g'(0)&-(1-\alpha)f'_\rho(1,x)\\
2g'(0)&-2g'(0)&1
\end{array}\right)
\]
Clearly, the vector $(1,1,0)$ is an eigenvector of this matrix and the corresponding eigenvalue is equal to one. 
More interestingly, the matrix has two other eigenvalues $\lambda_\pm(x)$, which  can be specified as follows
\[
\lambda_+(x)\lambda_-(x)=1\quad\text{and}\quad \lambda_+(x)+\lambda_-(x)=2(1+2(1-\alpha)f'_\rho(1,x)g'(0)).
\]
If the assumptions (Hf1-2) and (Hg1) hold, then we must have $f_{\rho}'(1,x) g'(0)\leq 0$, and hence $ \lambda_+(x)+\lambda_-(x)\leq 2$. As a consequence, in the plane transverse to $(1,1,0)$, the linear stability of $(x,x,1)$ falls into one of the following cases
\begin{itemize}
\item parabolic ({\sl ie.}\ $\lambda_\pm(x)=1$) if $f'_\rho(1,x) g'(0)=0$,
\item elliptic ({\sl ie.}\ $|\lambda_\pm(x)|=1$ but $\lambda_\pm(x)\neq 1$) if $f'_\rho(1,x) g'(0)\neq 0$ and $(1-\alpha)f'_\rho(1,x) g'(0)\geq -1$,
\item hyperbolic with one unstable direction ({\sl ie.}\  $\lambda_-(x)<-1<\lambda_+(x)<0$), if $f'_\rho(1,x) g'(0)\neq 0$ and $(1-\alpha)f'_\rho(1,x) g'(0)<-1$.
\end{itemize}
The expression of the Jacobian (actually, the resulting Jacobian after an appropriate change of variable, see equation \eqref{JACOB} below) shows that, if parabolic, the fixed point $(x,x,1)$ must be unstable when $g'(0)\neq 0$. This condition is satisfied generically when (Hg1) holds. Hence, we leave aside the case of parabolic fixed points and investigate the stability only in the elliptic case. 

Notice that when $f'_\rho(1,x) <0$, the point $(x,x,1)$ must be elliptic when the loyalty coefficient $\alpha$ is close enough to 1 or when the sensitivity $g'(0)$ of the sellers' reaction is sufficiently small (but does not vanish). 

\begin{figure}[ht]
\begin{center}
\includegraphics*[width=70mm]{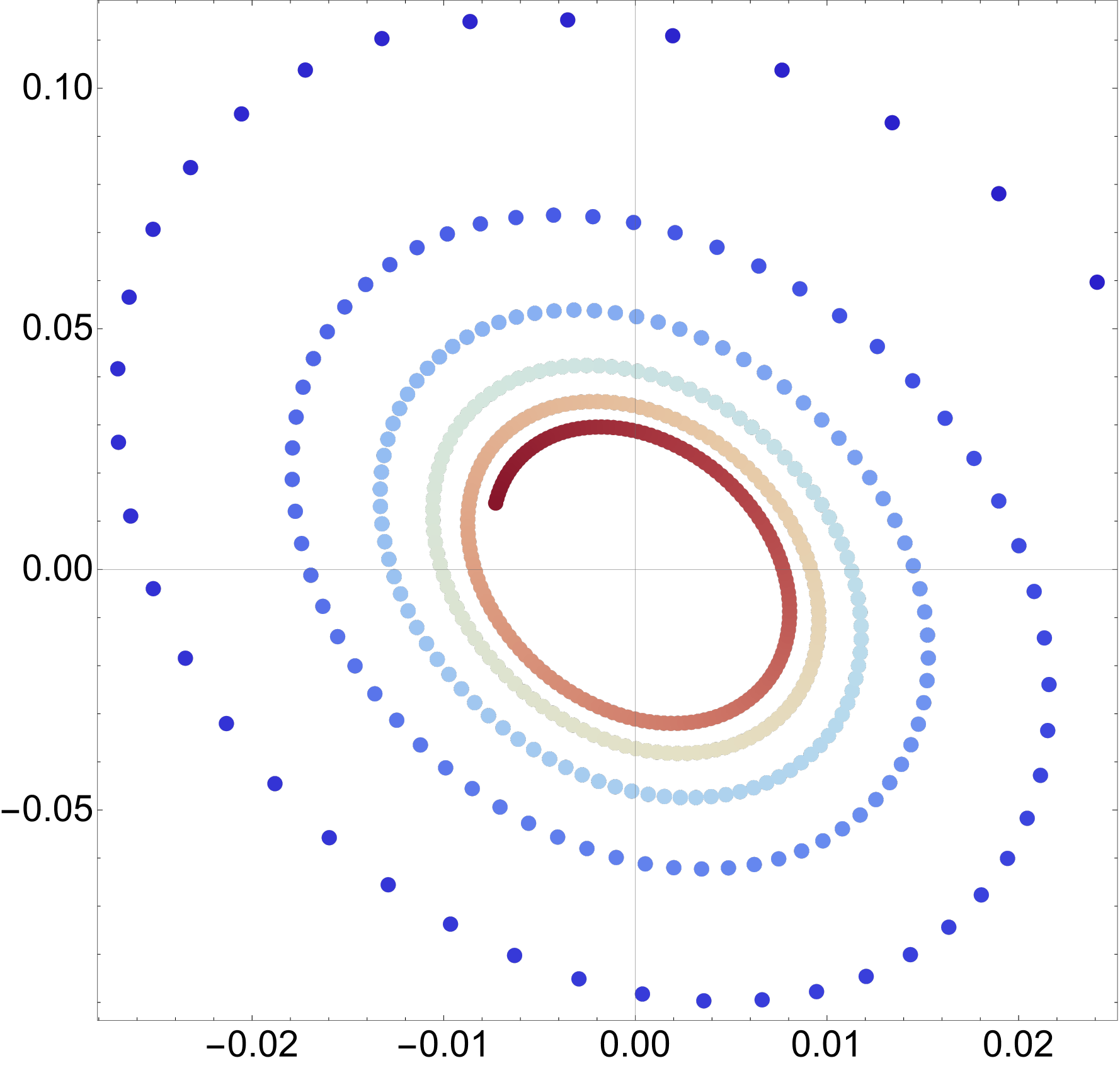}
\end{center}
\caption{Illustration of Theorem \ref{STABILITY}. Plot of fragments of an orbit for $f$ as in Fig.\ \ref{GRAPHF-SMOOTH} and $g(x)=\frac{x}2$. The parameter $\alpha$ has been chosen so that $\lambda_\pm(\frac12)=e^{\frac{i\pi}6}$ (NB: For such $f$, we actually have $\lambda_\pm(x)=\lambda_\pm(\frac12)$ for each $x\in [x_0,1-x_0]$). The dots represent points of coordinates $(x_1^{12 k}-x_2^{12 k},\ln \rho_1^{12 k})$ for $k\in [1,500]$. Initial condition: $x_1^0=0.51,x_2^0=0.48$ and $\rho_1^0=1.1$. Color code: Dark blue for $k=1$ to dark red for $k=500$. The picture clearly illustrates the convergence to $(0,0)$ as a consequence of the Theorem. Notice also that, due the nonlinear effects, the rotation angle slightly deviates from the one, namely $\frac{\pi}6$, of the linearized dynamics in the neighbourhood of the fixed points (NB: had the dynamics be that linearized one, the dots would have been aligned on a ray issued from the origin); yet, as expected, the spiralling drift is smaller as the dots come closer to the origin.}
\label{ILLUSTR-TRAJEC}
\end{figure}

Unlike when the spectrum of the Jacobian matrix lies inside the unit disk, the stability of an elliptic fixed point depends on higher-order terms of the corresponding Taylor expansion, and of the $3^\mathrm{rd}$-order coefficients of the normal form in particular \cite{L73,W78}. 
A normal-form-like computation of the map $F_\mathrm{skew}$ in the neighbourhood of a fixed point $(\frac12,\frac12,1)$ yields the following convergence result, which culminates our investigation of the dynamics for $N=2$.
\begin{Thm}
Let $N=2$ and assume that the following conditions holds

\begin{itemize}
\item there exist $\rho_0>1$ and $x_0\in (0,\frac12)$ such that $f\in C^4(U)$ and $|f'_x(\rho,x)|\leq 1$ for all $(\rho,x)\in U$, where $U=[\frac1{\rho_0},\rho_0]\times [x_0,1-x_0]$,
\item the map $f$ satisfies {\rm (Hf3)} and we have 
\begin{equation}
f'_x(1,\frac12)=1,\ f'_\rho(1,\frac12)<0,\ f'''_{\rho x^2}(1,\frac12)=0\quad\text{and}\quad f'''_{\rho^2 x}(1,\frac12)<0,
\label{CONDERIV}
\end{equation}
\item $g\in C^4([-2x_0,2x_0])$ and $g$ is an odd function that satisfies {\rm (Hg1)} and $g'(0)>0$,
\item the fixed point $(\frac12,\frac12,1)$ is transversally elliptic and $\lambda_\pm(\frac12)\neq -1$.
\end{itemize}
Then, there exists a neighbourhood of $(\frac12,\frac12,1)$ in $[0,1]^2\times \R_\ast^+$ such that for every initial condition in this neighbourhood, we have for the subsequent orbit of \eqref{BASIS}
\[
\lim_{t\to +\infty} (x_1^t-x_2^t,\rho_1^t)=(0,1).
\]
\label{STABILITY}
\end{Thm}
For the proof, see Section \ref{S-STABILITY}. An illustration of the long-time behaviours ensured by this statement is given on Figure \ref{ILLUSTR-TRAJEC}. 

Notice that Theorem \ref{STABILITY} is a local result that does not require the global assumptions (Hf1-2). Instead, it requires that $f$ satisfies a certain regularity together with some constraints on its derivatives, which appear to be generic when (Hf1-3) hold. For the sake of completeness, an example of map $f$ that satisfies all assumptions at once ({\sl ie.}\ (Hf1-3) and those in Theorem \eqref{STABILITY}) will be given immediately below. 

Notice also from the equation on $\lambda_\pm(x)$, that when $f'_\rho(1,\frac12)<0$ and $g'(0)>0$, the point $(\frac12,\frac12,1)$ is certainly elliptic and $\lambda_\pm(\frac12)\neq -1$ when $\alpha$ is close enough to 1, say when in some interval $(\alpha_\ast,1)$. 

Furthermore, the proof shows that the condition $f''_{\rho x}(1,\frac12)=0$ (which is a consequence of the symmetry (Hf3)) and $f'''_{\rho x^2}(1,\frac12)=0$ in \eqref{CONDERIV} are necessary for stability, {\sl ie.}\ even though elliptic, the point $(\frac12,\frac12,1)$ is unstable when (at least) one of these condition fails.


Finally, when all the assumptions are simultaneously satisfied, the combination of Lemma \ref{CONSTRAINTS-X-N2} and Theorem \ref{STABILITY} immediately imply that, when starting sufficiently close to $(\frac12,\frac11,1)$, the orbit of $F_\mathrm{skew}$ must asymptotically converge to a fixed point, as formally claimed now.
\begin{Cor}
Let $N=2$ and assume that f satisfies (Hf1-2) and $f$ and $g$ satisfy the conditions of Theorem \ref{STABILITY}. Then, there exists a neighbourhood of $(\frac12,\frac12,1)$ in $[0,1]^2\times \R_\ast^+$ such that for every initial condition in this neighbourhood, the subsequent orbit converges to a fixed point.
\label{CORTHM}
\end{Cor}

\paragraph{A map $f$ that complies all assumptions at once.}
Following Corollary \ref{CORTHM}, we aim at an example of map that simultaneously satisfies (Hf1-2) and the conditions in the Theorem. 

Evidently, a natural candidate would be the map $f(\rho,x)$ given by \eqref{LINEARMAP}. Provided that $c$ is itself differentiable, for every $x\in (0,1)$, such map $f(\rho,x)$ is a differentiable function of $\rho$ at all $\rho\neq 1$. Moreover, it is also differentiable at $\rho=1$ iff $c'(1)=0$. However, we have $f'_\rho(1,x)=0$ for all $x\in [0,1]$ in this case; thus every fixed point is transversally parabolic. Therefore, no map of the example \eqref{LINEARMAP} can satisfy the conditions in Theorem \ref{STABILITY}.

Nevertheless, one can amend this example, especially if one requires that $f(\rho,x)$ be differentiable at $\rho=1$ only for $x$ in a neighbourhood of $\frac12$, so that it simultaneously satisfies (Hf1-3) and the conditions in \eqref{CONDERIV}. The construction proceeds as follows.
\begin{itemize}
\item[1.] We determine $f(\rho,x)$ only for $\rho\geq 1$ and deduce the expression for $\rho\leq 1$ by symmetry. In doing so, we must ensure that, for $\rho=1$, the first expression (respectively its derivatives) match the second one (resp.\ the corresponding derivatives).
\item[2.] Let $\rho_0>1$ be arbitrary and let $c\in C^4([1,\rho_0])$ be a function as in the example \eqref{LINEARMAP} such that $c'(1)=0$ and $c''(1)<0$. For instance, one can choose $c(\rho)=e^{-(\ln \rho)^2}$.
\item[3.] For $\rho> \rho_0$, let $f(\rho,x)=c(\rho)x$ for all $x\in [0,1]$, as in \eqref{LINEARMAP}. 
\item[4.] Let $x_0\in (0,\frac12)$ be arbitrary. For $\rho\in [1,\rho_0]$ and $x\in [x_0,1]$, let $f(\rho,x)=\frac12-b(\rho)+c(\rho)(x-\frac12)$ where $b\in C^4([1,\rho_0])$ is increasing and satisfies $b(1)=0$, $b'(1)>0$, $b(\rho_0)=\frac{1-c(\rho_0)}2$, as well as the following inequality
\[
b'(\rho)+c'(\rho)x_0\geq 0,\ \forall \rho\in [1,\rho_0].
\] 
For instance, one can choose the linear map $b(\rho)=\frac{1-c(\rho_0)}2\frac{\rho-1}{\rho_0-1}$ provided that $x_0$ is sufficiently small, depending on $\rho_0$ and $c$. 
\item[5.] For $\rho\in [1,\rho_0)$ and $x\in [0,x_0]$, $f(\rho,x)$ is a smooth and increasing interpolation between 0 and $f\left(\rho,x_0\right)$ with derivative in $x$ not larger than 1. Such interpolation clearly exists. For $\rho=\rho_0$, the condition $b(\rho_0)=\frac{1-c(\rho_0)}2$ implies that we can choose $f(\rho_0,x)=c(\rho_0)x$ for all $x\in [0,1]$.
\end{itemize}
\begin{figure}[ht]
\begin{center}
\includegraphics*[width=90mm]{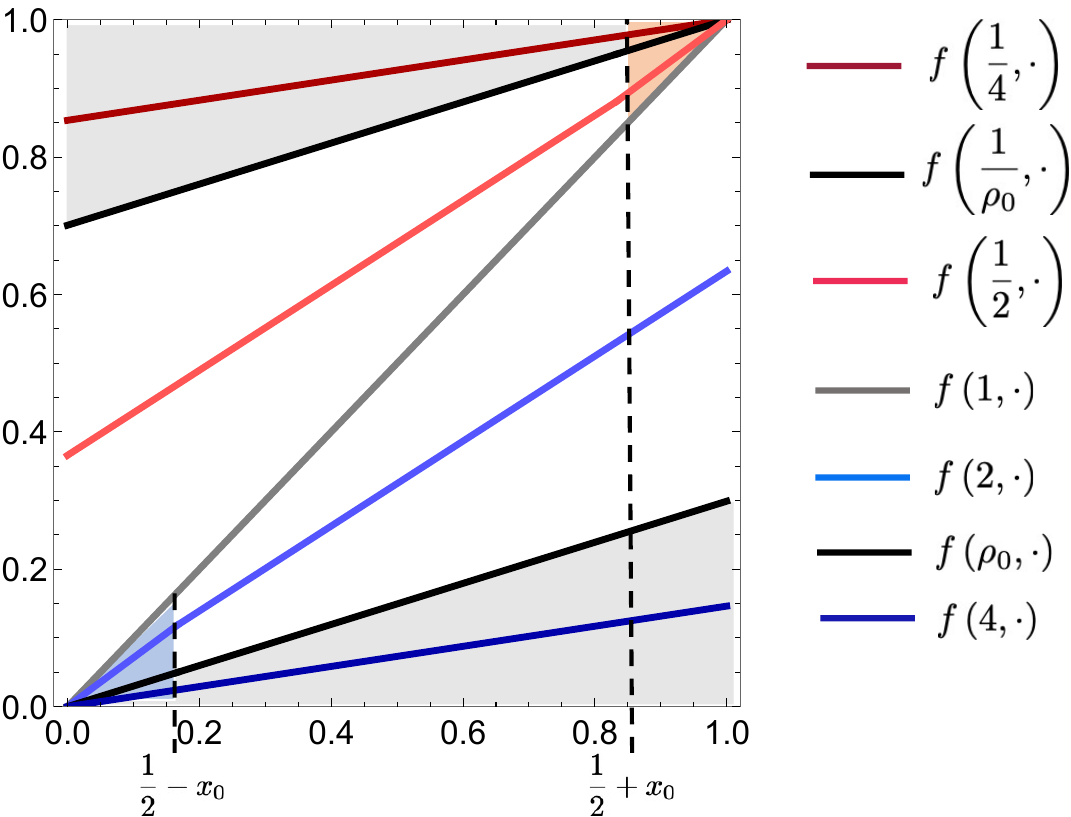}
\end{center}
\caption{Illustration of $f(\rho,\cdot)$ for the example that satisfies all conditions at once with $\rho_0=3$, $c(\rho)=e^{-(\ln \rho)^2}$, $x_0=\frac13$ and $b(\rho)=\frac{1-c(\rho_0)}4(\rho-1)$. The lower (resp.\ upper) grey region is where $f$ coincides with the linear map in the example \ref{LINEARMAP} for $\rho\geq 3$ (resp.\ $\rho\leq \frac13$). The left blue triangle is the domain $(\rho,x)\in [1,\rho_0)\times [0,\frac12-x_0]$ where the map smoothly interpolates between 0 and $f\left(\rho,\frac12-x_0\right)$ and similarly for the upper red triangle. In the remaining domains $[1,\rho_0)\times [\frac12-x_0,1]$ and $[\frac1{\rho_0},1]\times [0,\frac12+x_0]$ the map is affine and defined in a way so that it is of class $C^4$ in $U$.}
\label{GRAPHF-SMOOTH}
\end{figure}

We check that $f$ satisfy (Hf1-3) and the assumptions in the Theorem. Evidently, such a map satisfies (Hf1) and (Hf3). Moreover, the inequality on the derivative of $b$ and the fact that $c'(\rho)\leq 0$ for $\rho\geq 1$ ensure that $f$ is a decreasing function of $\rho\in [1,\rho_0]$ for $x\in [x_0,1]$. In addition, the interpolations for $x\in [0,x_0]$ can be chosen to be monotonous in $\rho$ too. For instance, one can define such interpolations for a finite set in $[1,\rho_0]$ and then linearly interpolate them for the remaining values of $\rho$. The other conditions of (Hf2) are easy to check.

As far as the conditions in the Theorem are concerned, the fact that $b$ and $c$ are of class $C^4$ and the linear dependence on $x$ ensure that $f$ is $C^4$ on $(1,\rho_0)\times [x_0,1-x_0]$. Moreover, the expression of $f$, the symmetries of $b$ and $c$ and the linear dependence on $x$ through $x-\frac12$ ensures that all derivatives match for $\rho=1$, {\sl ie.}\ $f\in C^4(U)$. Finally, using in particular that $c'(1)=0$, we have 
\[
f'_x(1,\frac12)=c(1)=1,\ f'_\rho(1,\frac12)=-b'(1)<0,\ f'''_{\rho x^2}(1,\frac12)=0\quad\text{and}\quad f'''_{\rho^2x}(1,\frac12)=c''(1)<0.
\]

\section{Proofs}
\subsection{Proof of Proposition \ref{BOUNDED-RHO}}\label{S-BOUNDED-RHO}
It suffices to prove that 
\[
\varlimsup_{t \to +\infty} \rho_i^t < +\infty,\ \forall i\in [1,N-1],
\]
because then, the permutation symmetry of the dynamics in the variables $(\mathbf x,\mathbf p)$ implies that we must have
\[
\varliminf_{t \to +\infty} \rho_i^t >0,\ \forall i\in [1,N-1].
\]
By contradiction, suppose that there exist $i' \in [1,N-1]$ and a subsequence $\{t_k\}_{k \in \N}$ such that ${\displaystyle\lim_{k \to +\infty}}\rho_{i'}^{t_k} = +\infty$. Let us show first that there exists $i'' \in [1,N-1]$ such that 
\[
\varlimsup_{k \to +\infty} \frac{\rho_{i''}^{t_k}}{\langle \boldsymbol\rho^{t_k}\rangle_{i''}^c} > 1.
\]
By contradiction, assume that for every $\epsilon>0$ there exists $k_\epsilon\in\N$ such that 
\[
\frac{\rho_{i}^{t_k}}{\langle \boldsymbol\rho^{t_k}\rangle_{i}^c} \leq 1+\epsilon,\ \forall i\in [1, N-1], k\geq k_\epsilon,
\]
that is to say
\[
(N-1)\rho_{i}^{t_k}\leq (1+\epsilon)\left(1+\sum_{j=1}^{N-1} \rho_j^{t_k}-\rho_{i}^{t_k}\right),\ \forall i\in [1, N-1], k\geq k_\epsilon.
\]
Summing over $i$ yields
\[
\sum_{i=1}^N \rho_i^{t_k}\leq \frac{(1+\epsilon)(N-1)}{N-1-(1+\epsilon)(N-2)},\ \forall k\geq k_\epsilon,
\]
provided that $N-1-(1+\epsilon)(N-2)>0$ {\sl viz.}\ $\epsilon<\frac1{N-2}$. However, this inequality on ${\displaystyle\sum_{i=1}^N} \rho_i^{t_k}$ is incompatible with the fact that ${\displaystyle\lim_{k \to +\infty}}\rho_{i'}^{t_k} = +\infty$; hence the contradiction.

Clearly, the assumptions (Hf1-2) imply that for every infinite sub-subsequence $\{t_{k_\ell}\}_{\ell\in\N}$ such that 
\[
\lim_{\ell\to+\infty} \frac{\rho_{i''}^{t_{k_\ell}}}{\langle \boldsymbol\rho^{t_{k_\ell}}\rangle_{i''}^c} > 1,
\]
we have 
\[
\varlimsup_{\ell\to +\infty}x_{i''}^{t_{k_\ell}}<1.
\]

Now, let us prove independently that ${\displaystyle \lim_{k \to +\infty}}\rho_{i'}^{t_k} = +\infty$ implies that for every infinite sub-subsequence $\{t_{k_\ell}\}_{\ell\in\N}$ such that $\{\mathbf x^{t_{k_\ell}}\}_{\ell\in\N}$ converges, we have
\begin{equation*}
\lim_{\ell \to +\infty} x_{i}^{t_{k_\ell}} = 1,\ \forall i\in [1,N-1],
\end{equation*}
which is obviously incompatible with the previous inequality and concludes the proof. First, the limit ${\displaystyle\lim_{k \to +\infty}}\rho_{i'}^{t_k} = +\infty$ implies
\[
\lim_{k \to +\infty}\langle \boldsymbol\rho^{t_k}\rangle_{N}^c= +\infty.
\]
and the assumptions (Hf1-2) impose that we must have 
\[
\varliminf_{k \to +\infty} x_{N}^{t_k} \geq \varliminf_{k \to +\infty}f_\alpha\left(\frac1{\langle \boldsymbol\rho^{t_k}\rangle_{N}^c},0\right)= 1,
\]
and thus ${\displaystyle \lim_{k \to +\infty}} x_{N}^{t_k}=1$. Moreover, by passing to a sub-subsequence if necessary, we may assume that 
\[
\langle \boldsymbol\rho^{t_k+1}\rangle_{N}^c \geq \langle \boldsymbol\rho^{t_k}\rangle_{N}^c,\ \forall k\in\N,
\]
(NB: If this was not the case, then the limit above could not be $+\infty$). The iteration rule for the $\rho_i^t$ then implies 
\begin{equation*}
\sum_{i =1}^{N-1} \rho_i^{t_k}\left(\frac{1+g\left(x_i^{t_k}-\langle \mathbf x^{t_k}\rangle_i^c\right)}{1+g\left(x_{N}^{t_k}-\langle \mathbf x^{t_k}\rangle_N^c\right)}-1\right)\geq 0,\ \forall k\in\N.
\end{equation*}
and trivially, when letting $\rho_\mathrm{max}^{t} = {\displaystyle\max_{i \in [1,N-1]}} \rho_i^{t}>0$,
\begin{equation}
\sum_{i =1}^{N-1}  \frac{\rho_i^{t_k}}{\rho_\mathrm{max}^{t_k}}\left(\frac{1+g\left(x_i^{t_k}-\langle \mathbf x^{t_k}\rangle_i^c\right)}{1+g\left(x_{N}^{t_k}-\langle \mathbf x^{t_k}\rangle_N^c\right)}-1\right)\geq 0,\ \forall k\in\N.
\label{RATIO-CONV}
\end{equation}
By compactness and by passing (again) to a sub-subsequence if necessary, we may assume that the following limits exist
\[
 \lim_{k \to +\infty} \frac{\rho_i^{t_k}}{\rho_\mathrm{max}^{t_k}}\leq 1\quad \mathrm{and}\quad \lim_{k \to +\infty} x_{i}^{t_k} \leq 1,\ \forall i\in [1,N-1].
\]
We separate two cases and first consider the case of those $i$ such that ${\displaystyle\lim_{k \to +\infty}} \frac{\rho_i^{t_k}}{\rho_\mathrm{max}^{t_k}}=0$. Then we have
\[
 \lim_{k \to +\infty} \frac{\rho_{i}^{t_k}}{\langle \boldsymbol\rho^{t_k}\rangle_{i}^c} \leq  (N-1)\lim_{k \to +\infty} \frac{\rho_i^{t_k}}{\rho_\mathrm{max}^{t_k}}=0
\]
and consequently ${\displaystyle\lim_{k \to +\infty}} x_{i}^{t_k} = 1$ as before. In the other case ${\displaystyle \lim_{k \to +\infty}} \frac{\rho_i^{t_k}}{\rho_\mathrm{max}^{t_k}}>0$, the limit ${\displaystyle\lim_{k \to +\infty}} x_{N}^{t_k} = 1$ and the strict monotonicity of $g$ impose that for such $i\in [1,N-1]$, we must have
\[
\lim_{k \to +\infty} \frac{\rho_i^{t_k}}{\rho_\mathrm{max}^{t_k}}\left(\frac{1+g\left(x_i^{t_k}-\langle \mathbf x^{t_k}\rangle_i^c\right)}{1+g\left(x_{N}^{t_k}-\langle \mathbf x^{t_k}\rangle_N^c\right)}-1\right)\leq 0,
\]
with strict inequality iff ${\displaystyle \lim_{k \to +\infty}} x_{i}^{t_k} < 1$. However, taking the limit $k\to +\infty$ in the inequality \eqref{RATIO-CONV} implies that such strict inequality is impossible; hence we must again have  ${\displaystyle \lim_{k \to +\infty}} x_{i}^{t_k} = 1$ as desired. 
 
\subsection{Proof of Proposition \ref{UNIFORM-BOUNDS-PRICESRAT}}\label{S-UNIFORM-BOUNDS-PRICESRAT}
\paragraph{Proof in the case $N=2$.} We are going to prove that for every orbit $\{(\mathbf x^t,\rho_1^t)\}_{t\in\N}$, there exists $t'\in\N$ so that 
\[
\rho_1^t\leq 2 S_g^{T_2},\ \forall t\geq t'.
\]
Applying the result to the orbit $\{(\pi \mathbf x^t,\pi \mathbf p^t)\}_{t\in\N}$ of \eqref{DEFDYNAM} with permuted coordinates implies that the same inequality also eventually holds for the variable $\frac{(\pi \mathbf p^t)_1}{(\pi \mathbf p^t)_2}=\frac{p_2^t}{p_1^t}=\frac1{\rho_1^t}$, completing the proof of the statement. 

The inequality above will be a direct consequence of the following property. 
\begin{Claim}
There exists $\gamma\in (0,1)$ such that when, in any given orbit, there exists $t\in\N$ such that 
\[
\rho_1^{t+s}>2\ \text{for}\ s\in [1,T_2],
\]
then we must have $\rho_1^{t+T_2+1}<\gamma\rho_1^{t+T_2}$. If, in addition, $\rho_1^{t+T_2+1}>1$, then we have 
\[
\rho_1^{t+T_2+s+1}<\gamma\rho_1^{t+T_2+s},
\]
for $s=1$ and this estimate prevails for all those $s\in\Z^+$ for which $\rho_1^{t+T_2+s}>1$.
\label{OSCILL2}
\end{Claim}
\noindent
{\sl Proof of the Claim.} The definition of $T_2$, the assumption of the claim and the monotonicity of $f$ imply that we have
\[
x_1^{t+T_2}< f_{\alpha}^{T_2}(2,1)<f_{\alpha}^{T_2}(\frac12,0)<x_2^{t+T_2},
\]
and then, using assumption (Hg1)
\[
\rho_1^{t+T_2+1}=\rho_1^{t+T_2}\frac{1+g(x_1^{t+T_2}-x_2^{t+T_2})}{1+g(x_2^{t+T_2}-x_1^{t+T_2})}<\gamma\rho_1^{t+T_2},
\]
as desired, where 
\[
\gamma=\frac{1+g(f_{\alpha}^{T_2}(2,1)-f_{\alpha}^{T_2}(\frac12,0))}{1+g(f_{\alpha}^{T_2}(\frac12,0)-f_{\alpha}^{T_2}(2,1))}<1.
\]
If, in addition, $\rho_1^{t+T_2+1}>1$, then we have
\[
x_1^{t+T_2+1}=f_{\alpha}(\rho_1^{t+T_2+1},x_1^{t+T_2})<x_1^{t+T_2}\quad \text{and}\quad x_2^{t+T_2}< f_{\alpha}(\frac1{\rho_1^{t+T_2+1}},x_2^{t+T_2})=x_2^{t+T_2+1},
\]
so that $\rho_1^{t+T_2+2}<\gamma\rho_1^{t+T_2+1}$ for the same reason as before. This argument repeats as long as $\rho_1^{t+T_2+s}>1$, completing the proof of the Claim. \hfill $\Box$

Claim \ref{OSCILL2} implies that in every orbit, there must exist $t'\in\N$ such that $\rho_1^{t'}\leq 2$. Then the definition of $S_g$ implies that we must have
\[
\rho_1^{t'+s}\leq 2S_g^{T_2}\ \text{for}\ s\in [1,T_2],
\]
as desired. Moreover, if there exists $k\in [1,T_2]$ for which $\rho_1^{t'+k}\leq 2$, then we can extend the inequality above to $t'+s$ for $s\in [k+1,k+T_2]$. Likewise, if there exists $k''\in [1,T_2]$ such that $\rho_1^{t'+k+k''}\leq 2$, we can extend the inequality further. This proves the desired conclusion unless there exists $t''\geq t'$ such that $\rho_1^{t''}\leq 2$ and $\rho_1^{t''+s}>2$ for all $s\in [1,T_2]$.

However, in that case, the fact that $S_g>1$ implies $\rho_1^{t''+s}\leq 2S_g^s\leq 2S_g^{T_2}$ for all $s\in [1,T_2]$. Also, Claim \ref{OSCILL2} implies that $\{\rho_1^{t''+s}\}_{s>T_2}$ must be exponentially decreasing until it passes below 1, {\sl viz.}\ there must exist $t^\dag>t''$ such that $\rho_1^{t^\dag}<1$; hence $\rho_1^{t}\leq 2S_g^{T_2}$ extends (at least) up to $t=t^\dag$. 

Repeating the argument for $\rho_1^{t^\dag}$, and for the potential subsequent values of $t$ at which $\rho_1^{t}<1$, we conclude that $\rho_1^t$ can never exceed $2S_g^{T_2}$ when $t>t'$, as desired. 

\paragraph{Proof in the case $N>2$.}
As before, we are going to prove that for every orbit $\{(\mathbf x^t,\boldsymbol\rho^t)\}_{t\in\N}$, there exists $t'\in\N$ so that 
\[
\max_{i\in [1,N-1]}\rho_i^t\leq N+1,\ \forall t\geq t'.
\]
Applying the result to the orbit $\{(\pi_{(i,N)} \mathbf x^t,\pi_{(i,N)} \mathbf p^t)\}_{t\in\N}$ of \eqref{DEFDYNAM}, where $i\in [1,N-1]$ is arbitrary and $\pi_{(i,N)}$ exchanges the coordinates $i$ and $N$, implies that we also have 
\[
\frac{(\pi_{(i,N)} \mathbf p^t)_i}{(\pi_{(i,N)} \mathbf p^t)_N}=\frac{p_N^t}{p_i^t}=\frac1{\rho_i^t}\leq N+1,
\]
provided that $t$ is sufficiently large, completing the proof of the statement. 

We use the notation
\[
\rho_\text{max}^t=\max_{i\in [1,N-1]}\rho_i^t.
\]
As in the proof for $N=2$, we shall rely on the following property. 
\begin{Claim}
There exists $\gamma_N\in (0,1)$ such that when, in any given orbit, there exist $t\in\N$ and $i\in [1,N-1]$ such that 
\[
\rho_i^{t+s}>N\quad \text{and}\quad \rho_\text{max}^{t+s}\leq N+1\quad \text{for}\ s\in [1,T_N],
\]
then we must have $\rho_i^{t+T_N+1}<\gamma_N\rho_i^{t+T_N}$. If, in addition, $\rho_i^{t+T_N+1}>1$, then we have 
\[
\rho_i^{t+T_N+s+1}<\gamma_N\rho_i^{t+T_N+s},
\]
for $s=1$ and this estimate prevails for all those $s\in\Z^+$ for which $\rho_i^{t+T_N+s}>1$.
\label{OSCILLN}
\end{Claim}
\noindent
{\sl Proof of the Claim.} The initial assumption implies 
\[
\frac{\rho_i^{t+s}}{\langle\boldsymbol\rho^{t+s}\rangle_i^c}\geq \frac{N(N-1)}{1+(N-2)(N+1)}>1\quad\text{and}\quad \frac1{\langle\boldsymbol\rho^{t+s}\rangle_N^c}\leq \frac{N-1}N,\ \text{for}\ s\in [1,T_N].
\]
The definition of $T_N$ and the monotonicity of $f$ imply that we have
\[
x_i^{t+T_N}< f_\alpha^{T_N}(\frac{N(N-1)}{1+(N-2)(N+1)},1)<f_\alpha^{T_N}(\frac{N-1}{N},0)<x_N^{t+T_N},
\]
and then, using (Hg1)
\[
\rho_i^{t+T_N+1}=\rho_i^{t+T_N}\frac{1+g\left(x_i^{t+T_N}-\langle \mathbf x^{t+T_N}\rangle_i^c\right)}{1+g\left(x_N^{t+T_N}-\langle \mathbf x^{t+T_N}\rangle_N^c\right)}<\gamma_N\rho_i^{t+T_N},
\]
as desired, where
\[
\gamma_N=\max_{x\in \left[0,\frac{N-2}{N-1}\right]}\frac{1+g\left(f_\alpha^{T_N}(\frac{N(N-1)}{1+(N-2)(N+1)},1)-\frac1{N-1}f_\alpha^{T_N}(\frac{N-1}N,0)-x\right)}{1+g\left(f_\alpha^{T_N}(\frac{N-1}N,0)-\frac1{N-1}f_\alpha^{T_N}(\frac{N(N-1)}{1+(N-2)(N+1)},1)-x\right)},
\]
and $\gamma_N<1$ because of strict monotonicity of $g$ and of the compactness of the interval in which  the maximum is taken. The rest of the proof is the same as in the proof of Claim \ref{OSCILL2}. \hfill $\Box$

In order to prove the statement in the case $N>2$, we first observe that if $\rho_\text{max}^{t}\leq N$ for some $t\in\N$, then the definition of $S_g$, the fact that $S_g>1$ and the condition in the statement imply that we must have
\[
\rho_\text{max}^{t+s}\leq NS_g^s\leq NS_g^{T_N}\leq N+1\ \text{for}\ s\in [1,T_N],
\]
as desired. Moreover, if $i\in [1,N-1]$ is such that there exists $k_i\in [1,T_N]$ for which $\rho_i^{t+k_i}\leq N$, then we are sure that $\rho_i^{t+T_N+1}\leq N+1$. Otherwise, if $\rho_i^{t+s}>N$ for $s\in [1,T_N]$, then Claim \ref{OSCILLN} implies that $\rho_i^{t+T_N+1}\leq N+1$ and the sequence $\{\rho_i^{t+s}\}_{s\geq T_N}$ must decay exponentially up until $\rho_i^{t+s}$ passes below 1. As before, we can propagate these estimates to obtain the desired conclusion.

It remains to prove that in every orbit, there exists $t\in\N$ such that $\rho_\text{max}^{t}\leq N$. By contradiction, assume that $\rho^t_\text{max}>N$ for all $t\in\N$. Then, we certainly have $\frac1{\langle \boldsymbol\rho^t\rangle_N^c}\leq \frac{N-1}{N}$ for all $t\in\N$ and hence $\{x_N^t\}_{t\in\N}$ converges exponentially fast to 1. 

Given $i\in [1,N-1]$, consider the subset of times defined by
\[
N_i=\left\{t\in\N\ :\ \rho_i^t=\rho_\text{max}^t>N\right\}.
\]
Let $i\in [1,N-1]$ be such that $\# N_i=+\infty$ and let $(\mathbf x^\infty,\boldsymbol\rho^\infty)$ be an accumulation point of the subsequence $\{(\mathbf x^t,\boldsymbol\rho^t)\}_{t\in N_i}$. We certainly have $x_N^\infty=1$ and $\rho_i^\infty=\rho_\text{max}^\infty>N$. Thus 
\[
\frac{\rho_i^\infty}{\langle\boldsymbol\rho^\infty\rangle_i^c}>\frac{N(N-1)}{1+(N-2)N}=\frac{N}{N-1},
\]
and therefore
\[
x_i^\infty\leq f_\alpha(\frac{N}{N-1},1)<1,
\]
meaning that for every $\epsilon>0$, there exists $t_\epsilon\in N_i$ such that we have 
\begin{equation}
x_i^t<f_\alpha(\frac{N}{N-1},1)+\epsilon,\ \forall t\in N_i,t>t_\epsilon.
\label{ACCUMUL}
\end{equation}

Now consider the orbit $\{(\mathbf \varkappa^t,\boldsymbol\varrho^t)\}_{t\in\N}$ with initial condition $(\mathbf \varkappa^0,\boldsymbol\varrho^0)=(\mathbf x^\infty,\boldsymbol\rho^\infty)$. Each coordinate of the map $F_\mathrm{skew}$ is a continuous function of each of its variables, and the derivative with respect to its proper coordinate is uniformly bounded (because so are the derivative of $x\mapsto f_\alpha(\rho,x)$ for every $\rho$ and the derivative of $\rho\mapsto \rho\frac{1+g(x)}{1+g(y)}$ for every $(x,y)$). Hence, each coordinate of $F_\mathrm{skew}$ is a continuous function of $(\mathbf x,\boldsymbol\varrho)$. Therefore, for every $t\in\N$, $(\mathbf \varkappa^t,\boldsymbol\varrho^t)$ is an accumulation point of the original orbit $\{(\mathbf x^t,\boldsymbol\rho^t)\}_{t\in\N}$. Therefore, we must have $\varkappa_N^t =1$ for all $t\in\N$.

Together with this equality, the iterations \eqref{BASIS} imply that $\{\varrho_i^t\}_{t\in\N}$ must be a decreasing sequence. Therefore, this sequence must converge and, by Proposition \ref{BOUNDED-RHO}, its limit must be positive. It follows that the ratio sequence $\{\frac{\varrho_i^{t+1}}{\varrho_i^t}\}_{t\in\N}$ must converge to 1, which in turn, by (Hg1), implies the following limit 
\[
\lim_{t\to +\infty}\varkappa_i^t=1.
\]
As a consequence, for every $\epsilon>0$, there exist $t'_\epsilon\in N_i$ and $k_\epsilon\in\N$ such that 
\[
x_i^{t+k}>1-\epsilon,\ \forall t\in N_i,t>t'_\epsilon,\ \text{and}\ k>k_\epsilon.
\]
Clearly, when $\epsilon$ is small enough and $t\in N_i,t>t'_\epsilon$ and $k>k_\epsilon$ are such that $t+k\in N_i$ and $t+k>t_\epsilon$, this inequality contradicts the inequality \eqref{ACCUMUL}. Hence, it is impossible that $\rho^t_\text{max}>N$ for all $t\in\N$. 

\subsection{Proof of Proposition \ref{REALIST-PRICES} and Remark \ref{REALIST-PRICES-N2} }\label{S-REALIST-PRICES}
\paragraph{Proof of Proposition \ref{REALIST-PRICES}.} Repeating {\sl mutatis mutandis} the arguments in the proof of statements {\em (i)} and {\em (ii)} of Proposition 3.6 in \cite{EF24} yields that the product ${\displaystyle\prod_{i=1}^N} p_i^t$ is non-increasing with $t$ and that if it does not converge to 0 as $t\to +\infty$, then we must have
\begin{equation}
\lim_{t\to +\infty}\max_{i,j\in [1,N]}|x_i^t-x_j^t|=0.
\label{SYNCHRO}
\end{equation}
Together with Proposition \ref{BOUNDED-RHO}, monotonicity implies that no sequence $\{p_i^t\}$ can diverge. Moreover, the same statement ensures that the condition ${\displaystyle\varlimsup_{t\to +\infty}\max_{i\in [1,N]}}p_i^t>0$ suffices to make sure that ${\displaystyle\prod_{i=1}^N} p_i^t$ converges to a positive number. It follows that 
\[
\varliminf_{t\to +\infty} p_i^t > 0,\ \forall i\in [1,N].
\]
Therefore, for every infinite subsequence $\{t_k\}_{k \in \N}$ such that 
\[
\lim_{k \to +\infty} \mathbf{p}^{t_k} = \mathbf{p}^{\infty},
\]
we must have $\mathbf p^\infty\in (\R_\ast^+)^N$. 
Moreover, the coordinates $p_i^\infty$ must be independent $i$, {\sl viz.}\ we must have 
\[
\lim_{k\to +\infty}\frac{p_i^{t_k}}{\langle \mathbf p^{t_k}\rangle_i^c}=1,\ \forall i\in [1,N],
\]
otherwise \eqref{SYNCHRO} could not hold, because one of these limits would have to be less than 1 and another one would have to be larger than 1. This proves that we must also have
\[
\lim_{t\to +\infty}\max_{i,j\in [1,N]}|p_i^t-p_j^t|=0,
\]
completing the proof of the Proposition.

\paragraph{Proof of Remark \ref{REALIST-PRICES-N2}.}  It is easy to check that the condition $b<\frac{a}2$ implies that $g$ increases on $(-1,1)$ and more generally that it satisfies (Hg1). That $b>0$ implies that it cannot satisfy (Hg2) for $N=2$. Moreover, the expression of $g$ implies that we have
\[
\left(1+g(x_1-x_2)\right)\left(1+g(x_2-x_1)\right)=1+\left(2b-a^2+b^2(x_1-x_2)^2\right)(x_1-x_2)^2.
\]
Hence, from the condition $\frac{a^2}2<b$, it follows that the product $p_1^tp_2^t$ is non-decreasing. As in the previous proof, it is therefore impossible that a sequence $\{p_i\}$ converges to 0 and the assumption ${\displaystyle\varliminf_{t\to +\infty}\min_{i\in [1,N]}}p_i^t<+\infty$ implies that this product converges to a finite limit.  Using the definition of the dynamics, we have in this case 
\[
\lim_{t\to +\infty}\frac{p_1^{t+1}p_2^{t+1}}{p_1^tp_2^t}= \lim_{t\to +\infty} \left(1+g(x_1^t-x_2^t)\right)\left(1+g(x_2^t-x_1^t)\right)=1,
\]
and the expression above shows that we must have ${\displaystyle \lim_{t\to +\infty}}x_1^t-x_2^t=0$, and the rest of the proof is as before for Proposition \ref{REALIST-PRICES}.

\subsection{Proof of Lemma \ref{CONSTRAINTS-X-N2} and Proposition \ref{CONSTRAINTS-X}}\label{S-CONSTRAINTS-X}
The convex combination in the first row of the iterations \eqref{DEFDYNAM}/\eqref{BASIS} imply that it suffices to prove the results for $\alpha=0$. 

\paragraph{Proof of Lemma \ref{CONSTRAINTS-X-N2}.} 
Using the symmetry assumption (Hf3), we obtain
 \begin{align*}
2\langle \mathbf x^{t+1}\rangle-1&=f(\rho_1^{t+1},x_1^t)+f(\frac1{\rho_1^{t+1}},x_2^t)-1=f(\rho_1^{t+1},x_1^t)-f(\rho_1^{t+1},1-x_2^t)\\
&=(2\langle \mathbf x^{t}\rangle-1)\frac{f(\rho_1^{t+1},x_1^t)-f(\rho_1^{t+1},1-x_2^t)}{x_1^t-(1-x_2^t)},
 \end{align*}
from where the Lemma immediately follows. 
\medskip

Throughout the proof of Proposition \ref{CONSTRAINTS-X}, the dependence on $t$ is dropped in the notations of the variables $\mathbf x^t$ and $\boldsymbol\rho^t$, for the sake of clarity. 
The symmetry of permutations of the coordinates and the invariance of ${\displaystyle \sum_{i=1}^N}x_i$ under this symmetry imply that we may assume w.l.o.g.\ that $\boldsymbol\rho\in R_N$ where $R_N\subset (0,1]^N$ is defined by
\[
R_N=\left\{\boldsymbol\rho \in (0,1]^N\ :\ \rho_1\leq \rho_2\leq \cdots\leq \rho_{N-1}\leq \rho_N=1\right\}.
\]
and at least one inequality $\rho_i\leq \rho_{i+1}$ is strict (otherwise there is nothing to prove because then $f_\frac{\rho_i}{\langle\boldsymbol\rho\rangle_i^c}=\text{Id}$ for all $i$). This implies that
\[
\frac{\rho_i}{\langle\boldsymbol\rho\rangle_i^c}\leq \frac{\rho_{i+1}}{\langle\boldsymbol\rho\rangle_{i+1}^c},\ \forall i\in [1,N-1]\quad\text{and}\quad \frac{\rho_1}{\langle\boldsymbol\rho\rangle_1^c}<1<\frac1{\langle\boldsymbol\rho\rangle_N^c}<N-1.
\]
Let then $N^-\in [1,N-1]$ be such that 
\[
\frac{\rho_i}{\langle\boldsymbol\rho\rangle_i^c}\leq 1,\ \forall i\in [1,N^-]\quad\text{and}\quad 1<\frac{\rho_i}{\langle\boldsymbol\rho\rangle_i^c},\ \forall i\in [N^-+1,N].
\]

\paragraph{Proof of statement {\em(i)} of Proposition \ref{CONSTRAINTS-X}.} This statement is a rephrasing of the following one, which specifies those conditions on the deviations $f_\mathrm{dev}$ that are required in the proof.
\begin{Lem}
Let $N>2$ be arbitrary, assume that $g$ satisfies {\rm (Hg1)} and $f$ is given by \eqref{SPEFAM} with $f_\mathrm{dev}(\rho,\cdot)\geq 0$ for all $\rho\in \R_\ast^+$ and it satisfies {\rm (Hf1-2)}. 

\noindent
(a) Assume that\footnote{Notice that
\[
\frac{(1-\rho)^2(N-2)}{N-1-\rho(N-2)}\in (0,1),\ \forall \rho<1.
\]}
\[
\sup_{x\in (0,1)}\frac{f_\mathrm{dev}(\rho,x)}{x}<\frac{(1-\rho)^2(N-2)}{N-1-(N-2)\rho},\ \forall \rho<1.
\eqno{(\mathrm{Hf}_\mathrm{dev}1)}
\] 
Then, for every orbit, we have $N\langle \mathbf x^{t+1}\rangle \geq \frac1{N-1}$ when $N\langle \mathbf x^{t}\rangle\geq \frac1{N-1}$.

\noindent
(b) Assume that either $N\in [3,4]$ and $f_\mathrm{dev}(\rho,\cdot)= 0$ for all $\rho\in \R_\ast^+$, or $N\geq 5$ and we have\footnote{Notice that
\[
\frac{(N-2)(\rho-1)}{\rho(1+(N-2)\rho)}\in (0,1),\ \forall \rho\in (1,N-1).
\]} 
\begin{equation*}
\sup_{x\in (0,1)}\frac{f_\mathrm{dev}(\rho,x)}{x}<\frac{(N-2)(\rho-1)}{\rho(1+(N-2)\rho)},\ \forall \rho\in (1,N-1).
\eqno{(\mathrm{Hf}_\mathrm{dev}2)}
\end{equation*}
and
\begin{equation*}
\sup_{x\in (0,1)}\frac{f_\mathrm{dev}(\rho,x)}{x}<(\rho-1)\left(\frac1{\rho}-\frac1{N-2}\right),\ \forall \rho\in (1,N-2).
\eqno{(\mathrm{Hf}_\mathrm{dev}3)}
\end{equation*}
Then, for every orbit, we have $N\langle \mathbf x^{t+1}\rangle\leq N-\frac1{N-1}$ when $N\langle \mathbf x^{t}\rangle\leq N-\frac1{N-1}$.
\label{STATEIP}
\end{Lem}

\begin{proof} {\em (a)} Since we may assume that $\alpha=0$, we aim at the following implication
\[
\sum_{i=1}^Nx_i\geq \frac1{N-1}\quad\Longrightarrow\quad \sum_{i=1}^Nf\left(\frac{\rho_i}{\langle\boldsymbol\rho\rangle_i^c},x_i\right)\geq \frac1{N-1}.
\]
The definitions of $N^-$ and of $f_\mathrm{dev}$ imply that we have
\begin{equation}
\sum_{i=1}^Nf\left(\frac{\rho_i}{\langle\boldsymbol\rho\rangle_i^c},x_i\right)=\sum_{i=1}^{N^-}1-\frac{\rho_i}{\langle\boldsymbol\rho\rangle_i^c}(1-x_i)-f_\mathrm{dev}\left(\frac{\rho_i}{\langle\boldsymbol\rho\rangle_i^c},1-x_i\right)+\sum_{i=N^-+1}^{N}\frac{\langle\boldsymbol\rho\rangle_i^c}{\rho_i}x_i+f_\mathrm{dev}\left(\frac{\rho_i}{\langle\boldsymbol\rho\rangle_i^c},x_i\right).
\label{EXPRESSIONF}
\end{equation}
Using that $f(\rho,x)\geq x$ when $\rho\leq 1$ and $f_\mathrm{dev}(\rho,x)\geq 0$, we obtain the following inequality
\[
\sum_{i=1}^Nf\left(\frac{\rho_i}{\langle\boldsymbol\rho\rangle_i^c},x_i\right)\geq \left(1-\frac{\rho_1}{\langle\boldsymbol\rho\rangle_1^c}-\sup_{x\in (0,1)}\frac{f_\mathrm{dev}\left(\frac{\rho_1}{\langle\boldsymbol\rho\rangle_1^c},x\right)}{x}\right)(1-x_1)+ \sum_{i=1}^{N^-}x_i+\sum_{i=N^-+1}^{N}\frac{\langle\boldsymbol\rho\rangle_i^c}{\rho_i}x_i,
\]
from where the conclusion immediately follows when $1-\frac{\rho_1}{\langle\boldsymbol\rho\rangle_1^c}-{\displaystyle\sup_{x\in (0,1)}}\frac{f_\mathrm{dev}\left(\frac{\rho_1}{\langle\boldsymbol\rho\rangle_1^c},x\right)}{x}\geq \frac1{N-1}$. Assume then that $1-\frac{\rho_1}{\langle\boldsymbol\rho\rangle_1^c}-{\displaystyle\sup_{x\in (0,1)}}\frac{f_\mathrm{dev}\left(\frac{\rho_1}{\langle\boldsymbol\rho\rangle_1^c},x\right)}{x}< \frac1{N-1}$. We claim that for every $r< 1$, we have
\begin{equation}
\frac1{N-1}\left(\min_{\boldsymbol\rho \in R_N\ :\ \frac{\rho_1}{\langle\boldsymbol\rho\rangle_1^c}=r}\langle\boldsymbol\rho\rangle_N^c-1\right)+1-r=\frac{(1-r)^2(N-2)}{N-1-(N-2)r},
\label{INEQCOEF2}
\end{equation}
which is equivalent to
\[
\min_{\boldsymbol\rho \in R_N\ :\ \frac{\rho_1}{\langle\boldsymbol\rho\rangle_1^c}=r}\langle\boldsymbol\rho\rangle_N^c=\frac{r}{N-1-(N-2)r}.
\]
In order to prove this equality, observe first that when $\frac{\rho_1}{\langle\boldsymbol\rho\rangle_1^c}=r$, we have 
\begin{equation}
\langle\boldsymbol\rho\rangle_N^c=\langle\boldsymbol\rho\rangle_1^c-\frac{1-\rho_1}{N-1}=\frac{\rho_1}{r}-\frac{1-\rho_1}{N-1}.
\label{REFORM}
\end{equation}
Besides, when $\boldsymbol\rho \in R_N$, we have 
\[
\langle\boldsymbol\rho\rangle_1^c\geq \frac{1+(N-2)\rho_1}{N-1}.
\]
The constraint $\frac{\rho_1}{\langle\boldsymbol\rho\rangle_1^c}=r$ then imposes that 
\[
r\leq \frac{(N-1)\rho_1}{1+(N-2)\rho_1}\quad\Longleftrightarrow\quad \rho_1\geq \frac{r}{N-1-(N-2)r},
\]
from where the relation \eqref{REFORM} immediately yields the desired minimum of $\langle\boldsymbol\rho\rangle_N^c$. 

The assumption $(\mathrm{Hf}_\mathrm{dev}1)$, together with \eqref{INEQCOEF2}, implies that for all $\boldsymbol\rho \in R_N$, we have 
\begin{equation*}
1-\frac{\rho_1}{\langle\boldsymbol\rho\rangle_1^c}-\sup_{x\in (0,1)}\frac{f_\mathrm{dev}\left(\frac{\rho_1}{\langle\boldsymbol\rho\rangle_1^c},x\right)}{x}> \frac1{N-1}\left(1-\langle\boldsymbol\rho\rangle_N^c\right)\geq \frac1{N-1}\left(1-\frac{\langle\boldsymbol\rho\rangle_i^c}{\rho_i}\right),\ \forall i\in [N^-+1,N],
\end{equation*}
which, in addition to $1-\frac{\rho_1}{\langle\boldsymbol\rho\rangle_1^c}-{\displaystyle\sup_{x\in (0,1)}}\frac{f_\mathrm{dev}\left(\frac{\rho_1}{\langle\boldsymbol\rho\rangle_1^c},x\right)}{x}\geq 0$, yields 
\begin{equation}
\sum_{i=1}^Nf\left(\frac{\rho_i}{\langle\boldsymbol\rho\rangle_i^c},x_i\right)\geq \left(1-\frac{\rho_1}{\langle\boldsymbol\rho\rangle_1^c}-\sup_{x\in (0,1)}\frac{f_\mathrm{dev}\left(\frac{\rho_1}{\langle\boldsymbol\rho\rangle_1^c},x\right)}{x}\right)\left(1-x_1-(N-1)\sum_{i=N^-+1}^{N}x_i\right)+\sum_{i=1}^{N}x_i.
\label{RHO1RHOI}
\end{equation}
Clearly, the RHS is an increasing function of $x_i$ for each $i\in [1,N^-]$. The assumption $1-\frac{\rho_1}{\langle\boldsymbol\rho\rangle_1^c}-{\displaystyle\sup_{x\in (0,1)}}\frac{f_\mathrm{dev}\left(\frac{\rho_1}{\langle\boldsymbol\rho\rangle_1^c},x\right)}{x}<\frac1{N-1}$ and $1-\frac{\rho_1}{\langle\boldsymbol\rho\rangle_1^c}-{\displaystyle\sup_{x\in (0,1)}}\frac{f_\mathrm{dev}\left(\frac{\rho_1}{\langle\boldsymbol\rho\rangle_1^c},x\right)}{x}\geq 0$ imply that  it is also an increasing function of $x_i$ for each $i\in [N^-+1,N]$. Therefore, its minimum in the domain ${\displaystyle\sum_{i=1}^N}x_i\geq \frac1{N-1}$ is attained for ${\displaystyle\sum_{i=1}^N}x_i= \frac1{N-1}$. When this equality holds, we certainly have 
\[
1-x_1-(N-1)\sum_{i=N^-+1}^{N}x_i\geq 1-x_1-(N-1)\sum_{i=2}^{N}x_i=(N-2)x_1\geq 0,
\]
which implies that whenever the inequality ${\displaystyle\sum_{i=1}^N}x_i\geq \frac1{N-1}$ holds, we have
\[
\sum_{i=1}^Nf\left(\frac{\rho_i}{\langle\boldsymbol\rho\rangle_i^c},x_i\right)\geq \sum_{i=1}^{N}x_i\geq \frac1{N-1},
\]
as desired.

\noindent
{\em Proof of (b) in the affine case.} We assume here that $f_\mathrm{dev}(\rho,\cdot)=0$ for all $\rho$, {\sl ie.}\ all maps $f(\rho,\cdot)$ are affine and given by \eqref{SPEFAM} with $f_\mathrm{dev}(\rho,\cdot)=0$ (NB: In practice, this proof only serves for $N\in [3,4]$ because for $N\geq 5$, the proof below holds when $f_\mathrm{dev}(\rho,\cdot)\neq 0$). We aim at proving that 
\begin{equation}
\sum_{i=1}^Nx_i\leq N-\frac1{N-1}\quad\Longrightarrow\quad \sum_{i=1}^Nf\left(\frac{\rho_i}{\langle\boldsymbol\rho\rangle_i^c},x_i\right)\leq N-\frac1{N-1}.
\label{PROOFB}
\end{equation}
We have 
\begin{align*}
\sum_{i=1}^Nf\left(\frac{\rho_i}{\langle\boldsymbol\rho\rangle_i^c},x_i\right)&=\sum_{i=1}^{N^-}1-\frac{\rho_i}{\langle\boldsymbol\rho\rangle_i^c}(1-x_i)+\sum_{i=N^-+1}^{N}\frac{\langle\boldsymbol\rho\rangle_i^c}{\rho_i}x_i\\
&\leq N^--\sum_{i=1}^{N^-}\frac{\rho_i}{\langle\boldsymbol\rho\rangle_i^c}(1-x_i)+\sum_{i=N^-+1}^{N-1}x_i+\langle\boldsymbol\rho\rangle_N^c x_N
\end{align*}
If $\langle\boldsymbol\rho\rangle_N^c \leq \frac{N-2}{N-1}$, then the conclusion immediately follows from the fact $\frac{\rho_i}{\langle\boldsymbol\rho\rangle_i^c}(1-x_i)\geq 0$ for $i\in [1,N^-]$ and $x_i\leq 1$ for $i\in [N^-+1,N]$. Assume then that $\langle\boldsymbol\rho\rangle_N^c > \frac{N-2}{N-1}$ and use the inequality\footnote{Indeed, the function $\rho_i\mapsto \langle\boldsymbol\rho\rangle_N^c-\frac1{N-1}\frac{\rho_i}{\langle\boldsymbol\rho\rangle_i^c}$ is decreasing on $(0,1]$; hence its supremum over $(0,1]$ is at most $\frac{N-2}{N-1}$. The displayed inequality follows suit.}
\[
-\frac{\rho_i}{\langle\boldsymbol\rho\rangle_i^c}\leq N-2-(N-1)\langle\boldsymbol\rho\rangle_N^c,\ \forall i\in [1,N^-],
\]
in order to obtain 
\[
\sum_{i=1}^Nf\left(\frac{\rho_i}{\langle\boldsymbol\rho\rangle_i^c},x_i\right)\leq N^-+(N-2)\sum_{i=1}^{N^-}(1-x_i)+\sum_{i=N^-+1}^{N-1}x_i+\langle\boldsymbol\rho\rangle_N^c\left(x_N-(N-1)\sum_{i=1}^{N^-}(1-x_i)\right).
\]
In the case where $x_N-(N-1){\displaystyle\sum_{i=1}^{N^-}}(1-x_i)\geq 0$, we use that $\langle\boldsymbol\rho\rangle_N^c\leq 1$ in order to obtain
\[
\sum_{i=1}^Nf\left(\frac{\rho_i}{\langle\boldsymbol\rho\rangle_i^c},x_i\right)\leq \sum_{i=1}^N x_i,
\]
from where the conclusion follows. Otherwise, the inequality $\langle\boldsymbol\rho\rangle_N^c > \frac{N-2}{N-1}$ implies that we must have
\[
\sum_{i=1}^Nf\left(\frac{\rho_i}{\langle\boldsymbol\rho\rangle_i^c},x_i\right)\leq N^-+\sum_{i=N^-+1}^{N-1}x_i+\frac{N-2}{N-1}x_N\leq N-\frac1{N-1},
\]
as desired. 

\noindent
{\em Proof of (b) for $N\geq 5$.} The implication \eqref{PROOFB} is obvious when there exists $i\in [N^-+1,N]$ such that 
\[
f\left(\frac{\rho_i}{\langle\boldsymbol\rho\rangle_i^c},x_i\right)\leq 1-\frac{1}{N-1},
\]
because each of the other $N-1$ terms in the sum does not exceed $1$. Assume then that 
\[
f\left(\frac{\rho_i}{\langle\boldsymbol\rho\rangle_i^c},x_i\right)> 1-\frac{1}{N-1},\ \forall i\in [N^-+1,N],
\]
that is to say, given that $f(\rho,\cdot)$ is increasing and $f(\rho,x)\leq x$ when $\rho>1$,
\[
\min\left\{f\left(\frac{\rho_i}{\langle\boldsymbol\rho\rangle_i^c},1\right),x_i\right\}> 1-\frac{1}{N-1},\ \forall i\in [N^-+1,N].
\]
Given that $(\mathrm{Hf}_\mathrm{dev}2)$ imposes $f_\mathrm{dev}\left(\rho,1\right)<\frac{(N-2)(\rho-1)}{\rho(1+(N-2)\rho)}<\frac1{\rho}$, the assumption above requires 
\[
\frac{\rho_i}{\langle\boldsymbol\rho\rangle_i^c}<\frac{2(N-1)}{N-2},\ \forall i\in [N^-+1,N],
\]
(and we actually have $\frac{2(N-1)}{N-2}\leq N-1$ for all $N\geq 4$). 

Moreover, we claim that for every $r<1$, we have
\begin{equation}
\min_{\boldsymbol\rho \in [0,1]^N\ :\ \rho_N=1, \langle \boldsymbol\rho\rangle_N^c=r}\sum_{i=1}^{N-1}\frac{\rho_i}{\langle \boldsymbol\rho\rangle_i^c}=\frac{(N-1)^2r}{1+(N-2)r}=(N-1)\left(r+\frac{(N-2)r(1-r)}{1+(N-2)r}\right),
\label{INEQCOEF}
\end{equation}
where the first equality is an immediate consequence of the following technical statement.
\begin{Claim}
Let $N\in\Z^+$, $S\in (0,N)$ and $L\in\R^+$ be arbitrary. We have 
\[
\min_{\mathbf x\in [0,1]^N\ :\ \sum_{i=1}^Nx_i=S}\sum_{i=1}^{N}\frac{x_i}{L+S-x_i}=\frac{NS}{NL+(N-1)S}.
\]
\end{Claim}
\noindent
{\em Proof of the Claim.} By induction. For $N=1$, the LHS reduces to $\frac{S}{L}$, which is equal to the RHS in this case.  Let now $N\in\Z^+$ be arbitrary. By isolating the last term of the sum, one can separate the minimisation of the sum over of the first terms to obtain
\begin{align*}
\min_{\mathbf x\in [0,1]^{N+1}: \sum_{i=1}^{N+1}x_i=S}\sum_{i=1}^{N+1}\frac{x_i}{L+S-x_i}&=\min_{x_{N+1}\in [0,1\wedge S]}\left\{\min_{\mathbf x\in [0,1]^N:\sum_{i=1}^Nx_i=S-x_{N+1}}\left(\frac{x_{N+1}}{L+S-x_{N+1}}+\sum_{i=1}^{N}\frac{x_i}{L+S-x_i}\right)\right\}\\
&=\min_{x_{N+1}\in [0,1\wedge S]}\left\{\frac{x_{N+1}}{L+S-x_{N+1}}+\min_{\mathbf x\in [0,1]^N:\sum_{i=1}^Nx_i=S-x_{N+1}}\sum_{i=1}^{N}\frac{x_i}{L+S-x_i}\right\},
\end{align*}
and then use the induction assumption to obtain 
\[
\min_{\mathbf x\in [0,1]^{N+1}\ :\ \sum_{i=1}^{N+1}x_i=S}\sum_{i=1}^{N+1}\frac{x_i}{L+S-x_i}=\min_{x_{N+1}\in [0,1\wedge S]}\left\{\frac{x_{N+1}}{L+S-x_{N+1}}+\frac{N(S-x_{N+1})}{N(L+x_{N+1})+(N-1)(S-x_{N+1})}\right\}.
\]
Now, the study of the variations of the function $x\mapsto \frac{x}{L+S-x}+\frac{N(S-x)}{N(L+x)+(N-1)(S-x)}$ (which can be achieved by investing the variations of the two terms separately) easily shows that it has a minimum on $[0,1\wedge S]$ at $x=\frac{S}{N+1}$, whose value is equal to $\frac{(N+1)S}{(N+1)L+NS}$ as desired. \hfill $\Box$

The assumption $(\mathrm{Hf}_\mathrm{dev}2)$, together with \eqref{INEQCOEF}, implies that for all $\boldsymbol \rho\in R_N$ and $x_N\in (0,1)$, we have since $\langle\boldsymbol\rho\rangle_N^c<1$
\begin{equation}
(N-1)\left(\langle\boldsymbol\rho\rangle_N^c+\frac{f_\mathrm{dev}\left(\frac{1}{\langle\boldsymbol\rho\rangle_N^c},x_N\right)}{x_N}\right)<\sum_{i=1}^{N-1}\frac{\rho_i}{\langle\boldsymbol\rho\rangle_i^c}.
\label{CONTROLRHON}
\end{equation}
By inserting this inequality into \eqref{EXPRESSIONF}, we get using that $f_\mathrm{dev}(\rho,\cdot)\geq 0$
\[
\sum_{i=1}^Nf\left(\frac{\rho_i}{\langle\boldsymbol\rho\rangle_i^c},x_i\right)\leq N^-+\sum_{i=1}^{N^-}\frac{\rho_i}{\langle\boldsymbol\rho\rangle_i^c}\left(\frac{x_N}{N-1}-(1-x_i)\right)+\sum_{i=N^-+1}^{N-1}\frac{\langle\boldsymbol\rho\rangle_i^c}{\rho_i}x_i+f_\mathrm{dev}\left(\frac{\rho_i}{\langle\boldsymbol\rho\rangle_i^c},x_i\right)+\frac{\rho_i}{\langle\boldsymbol\rho\rangle_i^c}\frac{x_N}{N-1}.
\]
Now we use the fact that $\frac{\rho_i}{\langle\boldsymbol\rho\rangle_i^c}\in (1,\frac{2(N-1)}{N-2})$ for $i\in [N^-+1,N-1]$, the inequality $\frac{2(N-1)}{N-2}\leq N-2$ for all $N\geq 5$ and the assumption $(\mathrm{Hf}_\mathrm{dev}3)$ in order to show that , for each $ i\in [N^-+1,N-1]$, we have 
\[
\left(\frac{\langle\boldsymbol\rho\rangle_i^c}{\rho_i}+\frac{f_\mathrm{dev}\left(\frac{\rho_i}{\langle\boldsymbol\rho\rangle_i^c},x_i\right)}{x_i}\right)x_i+\frac{\rho_i}{\langle\boldsymbol\rho\rangle_i^c}\frac{x_N}{N-1}<
x_i+\frac{x_N}{N-1},
\]
if $x_N<\frac{N-1}{N-2}x_i$. This inequality holds because we assume that $x_i>1-\frac1{N-1}$. Therefore, we obtain  
\[
\sum_{i=1}^Nf\left(\frac{\rho_i}{\langle\boldsymbol\rho\rangle_i^c},x_i\right)\leq N^-+\sum_{i=1}^{N^-}\frac{\rho_i}{\langle\boldsymbol\rho\rangle_i^c}\left(\frac{x_N}{N-1}-(1-x_i)\right)+\sum_{i=N^-+1}^{N-1}x_i+\frac{(N-N^--1)x_N}{N-1}.
\]
Letting ${\cal I}=\left\{i\in [1,N^-]\ :\ \frac{x_N}{N-1}-(1-x_i)\geq 0\right\}$, we get
\[
\sum_{i=1}^Nf\left(\frac{\rho_i}{\langle\boldsymbol\rho\rangle_i^c},x_i\right)\leq N^--\#{\cal I}+\sum_{i\in {\cal I}}x_i+\sum_{i=N^-+1}^{N-1}x_i+\frac{(N-N^-+\#{\cal I}-1)x_N}{N-1}.
\]
Basic estimates then yield
\[
\sum_{i=1}^Nf\left(\frac{\rho_i}{\langle\boldsymbol\rho\rangle_i^c},x_i\right)\leq\left\{\begin{array}{ccl}
\sum_{i=1}^Nx_i&\text{if}&\#{\cal I}=N^-\\
N-1+x_N\frac{N-2}{N-1}&\text{if}&\#{\cal I}<N^-
\end{array}\right.
\]
from where the desired conclusion immediately follows.
\end{proof}

\paragraph{Proof of statement {\em (ii)} of Proposition \ref{CONSTRAINTS-X}.} As before, this statement is a rephrasing of the following one, which specifies the conditions on the deviations $f_\mathrm{dev}$.
\begin{Lem}
Let $N>2$ be arbitrary, assume that $g$ satisfies {\rm (Hg1)} and $f$ is given by \eqref{SPEFAM} with $f_\mathrm{dev}(\rho,\cdot)\geq 0$ for all $\rho\in \R_\ast^+$ and it satisfies {\rm (Hf1-2)}. Assume also that $(\mathrm{Hf}_\mathrm{dev}1)$ holds. Then, for every orbit, we have $\langle \mathbf x^{t+1}\rangle>\langle \mathbf x^{t}\rangle$ when $N\langle \mathbf x^{t}\rangle<\frac1{N-1}$ and $\boldsymbol\rho^t\neq \mathbf 1$.
\label{MONOT}
\end{Lem}
\begin{proof}
The proof is an immediate consequence of the proof of statement (a) of Lemma \ref{STATEIP} above. Indeed, the inequality \eqref{RHO1RHOI} is equivalent to the following one
\[
\sum_{i=1}^Nf\left(\frac{\rho_i}{\langle\boldsymbol\rho\rangle_i^c},x_i\right)-x_i>\left(1-\frac{\rho_1}{\langle\boldsymbol\rho\rangle_1^c}-\sup_{x\in (0,1)}\frac{f_\mathrm{dev}\left(\frac{\rho_1}{\langle\boldsymbol\rho\rangle_1^c},x\right)}{x}\right)\left(1-x_1-(N-1)\sum_{i=N^-+1}^Nx_i\right),
\]
and the RHS must be positive when $N\langle \mathbf x^{t}\rangle<\frac1{N-1}$ and $\boldsymbol\rho^t\neq \mathbf 1$. 
\end{proof}

\paragraph{Proof of statement {\em (iii)}  of Proposition \ref{CONSTRAINTS-X}.} Here, we prove the following statement which specifies additional conditions on $f_\mathrm{dev}$.
 \begin{Lem}
Let $N>2$ be arbitrary, assume that $g$ satisfies {\rm (Hg1)} and $f$ is given by \eqref{SPEFAM} with $f_\mathrm{dev}(\rho,\cdot)\geq 0$ for all $\rho\in \R_\ast^+$ and it satisfies {\rm (Hf1-2)}. Assume also that  $(\mathrm{Hf}_\mathrm{dev}2)$ and the following assumption\footnote{We have 
\[
\frac{(N-\rho)\rho-(N-1)}{(N-1)\rho}\in (0,1),\ \forall \rho\in (1,N-1).
\]} also hold
\begin{equation*}
\sup_{x\in (0,1)}\frac{f_\mathrm{dev}(\rho,x)}{x}<\frac{(N-\rho)\rho-(N-1)}{(N-1)\rho},\ \forall \rho\in (1,N-1).
\eqno{(\mathrm{Hf}_\mathrm{dev}4)}
\end{equation*}
Then, for every orbit, we have $\langle \mathbf x^{t+1}\rangle<\langle \mathbf x^{t}\rangle$ when $N\langle \mathbf x^{t}\rangle>N-\frac1{N-1}$ and $\boldsymbol\rho^t\neq \mathbf 1$. 
\end{Lem}
\begin{proof} 
We aim at proving the inequality ${\displaystyle\sum_{i=1}^N}f\left(\frac{\rho_i}{\langle\boldsymbol\rho\rangle_i^c},x_i\right)-x_i< 0$ when ${\displaystyle\sum_{i=1}^N}x_i> N-\frac1{N-1}$ and $\boldsymbol\rho\neq \mathbf 1$.  The setting is the same as in the proof of statement (b), nonlinear case, of Lemma \ref{STATEIP}. Using the expression \eqref{EXPRESSIONF} and the inequality \eqref{CONTROLRHON}, we get
\begin{align*}
\sum_{i=1}^Nf\left(\frac{\rho_i}{\langle\boldsymbol\rho\rangle_i^c},x_i\right)-x_i&< \sum_{i=1}^{N^-}\left(1-\frac{\rho_i}{\langle\boldsymbol\rho\rangle_i^c}\right)(1-x_i)+\sum_{i=N^-+1}^{N-1}\left(\frac{\langle\boldsymbol\rho\rangle_i^c}{\rho_i}-1\right)x_i+f_\mathrm{dev}\left(\frac{\rho_i}{\langle\boldsymbol\rho\rangle_i^c},x_i\right)\\
&+\frac{x_N}{N-1}\sum_{i=1}^{N-1}\left(\frac{\rho_i}{\langle\boldsymbol\rho\rangle_i^c}-1\right)
\end{align*}
From $(\mathrm{Hf}_\mathrm{dev}4)$, we have for each $i\in [N^-+1,N-1]$ since $\frac{\rho_i}{\langle\boldsymbol\rho\rangle_i^c}\in (1,N-1)$,
\[
\frac{f_\mathrm{dev}\left(\frac{\rho_i}{\langle\boldsymbol\rho\rangle_i^c},x_i\right)}{x_i}<\frac{(N-\frac{\rho_i}{\langle\boldsymbol\rho\rangle_i^c})\frac{\rho_i}{\langle\boldsymbol\rho\rangle_i^c}-(N-1)}{(N-1)\frac{\rho_i}{\langle\boldsymbol\rho\rangle_i^c}}
=1-\frac{\langle\boldsymbol\rho\rangle_i^c}{\rho_i}+\frac1{N-1}\left(1-\frac{\rho_i}{\langle\boldsymbol\rho\rangle_i^c}\right),
\]
which yields
\[
\sum_{i=N^-+1}^{N-1}\left(\frac{\langle\boldsymbol\rho\rangle_i^c}{\rho_i}-1\right)x_i+f_\mathrm{dev}\left(\frac{\rho_i}{\langle\boldsymbol\rho\rangle_i^c},x_i\right)<\frac1{N-1}\sum_{i=N^-+1}^{N-1}\left(1-\frac{\rho_i}{\langle\boldsymbol\rho\rangle_i^c}\right)x_i.
\]

Let now $\mathbf x\in [0,1]^N$ be such that ${\displaystyle\sum_{i=1}^N}x_i> N-\frac1{N-1}$. Clearly, there exists $\epsilon>0$ such that 
\[
\max_{i\in [1,N^-]}1-x_i<\frac{1-\epsilon}{N-1}\quad\text{and}\quad 1-\epsilon<\min_{i\in [N^-+1,N]}x_i.
\]
From the definition of $N^-$, it follows that 
\[
\sum_{i=1}^{N^-}\left(1-\frac{\rho_i}{\langle\boldsymbol\rho\rangle_i^c}\right)(1-x_i)<\frac{1-\epsilon}{N-1}\sum_{i=1}^{N^-}\left(1-\frac{\rho_i}{\langle\boldsymbol\rho\rangle_i^c}\right)
\]
and
\[
\frac1{N-1}\sum_{i=N^-+1}^{N-1}\left(1-\frac{\rho_i}{\langle\boldsymbol\rho\rangle_i^c}\right)x_i<\frac{1-\epsilon}{N-1}\sum_{i=N^-+1}^{N-1}\left(1-\frac{\rho_i}{\langle\boldsymbol\rho\rangle_i^c}\right).
\]
Besides, we are going to prove that ${\displaystyle\sum_{i=1}^{N-1}}\left(1-\frac{\rho_i}{\langle\boldsymbol\rho\rangle_i^c}\right)\geq 0$. Altogether, this implies that
\[
\sum_{i=1}^Nf\left(\frac{\rho_i}{\langle\boldsymbol\rho\rangle_i^c},x_i\right)-x_i<\frac{1-\epsilon}{N-1}\sum_{i=1}^{N^-}\left(1-\frac{\rho_i}{\langle\boldsymbol\rho\rangle_i^c}\right)+\frac{1-\epsilon}{N-1}\sum_{i=N^-+1}^{N-1}\left(1-\frac{\rho_i}{\langle\boldsymbol\rho\rangle_i^c}\right)+\frac{1-\epsilon}{N-1}\sum_{i=1}^{N-1}\left(\frac{\rho_i}{\langle\boldsymbol\rho\rangle_i^c}-1\right)=0,
\]
as desired.

In order to prove that ${\displaystyle\sum_{i=1}^{N-1}}\left(1-\frac{\rho_i}{\langle\boldsymbol\rho\rangle_i^c}\right)\geq 0$, observe first that for every $i,k\in [1,N]$, the derivative of 
\[
x_k\mapsto \sum_{i=1}^N\frac{x_i}{1-x_i+\sum_{j=1}^Nx_j},
\]
is an increasing function on $[0,1]$. Hence we certainly have ${\displaystyle\max_{x_k\in [0,1]}\sum_{i=1}^N}\frac{x_i}{1-x_i+\sum_{j=1}^Nx_j}={\displaystyle\max_{x_k\in \{0,1\}}\sum_{i=1}^N}\frac{x_i}{1-x_i+\sum_{j=1}^Nx_j}$, that is to say
\[
\max_{\mathbf x\in [0,1]^N}\sum_{i=1}^N\frac{x_i}{1-x_i+\sum_{j=1}^Nx_j}=\max_{\mathbf x\in \{0,1\}^N}\sum_{i=1}^N\frac{x_i}{1-x_i+\sum_{j=1}^Nx_j}.
\]
Now, in order to compute the RHS, thanks to the invariance under permutation, we may assume that there exists $N_0\in [0,N-1]$ such that 
\[
x_i=0,\ \forall i\in [1,N_0]\ \text{and}\ x_i=1, \forall i\in [N_0+1,N],
\]
that is to say
\[
\max_{\mathbf x\in \{0,1\}^N}\sum_{i=1}^N\frac{x_i}{1-x_i+\sum_{j=1}^Nx_j}=\max_{N_0\in [0,N-1]}\sum_{i=N_0+1}^N\frac{1}{N-N_0}=1,
\]
from where the desired inequality immediately follows.
\end{proof}

\paragraph{Expression of $C_N(\rho)$.}
For $\rho<1$, the expression is a combination of the original condition on $f_\mathrm{dev}$ and the assumption $(\mathrm{Hf}_\mathrm{dev}1)$, namely we have
\[
C_N(\rho)=\min\left\{1-\rho,\frac{(1-\rho)^2(N-2)}{N-1-(N-2)\rho}\right\},\ \forall \rho<1.
\]
For $\rho\in (1,N-1),$ we get a combination of the original condition on $f_\mathrm{dev}$ and the conditions $(\mathrm{Hf}_\mathrm{dev}2-4)$
\[
C_N(\rho)=\left\{\begin{array}{ccl}
\min\{1-\frac1{\rho},\frac{(N-2)(\rho-1)}{\rho(1+(N-2)\rho)},(\rho-1)\left(\frac1{\rho}-\frac1{N-2}\right),\frac{(N-\rho)\rho-(N-1)}{(N-1)\rho}\}&\text{if}&\rho\in (1,N-2)\\
\min\{1-\frac1{\rho},\frac{(N-2)(\rho-1)}{\rho(1+(N-2)\rho)},\frac{(N-\rho)\rho-(N-1)}{(N-1)\rho}\}&\text{if}&\rho\in [N-2,N-1)
\end{array}\right.
\]
The proof of Proposition \ref{CONSTRAINTS-X} is complete.

\subsection{Proof of Corollary \ref{BOUNDED-X}}\label{S-BOUNDED-X}
We use contradiction based on the following claim.
\begin{Claim}
If there exists $i\in [1,N]$ such that 
\[
\varliminf_{t\to +\infty}x_i^{t}=0,
\]
then we must have 
\[
\varliminf_{t\to +\infty}\max_{i\in [1,N]}x_i^{t}=0.
\]
\label{CLAIM1}
\end{Claim}
If Claim \ref{CLAIM1} is true, then there must be a diverging  subsequence $\{t_k\}_{k\in\N}$ such that ${\displaystyle\lim_{k\to+\infty}} \langle \mathbf x^{t_k}\rangle=0$. Hence, there must be a diverging sub-subsequence, also denoted $\{t_k\}_{k\in\N}$, such that 
\[
\langle \mathbf x^{t_k+1}\rangle< \langle \mathbf x^{t_k}\rangle< \frac1{N(N-1)}\ \text{and}\ \boldsymbol\rho^{t_k}\neq \mathbf 1,\ \forall k\in\N.
\]
However, this is clearly with  incompatible with Lemma \ref{CONSTRAINTS-X-N2} when $N=2$ (resp.\ Proposition \ref{CONSTRAINTS-X} when $N>2$). Hence, we must have 
\[
\varliminf_{t\to+\infty}x_i^{t}>0,\ \forall i\in [1,N],
\]
from where the lower bound of the Corollary immediately follows. The similar proof of the upper bound is left to the reader.

In order to prove Claim \ref{CLAIM1}, we also use contradiction based on the following additional claim. 
\begin{Claim}
Assume that $i\in [1,N]$ is such that
\[
\varliminf_{t\to+\infty}x_i^{t}=0.
\]
Then for every $n\in\N$, we have 
\[
\varliminf_{t\to+\infty}x_i^{t-\ell}=0,\ \forall \ell\in [0,n].
\]
\label{CLAIM2}
\end{Claim}
\begin{proof}
Let $\{t_k\}_{k\in\N}$ be a diverging subsequence such that 
\[
\lim_{k\to +\infty}x_i^{t_k}=0.
\]
There must be infinitely many instances of $k$ such that $\frac{\rho_i^{t_k}}{\langle \boldsymbol\rho^{t_k}\rangle_i^c}\geq 1$. Up to passing to a sub-subsequence, we may assume that $\frac{\rho_i^{t_k}}{\langle \boldsymbol\rho^{t_k}\rangle_i^c}\geq 1$ for all $k\in\N$.

Now, recall that the map $F_\mathrm{skew}$ defined by \eqref{BASIS} is invertible; hence the sequence $\{x_i^{t_k-1}\}_{k\in\N}$ is well defined. We claim that ${\displaystyle\lim_{k\to +\infty}}x_i^{t_k-1}=0$. Indeed, by compactness, consider any convergent subsequence of $\{x_i^{t_k-1}\}_{k\in\N}$, which we assume to coincide with the sequence itself, for the sake of notation. Together with separate continuity, the fact that ${\displaystyle\sup_{(\rho,x)\in \R_\ast^+\times [0,1]}}|f'_x(\rho,x)|\leq 1$ implies that the map $f$ is continuous in the variable $(\rho,x)$. Therefore, the (sub-)sequence $\{f_\alpha\left(\frac{\rho_i^{t_k}}{\langle \boldsymbol\rho^{t_k}\rangle_i^c},x_i^{t_k-1}\right)\}_{k\in\N}$ converges. However this sequence is nothing but $\{x_i^{t_k}\}_{k\in\N}$ whose limit is 0; hence, the (sub-)sequence $\{x_i^{t_k-1}\}_{k\in\N}$ must converge to the pre-image of $0$ under $f_\alpha\left(\frac{\rho_i^{t_k}}{\langle \boldsymbol\rho^{t_k}\rangle_i^c},\cdot\right)$, which is equal to 0, as claimed.

Repeating the argument for the consecutive pre-images $x_i^{t_k-\ell}$, it follows that for every $n\in\N$, there exists a diverging sub-subsequence $\{t_k\}_{k\in\N}$ such that 
\[
\lim_{k\to +\infty}x_i^{t_k-\ell}=0,\ \forall \ell\in [1,n].
\]
\end{proof}

\noindent
{\bf Proof of Claim \ref{CLAIM1}.} Let $\{t_k\}_{k\in\N}$ be a diverging subsequence such that 
\[
\lim_{k\to +\infty}x_N^{t_k}=0,
\]
and let $n\in\N$ be arbitrary. For all $k$ sufficiently large (depending on $n$), there exists $\epsilon>0$ (not depending on $n$) such that have
\[
\max_{i\in [1,N-1]}x_i^{t_k-\ell}-x_N^{t_k-\ell}>\epsilon,\ \forall \ell\in [0,n-1].
\]
Naturally, the index $i$ for which this inequality holds may vary with $\ell$. However, for $n=m(N-1)$, the fraction of those $\ell$ for which the inequality fails cannot be smaller than $m$ for every $i$. This means that there exists $i\in [1,N-1]$ such that 
\[
\#S_i^{t_k}\geq m\quad \text{where}\quad S_i^{t_k}=\left\{\ell\in [0,m(N-1)-1]\ :\  x_i^{t_k-\ell}-x_N^{t_k-\ell}>\epsilon\right\}
\]
provided that $k$ is sufficiently large. Let $\ell_k=\max\{\ell\in [0,m(N-1)-1]\ :\  x_i^{t_k-\ell}-x_N^{t_k-\ell}>\epsilon\}$. The fact that 
\[
\lim_{k\to +\infty}x_N^{t_k-\ell_k}=0
\]
implies that 
\[
\varliminf_{k\to +\infty}x_i^{t_k-\ell_k}>0.
\]
Consequently, Claim \ref{CLAIM2} implies that 
\[
\varliminf_{k\to +\infty}x_i^{t_k-\ell}>0,\ \forall \ell \in [0,\ell_k],
\]
and then 
\[
x_i^{t_k-\ell}-x_N^{t_k-\ell}>0,\ \forall \ell \in [0,\ell_k],
\]
provided that $k$ is sufficiently large. Accordingly, we have 
\begin{equation}
\rho_i^{t_k-\ell+1}\geq \rho_i^{t_k-\ell},\ \forall \ell \in [0,\ell_k].
\label{DIFFIN}
\end{equation}

%
%
Now, notice that 
\[
x_i-\langle \mathbf x\rangle_i^c=x_N-\langle \mathbf x\rangle_N^c+\frac{N}{N-1}(x_i-x_N),
\]
and by (Hg1) and compactness of the interval under consideration, let $\gamma>1$ be defined by
\[
\gamma=\inf_{x\in (-1,1-\frac{N}{N-1}\epsilon]}\frac{1+g(x+\frac{N}{N-1}\epsilon)}{1+g(x)}
\]
The definition of $S_i^{t_k}$ implies that for all $k$ sufficiently large, we have 
\[
\frac{1+g(x_i^{t_k-\ell}-\langle \mathbf x^{t_k-\ell}\rangle_i^c)}{1+g(x_N^{t_k-\ell}-\langle \mathbf x^{t_k-\ell}\rangle_N^c)}\geq \gamma,\ \forall \ell\in S_i^{t_k},
\]
and then
\[
\rho^{t_k-\ell+1}_i\geq \rho^{t_k-\ell}_i\gamma,\ \forall \ell\in S_i^{t_k}.
\]
Together with the inequalities \eqref{DIFFIN} and $\#S_i^{t_k}\geq m$, we conclude that we must have
\[
\rho^{t_k+1}_i\geq \rho^{t_k-\ell_k}_i\gamma^m
\]
However, for $m$ large enough such that $\frac{\gamma^m}{M}>M$, where $M$ is the bound on ${\displaystyle\max_{i\in [1,N-1]}}\max\left\{\rho_i^t,\frac1{\rho_i^t}\right\}$ in Proposition \ref{BOUNDED-RHO}, this inequality is clearly incompatible with that Proposition; hence the contradiction. \hfill $\Box$

\subsection{Proof of Proposition \ref{UNIFORM-BOUNDS}}\label{S-UNIFORM-BOUNDS}
The statement will be a direct consequence of the following properties, which are proved below:
\begin{itemize}
\item[{\em (a)}] For every $(\epsilon,m) \in (0,1)^2$, there exists $\epsilon'\in (0,1)$ such that if $|x_1-x_2|\leq \epsilon$ and ${\displaystyle\max_{i\in [1,2]}}\left\{\rho_i,\tfrac1{\rho_i}\right\}\leq \frac1{m}$, then we have
\[
\left|f_\alpha(\rho_1,x_1)- f_\alpha(\rho_2,x_2)\right|\leq \epsilon'.
\]
\item[{\em (b)}] For every $m\in (0,1)$, there exists $\epsilon \in (0,1)$ such that if $x_1-x_2> \epsilon$ (resp.\ $x_1-x_2< -\epsilon$) and ${\displaystyle\max_{i\in [1,2]}}\left\{\rho_i,\tfrac1{\rho_i}\right\}\leq \frac1{m}$, then we have
\[
f_\alpha(\rho_1,x_1)- f_\alpha(\rho_2,x_2)>-\epsilon\ \left(\text{resp.}\ f_\alpha(\rho_1,x_1)- f_\alpha(\rho_2,x_2)<\epsilon\right).
\]
\item[{\em (c)}] Let $\{(\mathbf x^t,\boldsymbol\rho^t)\}_{t\in\N}$ be an orbit of \eqref{BASIS} such that ${\displaystyle\max_{i\in [1,N-1]}}\left\{\rho_i^t,\frac1{\rho_i^t}\right\}\leq \frac1{m}$ for all $t\in\N$, for some $m\in (0,1)$. Then for every $\epsilon>0$, there exists $T_\epsilon\in \Z^+$ such that for every $t\in\N$ and $i\in [1,N-1]$ such that $|x_i^t-x_N^t|>\epsilon$, there exists $s\in [1,T_\epsilon]$ such that
\[
\left|x_i^{t+s}- x_N^{t+s}\right|\leq \epsilon.
\]
\end{itemize}
In order to prove the Proposition, take an arbitrary orbit $\{(\mathbf x^t,\boldsymbol\rho^t)\}_{t\in\N}$, and let $t'$ be the instant given by Proposition \ref{UNIFORM-BOUNDS-PRICESRAT} (either for $N=2$ or $N>2$). Consider the orbit with initial condition $(\mathbf x^{t'},\boldsymbol \rho^{t'})$. Let also $\epsilon^0_N$ be the value of $\epsilon$ that results from {\em (b)} for $m$ given by the inverse of the bound in Proposition \ref{UNIFORM-BOUNDS-PRICESRAT}, and let $T_{\epsilon^0_N}$ be given by {\em (c)}. For $k\in\N$, define $\epsilon^{k+1}_N$ by induction, as the $\epsilon'$ that results from property {\em (a)} with $\epsilon_N^k=\epsilon$ and $m$ given by the inverse of the bound in Proposition \ref{UNIFORM-BOUNDS-PRICESRAT}. Clearly, Proposition \ref{UNIFORM-BOUNDS} follows with $\epsilon_N=\epsilon_N^{T_{\epsilon^0_N}}\in (0,1)$.
\medskip

\noindent
{\em Proof of (a).} We separate the cases $x_1<x_2<x_1+\epsilon$ and $x_2<x_1<x_2+\epsilon$. Assume the first case. The second case can be treated similarly. The assumptions (Hf1-2) and the condition ${\displaystyle\max_{i\in [1,2]}}\left\{\rho_i,\tfrac1{\rho_i}\right\}\leq \frac1{m}$ imply
\[
f_\alpha(\rho_1,x_1)- f_\alpha(\rho_2,x_2)\leq \max_{x\in [0,1]}f_\alpha(m,x)- f_\alpha(\frac1{m},x)<1.
\]
On the other hand, we have
\[
f_\alpha(\rho_1,x_1)- f_\alpha(\rho_2,x_2)\geq  f_\alpha(\frac1{m},\max\{x_2-\epsilon,0\})- f_\alpha(m,x_2)\geq \max\{x_2-\epsilon,0\}-x_2\geq -\epsilon,
\]
and the desired estimate holds with $\epsilon'=\max\{\epsilon,{\displaystyle\max_{x\in [0,1]}}f_\alpha(m,x)- f_\alpha(\frac1{m},x)\}$.
\medskip

\noindent
{\em Proof of (b).} Using similar arguments as in {\em (a)}, we have when $x_1-x_2> \epsilon$ and ${\displaystyle\max_{i\in [1,2]}}\left\{\rho_i,\tfrac1{\rho_i}\right\}\leq \frac1{m}$
\[
f_\alpha(\rho_1,x_1)- f_\alpha(\rho_2,x_2)> f_\alpha(\frac1{m},\epsilon)- f_\alpha(m,1-\epsilon),
\]
and the conclusion follows from the facts that ${\displaystyle\lim_{\epsilon\to 1}}f_\alpha(\frac1{m},\epsilon)>0$ and  ${\displaystyle\lim_{\epsilon\to 1}}f_\alpha(m,1-\epsilon)<1$. The proof in the case $x_1-x_2<-\epsilon$ is similar.
\medskip

\noindent
{\em Proof of (c).} By contradiction, assume that $|x_1^{t+s}-x_N^{t+s}|>\epsilon$ for all $s\in\N$. Then property {\em (b)} implies that we must have
\[
\text{either}\ x_1^{t+s}-x_N^{t+s}>\epsilon,\ \forall s\in\N,\ \text{or}\ x_1^{t+s}-x_N^{t+s}<-\epsilon,\ \forall s\in\N.
\]
Assume the first case. The second case can be treated similarly, by symmetry. It follows from (Hg1) and (Hg3) that there exists $\gamma>1$ such that 
\[
\rho_i^{t+s+1}>\gamma \rho_i^{t+s},\ \forall s\in\N,
\]
hence ${\displaystyle\lim_{s\to +\infty}}\rho_i^{t+s}=+\infty$, since $\rho_i^t\geq m$. However, this divergence is incompatible with the fact that $\rho_i^t\leq \frac1{m}$ for all $t\in\N$.
 
\subsection{Proof of Proposition \ref{ALTERNATIONS}}\label{S-ALTERNATIONS}

\paragraph{Proof of infinite changes of $\underline{i}^t_{\mathbf x}$.} By contradiction, we are going to show that it is impossible that $\underline{i}^t_{\mathbf x}$ remain the same for all $t\in\N$. The conclusion then follows by applying repeatedly the argument to the orbit starting from the first iterate at which this quantity changes. 
 
Up to a permutation of the coordinates, assume then that
 \[
 x_N^t=\min_{i\in [1,N]}x_i^t,\ \forall t\in\N.
 \]
The iteration rule for the $\rho_i^t$, the fact that $x_i^t>x_N^t$ and strict monotonicity of $g$ then imply 
\[
\rho_i^{t+1}=\rho_i^t\frac{1+g\left(x_i^{t}-\langle \mathbf x^{t}\rangle_i^c\right)}{1+g\left(x_{N}^{t}-\langle \mathbf x^{t}\rangle_N^c\right)}>\rho_i^t,\ \forall i\in [1,N-1].
\]
From Proposition \ref{BOUNDED-RHO}, it results that the following limits exist
\[
\lim_{t\to +\infty}\rho_i^t=\rho_i^\infty,\ \forall i\in [1,N-1],
\]
(and by definition of the $\rho_i$, we obviously have $\rho_N^\infty=1$). From the expression above, it results that the following limits exist
\[
\lim_{t\to +\infty}\frac{1+g\left(x_i^{t}-\langle \mathbf x^{t}\rangle_i^c\right)}{1+g\left(x_{N}^{t}-\langle \mathbf x^{t}\rangle_N^c\right)}=1,\ \forall i\in [1,N-1].
\]
By compactness, let $\{t_k\}_{k\in\N}$ be any infinite sequence such that the following limit exists
\[
\lim_{k\to +\infty}\mathbf x^{t_k}=\mathbf x^\infty.
\]
Together with the assumption (Hg1), the limit above of the ratio implies that $x_i^\infty=x_N^\infty$ for all $i\in [1,N-1]$. Since this holds for every subsequence, we conclude that 
\[
\lim_{t\to +\infty}x_i^t-x_N^t=0,\ \forall i\in [1,N-1].
\]
In particular, we have for every subsequence as above,
\[
\lim_{k\to +\infty}x_i^{t_k+1}-x_N^{t_k+1}=0,\ \forall i\in [1,N-1],
\]
which, using also the iteration rule for the $x_i^t$, results in the following equality
\[
f\left(\frac{\rho_{i}^{\infty}}{\langle \boldsymbol\rho^{\infty}\rangle_{i}^c},x_i^\infty\right)=f\left(\frac{1}{\langle \boldsymbol\rho^{\infty}\rangle_{N}^c},x_N^\infty\right),\ \forall i\in [1,N-1],
\]
and then since $x_i^\infty=x_N^\infty$
\[
f\left(\frac{\rho_{i}^{\infty}}{\langle \boldsymbol\rho^{\infty}\rangle_{i}^c},x_N^\infty\right)=f\left(\frac{1}{\langle \boldsymbol\rho^{\infty}\rangle_{N}^c},x_N^\infty\right),\ \forall i\in [1,N-1].
\]
We claim that this imposes 
\begin{equation}
\rho_i^\infty=1\ (=\rho_N^\infty),\ \forall i\in [1,N-1].
\label{EQUAL-RHO}
\end{equation}
In order to see this, let us first show that if the coordinates $\{\rho_i^\infty\}_{i=1}^{N}$ were not all equal, then there would exist $i',i''\in [1,N]$ such that 
 \[
\frac{\rho_{i'}^{\infty}}{\langle \boldsymbol\rho^{\infty}\rangle_{i'}^c} <1<\frac{\rho_{i''}^{\infty}}{\langle \boldsymbol\rho^{\infty}\rangle_{i''}^c}.
 \]
 Indeed, by contradiction, suppose that 
 \[
 (N-1)\rho_{i}^\infty\geq \sum_{j=1}^N\rho_j^\infty-\rho_{i}^\infty,\ \forall i\in [1,N],
 \]
and one inequality at least must be strict when the coordinates $\{\rho_i^\infty\}_{i=1}^{N}$ are not all equal. Summing over $i$ yields $N{\displaystyle\sum_{i=1}^N}\rho_i^\infty>N{\displaystyle\sum_{i=1}^N}\rho_i^\infty$  which is impossible. Hence, the existence of $i'$ as desired. The proof of existence of $i''$ is the same.

In addition, the facts that $f(\rho,x)>x$ for $\rho<1$ and $x\in [0,1)$, and respectively that $f(\rho,x)<x$ for $\rho>1$ and $x\in (0,1]$ make it impossible that 
\[
f\left(\frac{\rho_{i'}^{\infty}}{\langle \boldsymbol\rho^{\infty}\rangle_{i'}^c},x_N^\infty\right)=f\left(\frac{\rho_{i''}^{\infty}}{\langle \boldsymbol\rho^{\infty}\rangle_{i''}^c},x_N^\infty\right),
\]
meaning that all the coordinates of  $\{\rho_i^\infty\}_{i=1}^{N}$ must be equal, proving the equalities \eqref{EQUAL-RHO}. 

In order to complete the proof, we show independently that each sequence $\{\frac{x_i^t}{x_N^t}\}_{t\in\N}$ is increasing, leading to a contradiction with the limits of $x_i^t-x_N^t$. We have proved that the sequences $\{\rho_i^t\}_{t\in\N}$ ($i\in [1,N-1]$) are all increasing and converge to 1. In particular, we have 
\[
\frac1{\langle \boldsymbol\rho^{t}\rangle_{N}^c}>1,\ \forall t\in\N,
\]
implying that the sequence $\{x_N^t\}_{t\in\N}$ must be decreasing. Let $t\in\N$ and $i\in [1,N-1]$ be arbitrary. We separate the cases $\frac{\rho_{i}^{t+1}}{\langle \boldsymbol\rho^{t+1}\rangle_{i}^c}<1$ and $\frac{\rho_{i}^{t+1}}{\langle \boldsymbol\rho^{t+1}\rangle_{i}^c}\geq 1$. In the first case, we certainly have $x_i^{t+1}>x_i^t$ and hence $\frac{x_i^{t+1}}{x_N^{t+1}}>\frac{x_i^{t}}{x_N^{t}}$, as desired. 

In the second case, which can only happen if $N>2$, the fact that $\rho^{t+1}_i<1$ for all $i$ implies the inequality $\frac{\rho_{i}^{t+1}}{\langle \boldsymbol\rho^{t+1}\rangle_{i}^c}<\frac1{\langle \boldsymbol\rho^{t+1}\rangle_{N}^c}$. Moreover, the assumption that the maps $x\mapsto \frac{f_\mathrm{dev}(\rho,x)}{x}$ are non-decreasing for every $\rho$, monotonicity in $\rho$ and the fact that we must have $x_i^t>x_N^t$, yield
\[
\frac{f_{\frac{\rho_{i}^{t+1}}{\langle \boldsymbol\rho^{t+1}\rangle_{i}^c}}(x_i^t)}{x_i^t}>  \frac{f_{\frac{1}{\langle \boldsymbol\rho^{t+1}\rangle_{N}^c}}(x_i^t)}{x_i^t}\geq \frac{f_{\frac{1}{\langle \boldsymbol\rho^{t+1}\rangle_{N}^c}}(x_N^t)}{x_N^t},
\]
from where $\frac{x_i^{t+1}}{x_N^{t+1}}>\frac{x_i^{t}}{x_N^{t}}$ immediately follows. The proof is complete. 

\noindent
The proof that $\overline{i}^t_{\mathbf x}$ changes infinitely often is similar and relies on the property that the maps $x\mapsto \frac{1-f(\rho,x)}{1-x}$ are non-decreasing for every $\rho< 1$.

\paragraph{Proof of infinite changes of $\underline{i}^t_{\mathbf p}$.} 
Again by contradiction. Assume w.l.o.g.\ that $\underline{i}^t_{\mathbf p}=N$ for all $t\in\N$. Then, we have 
\[
\frac{\rho_{i}^{t}}{\langle \boldsymbol\rho^{t}\rangle_{i}^c}>\frac1{\langle \boldsymbol\rho^{t}\rangle_{N}^c},\ \forall t\in\N.
\]
The monotonicity of $(x,\rho)\mapsto f(\rho,x)$ implies that for any given $i\in [1,N-1]$, if there exists $t_i\in \N$ such that $x_i^{t_i}<x_N^{t_i}$, then we must have 
\[
x_i^{t}<x_N^{t},\ \forall t>t_i.
\]
Let us show that such $t_i$ must indeed exists for each $i$. By contradiction, assume that $x_N^t< x_i^t$ for all $t\in\N$. Then $\{\rho_i^t\}_{t\in\N}$ must be an increasing sequence that converges to $\rho_i^\infty>\rho_i^0>1$.  However, the same reasoning as above that yields the equality \eqref{EQUAL-RHO} implies 
\[
\lim_{t\to +\infty}x_i^t-x_N^t=0,
\]
and then $\frac{\rho_{i}^{\infty}}{\langle \boldsymbol\rho^{\infty}\rangle_{i}^c}=\frac{1}{\langle \boldsymbol\rho^{\infty}\rangle_{N}^c}$ for every accumulation point of the sequence $\{\boldsymbol\rho^t\}_{t\in\N}$, which imposes in turn that $\rho_i^\infty=1$. However, this conclusion is incompatible with  $\rho_i^\infty>1$.

Therefore,  $t_i$ exists for each $i\in [1,N-1]$. But then the following equality results
\[
\overline{i}^t_{\mathbf x}=N,\ \forall t\geq \max_{i\in [1,N-1]}t_i,
\]
which we previously showed to be impossible. The proof is complete.

\subsection{Proof the Theorem \ref{STABILITY}}\label{S-STABILITY}
Following \cite{L73,W78}, we first compute the normal form of $F_\mathrm{skew}$. To that goal, we proceed to a series of changes of variables in order to obtain an appropriate setting in which the variable transverse to $\langle \mathbf x\rangle$ is a complex number. In the second part, we analyse this normal form and its coefficients in order to establish the desired convergence.

\paragraph{Changes of variables.} 
The first change of variable $(\mathbf x,\rho_1)\mapsto (\mu,\Delta,\epsilon)$ is defined by  
\[
\mu=\langle\mathbf x\rangle=\frac{x_1+x_2}2,\quad \Delta=\frac{x_1-x_2}2\quad\text{and}\quad\epsilon=\log\rho_1.
\]
This transformation conjugates $F_\mathrm{skew}$ to the mapping those coordinates are defined by
\[
\left\{\begin{array}{l}
\overline{\mu}(\mu,\Delta,\epsilon)=\alpha \mu+\frac{1-\alpha}2\left(f(e^{\overline{\epsilon}},\mu+\Delta)+f(e^{-\overline{\epsilon}},\mu-\Delta)\right)\\
\overline{\Delta}(\mu,\Delta,\epsilon)=\alpha \Delta+\frac{1-\alpha}2\left(f(e^{\overline{\epsilon}},\mu+\Delta)-f(e^{-\overline{\epsilon}},\mu-\Delta)\right)\\
\overline{\epsilon}(\Delta,\epsilon)=\epsilon+\log\frac{1+g(2\Delta)}{1+g(-2\Delta)}
\end{array}\right.
\]
The Jacobian matrix at any point $(\mu,0,0)$ for which $f'_x(1,\mu)=1$ writes
\begin{equation}
\left(\begin{array}{ccc}
1&0&0\\
0&1+4(1-\alpha)f_p(\mu)g_p&(1-\alpha)f_p(\mu)\\
0&4g_p&1
\end{array}\right)
\label{JACOB}
\end{equation}
where 
\[
f_p(\mu)=f_{\rho}'(1,\mu)\quad \text{and}\quad g_p=g'(0).
\]
Notice that if $(\mu,\mu,1)$ is a (transversally) parabolic fixed point of $F_\mathrm{skew}$ (in the sense defined before the Theorem) with $f_p(\mu)=0$ and $g_p>0$, then expression \eqref{JACOB} implies that it must be linearly unstable. 

The eigenvalues $\lambda_\pm(\mu)$ of the lower right minor $2\times 2$ matrix of \eqref{JACOB} write
\[
\lambda_\pm(\mu)=1+2(1-\alpha)f_p(\mu)g_p\pm 2\sqrt{(1-\alpha)f_p(\mu)g_p(1+(1-\alpha)f_p(\mu)g_p)},
\]
and the corresponding eigenvectors write (assuming that $g_p\neq 0$)
\[
e_\pm(\mu)=\left(\begin{array}{c}
\frac{(1-\alpha)f_p(\mu)}{2}\pm \frac12\sqrt{\frac{(1-\alpha)f_p(\mu)(1+(1-\alpha)f_p(\mu)g_p)}{g_p}}\\
1\end{array}\right)
\]
Notice that when $(\mu,\mu,1)$ is an elliptic fixed point, {\sl ie.}\ when $(1-\alpha)f_p(\mu)g_p\in [-1,0)$, we have $\lambda_\pm(\mu)=e^{\pm i\theta(\mu)}$ where $\theta(\mu)\in (0,\pi]$ is defined by 
\begin{equation}
\cos\theta(\mu)=1+2(1-\alpha)f_p(\mu)g_p.
\label{DEFCOS}
\end{equation}
More generally, for every $\mu\in [0,1]$ such that $f_p(\mu)< 0$ and $(1-\alpha)f_p(\mu)g_p>-1$, this expression defines a unique $\theta(\mu)\in (0,\pi]$. 

Assuming that $\mu$ is such that the previous inequalities hold, we consider a second change of variables $(\mu,\Delta,\epsilon)\mapsto (\mu,x,y)$ in which the variables $(x,y)$ are defined through the coordinates of $e_-(\mu)$ by inverting the following system (NB: $\sin\theta(\mu)=2\sqrt{-(1-\alpha)f_p(\mu)g_p(1+(1-\alpha)f_p(\mu)g_p)}$)
\[
\left(\begin{array}{c}
\Delta\\
\epsilon\end{array}\right)=\left(\begin{array}{cc}
\frac{\cos\theta(\mu)-1}{4g_p}&-\frac{\sin\theta(\mu)}{4g_p}\\
1&0\end{array}\right)\left(\begin{array}{c}
x\\
y\end{array}\right)
\quad\text{viz.}\quad 
\left\{\begin{array}{ccl}
\Delta(\mu,x,y)&=&\frac{\cos\theta(\mu)-1}{4g_p} x-\frac{\sin\theta(\mu)}{4g_p}y\\
\epsilon(\mu,x,y)&=&x
\end{array}\right.
\]
which yields
\[
\left\{\begin{array}{l}
x(\mu,\Delta,\epsilon)=\epsilon\\
y(\mu,\Delta,\epsilon)=\frac{\cos\theta(\mu)-1}{\sin\theta(\mu)}\epsilon-\frac{4g_p}{\sin\theta (\mu)}\Delta
\end{array}\right.
\]
In these new variables, the dynamics writes $(\mu,x,y)\mapsto (\tilde\mu(\mu,x,y),X(\mu,x,y),Y(\mu,x,y))$ where  
\[
\begin{array}{l}
\tilde\mu(\mu,x,y)=\overline{\mu}(\mu,\Delta(\mu,x,y),x)\\
X(\mu,x,y)=\overline{\epsilon}(\Delta(\mu,x,y),x)\\
Y(\mu,x,y)=\frac{\cos\theta(\mu)-1}{\sin\theta(\mu)}X(\mu,x,y)-\frac{4g_p}{\sin\theta(\mu)}\tilde{\Delta}(\mu,x,y)\quad\text{and}\quad \tilde{\Delta}(\mu,x,y)=\overline{\Delta}(\mu,\Delta(\mu,x,y),x)
\end{array}
\]
As expected, in these new variables, the Jacobian matrix writes, using the definition of $\cos\theta(\mu)$
\[
\left(\begin{array}{ccc}
1&0&0\\
0&\cos\theta(\mu)&-\sin\theta(\mu)\\
0&\sin\theta(\mu)&\cos\theta(\mu)
\end{array}\right)
\]
We finally consider a complexification of the transverse variable $(x,y)$. Letting $z=x+iy$, we define the map $F_\C$ acting in $(\mu,z)\in [0,1]\times \C$ by
\[
F_\C(\mu,z)=(\widehat{\mu}(\mu,z),Z(\mu,z)),
\]
where
\[
\widehat{\mu}(\mu,z)= \tilde\mu(\mu,\mathrm{Re}(z),\mathrm{Im}(z))\quad \text{and}\quad Z(\mu,z)=X(\mu,\mathrm{Re}(z),\mathrm{Im}(z))+i Y(\mu,\mathrm{Re}(z),\mathrm{Im}(z)).
\]
In these variables, the Jacobian matrix writes (still under the assumptions $f'_x(1,\mu)=1$, $f_p(\mu)< 0$ and $(1-\alpha)f_p(\mu)g_p>-1$)
\[
\left(\begin{array}{cc}
1&0\\
0&e^{i\theta(\mu)}
\end{array}\right)
\]
When $f$ satisfies the symmetry (Hf3), Lemma \ref{CONSTRAINTS-X-N2} implies that we have $\widehat{\mu}(\frac12,z)=\frac12$ for all $z$. In particular, $(\frac12,0)$ is a fixed point of $F_\C$. We aim at determining the behaviours of orbits in a sufficiently small neighbourhood of $(\frac12,0)$ in $\R\times \C$, assuming that this point is elliptic.

As commented after Lemma \ref{CONSTRAINTS-X-N2}, the assumptions on the components of the dynamics do not suffice to determine the limit of the first component of $\{F^t_\C(\mu,z)\}_{t\in\N}$, namely the variable $\mu^t$. In particular, while we can make sure that this component must remain close to $\frac12$, its limit as $t\to +\infty$ may differ from that value. 

Instead, focus here will be made on the long term behaviour of the second component of $\{F^t_\C(\mu,z)\}_{t\in\N}$, namely the variable $z^t$, when starting sufficiently close to $(\frac12,0)$. To that goal, we begin by establishing the expansion of the map $Z$ in $(\mu+\delta,z)$ for $(\delta,z)$ in the neighbourhood of the origin in $\R\times \C$.

\paragraph{Expansion of the map $Z$.} 
Evidently, the map $(\mu,\Delta,\epsilon)\mapsto f(e^{\epsilon},\mu\pm\Delta)$ is of class $C^4$ in $[x_0,1-x_0]\times [-2x_0,2x_0]\times [-\ln \rho_0,\ln\rho_0]$. Besides, the set 
\[
\left\{(\Delta,\epsilon)\ :\ \overline{\epsilon}(\Delta,\epsilon)\in [-\ln\rho_0,\ln\rho_0]\right\}
\]
contains a neighbourhood of the origin in $\R^2$. The fact that $g\in C^4([-2x_0,2x_0])$ implies that the map $(\Delta,\epsilon)\mapsto \overline{\epsilon}(\Delta,\epsilon)$ is also of class $C^4$ in this neighbourhood. Altogether, this implies that for every $\mu\in (x_0,1-x_0)$, there exists a neighbourhood $U_\mu$ of the origin in $\C$ such that the map $z\mapsto Z(\mu,z)\in C^4(U_\mu)$. 

Furthermore, the definitions of $\overline{\Delta}$ and $\overline{\epsilon}$ and the symmetry of $g$ imply the following symmetries
\[
\left\{\begin{array}{l}
\overline{\Delta}(\mu,-\Delta,-\epsilon)=-\overline{\Delta}(\mu,\Delta,\epsilon)\\
\overline{\epsilon}(\mu,-\Delta,-\epsilon)=-\overline{\epsilon}(\mu,\Delta,\epsilon)
\end{array}\right.
\]
and then 
\[
\left\{\begin{array}{l}
X(\mu,-x,-y)=-X(\mu,x,y)\\
Y(\mu,-x,-y)=-Y(\mu,x,y)
\end{array}\right.
\]
%
%
%
This symmetry transfers to all derivatives wrt to the $\mu$-variable, and also to all derivatives that are even in $(x,y)$ and arbitrary in $\mu$, {\em viz.}\ we have
\[
X^{(k)}_{\mu^{k-2\ell}x^{2\ell-m}y^m}(\mu,0,0)=0,\ \forall k\in \N,\ \ell\in \left\{0,\cdots ,\lfloor\frac{k}2\rfloor\right\},\ m\in \{0,\cdots ,2\ell\}.
\]
Accordingly, the 3rd-order expansion of $X$ at $(\mu,0,0)$ - which we denote by 0 for the sake of notation - writes
\begin{align*}
X(\mu+\delta,x,y)=&\left(X'_x(0)+X''_{\mu x}(0)\delta +\frac12X'''_{\mu^2 x}(0)\delta^2\right)x+\frac16 X'''_{x^3}(0)x^3+\frac12 X'''_{x^2y}(0)x^2y+\frac12 X'''_{xy^2}(0)xy^2\\
&+\left(X'_y(0)+X''_{\mu y}(0)\delta +\frac12X'''_{\mu^2 y}(0)\delta^2\right)y+\frac16 X'''_{y^3}(0)y^3+r_4(\delta,x).
\end{align*}
where, given $n\in \N$, the symbol $r_{n+1}$ denotes a generic remainder that satisfies 
\[
\lim_{\max\{|\delta|,|z|\}\to 0}\frac{|r_{n+1}(\delta,z)|}{(\max\{|\delta|,|z|\})^n}=0.
\]
A similar expansion holds for the expansion of $Y$.

By differentiating the relation \eqref{DEFCOS} one gets the following expressions
\[
f''_{\rho x}(1,\mu)=-\frac{\theta'(\mu)\sin\theta(\mu)}{2(1-\alpha)g_p}\quad\text{and}\quad f'''_{\rho x^2}(1,\mu)=-\frac{(\theta'(\mu))^2\cos\theta(\mu)+\theta''(\mu)\sin\theta(\mu)}{2(1-\alpha)g_p}.
\] 
Altogether, using these derivatives, the relation \eqref{DEFCOS} and also the equalities $f'_x(1,\mu)=1$ and $f''_{x^2}(1,\mu)=f'''_{x^3}(1,\mu)=0$, explicit computations yield the following expressions for the derivatives involved in the expansions
\begin{itemize}
\item $X''_{\mu x}(0)=-\theta'(\mu)\sin\theta(\mu)$ and $X'''_{\mu^2 x}(0)=-(\theta'(\mu))^2\cos\theta(\mu)-\theta''(\mu)\sin\theta(\mu)$,
\item $X''_{\mu y}(0)=-\theta'(\mu)\cos\theta(\mu)$ and $X'''_{\mu^2 y}(0)=(\theta'(\mu))^2\sin\theta(\mu)-\theta''(\mu)\cos\theta(\mu)$,
\item $Y''_{\mu x}(0)=\theta'(\mu)\cos\theta(\mu)$ and $Y'''_{\mu^2 x}(0)=-(\theta'(\mu))^2\sin\theta(\mu)+\theta''(\mu)\cos\theta(\mu)$,
\item $Y''_{\mu y}(0)=-\theta'(\mu)\sin\theta(\mu)$ and $Y'''_{\mu^2 y}(0)=-(\theta'(\mu))^2\cos\theta(\mu)-\theta''(\mu)\sin\theta(\mu)$.
\end{itemize}
Passing to the complexification, we finally get the following expansion of the map $Z$
\begin{equation*}
Z(\mu+\delta,z)=e^{i\theta(\mu)}\left(1+i\theta'(\mu) \delta+\frac12\left(-(\theta'(\mu))^2+i\theta''(\mu)\right)\delta^2\right)z +A(\mu,z)+ r_4(\delta,z),
\end{equation*}
where the $3^\mathrm{rd}$-order term $A(\mu,z)$ involves combinations of the 3rd order derivatives of $X$ and $Y$ in $(x,y)$ and it generically writes
\[
A(\mu,z)=\frac{c_{3}(\mu,0)}6z^3+\frac{c_{2}(\mu,0)}2z^2\bar z+\frac{c_{1}(\mu,0)}2z\bar z^2+\frac{c_{0}(\mu,0)}6\bar z^3,
\]
where the coefficients $c_i(\mu,0)$ depend continuously on $\mu$.

\paragraph{Analysis of the dynamics in the neighbourhood of $(\frac12,0)$.} The iterations of the variable $z$ are governed by the maps $Z^t$ defined by the following induction
\[
Z^1(\mu,z)=Z(\mu,z)\quad \text{and}\quad Z^{t+1}(\mu,z)=Z^t(\widehat{\mu}(\mu,z),Z(\mu,z)),\ \forall t\in\Z^+.
\]
assuming (Hf3) and that the fixed point $(\frac12,0)$ is elliptic, we aim at conditions on the coefficients involved in the expansion of $Z$ that ensures that ${\displaystyle\lim_{t\to +\infty}}|Z^t(\mu,z)|=0$ when $|z|$ is sufficiently small and $\mu$ is sufficiently close to $\frac12$. 

The above expansion of $Z$ indicates that a necessary condition for the damping of $Z^t$ is that $\theta'(\frac12)=\theta''(\frac12)=0$, namely that we have $f''_{\rho x}(1,\frac12)=f'''_{\rho x^2}(1,\frac12)=0$. Indeed, otherwise, we certainly have
\[
\left|1+i\theta'(\frac12) \delta+\frac12\left(-(\theta'(\frac12))^2+i\theta''(\frac12)\right)\delta^2\right|>1
\]
when $\delta$ is sufficiently small, implying that $|Z(\mu,z)|>|z|$ for all $z$ such that $|z|$ is small enough. 

From now on, we assume that $f''_{\rho x}(1,\frac12)=f'''_{\rho x^2}(1,\frac12)=0$ and consider the dynamics in the neighbourhood of $(\frac12,0)$. The fact that $(\frac12,0)$ is elliptic suggests that the instant behaviour of $|Z(\mu,z)|$ might not be isotropic, {\sl ie.}\ $|Z(\mu,z)|<|z|$ may fail depending on $\mathrm{Arg}(z)$. 
In order to address this issue, 
we instead aim at a criterion such that 
\begin{equation}
\lim_{t\to+\infty}|Z^{nt}(\frac12+\delta,z)|=0
\label{LIMITZ}
\end{equation}
for some $n$ suitably chosen, and provided that $|\delta|$ and $|z|$ are sufficiently small.

Dropping the explicit dependence on $\delta$ for simplicity, using the assumption $f''_{\rho x}(1,\frac12)=f'''_{\rho x^2}(1,\frac12)=0$ and (Hf3) in an easy induction yields that we have for every $n\in\Z^+$
\[
 Z^n(z)=e^{in\theta}z+e^{i(n-1)\theta}\sum_{k=0}^{n-1}e^{-ik\theta}A(e^{ik\theta}z)+r_4(z),
 \]
where
 \begin{align*}
\sum_{k=0}^{n-1}e^{-ik\theta}A(e^{ik\theta}z)=&\frac{c_{3}(0)}6z^3\sum_{k=0}^{n-1}e^{2ik\theta}+n\frac{c_{2}(0)}2z^2\bar z+\frac{c_{1}(0)}2z\bar z^2\sum_{k=0}^{n-1}e^{-2ik\theta}
+\frac{c_{0}(0)}6\bar z^3\sum_{k=0}^{n-1}e^{-4ik\theta}
\end{align*}
In particular, the assumption that the fixed point is elliptic and $\theta\not\in \pi\Z$ imply that the coefficients in the sums do not vanish; hence we have 
\[
 Z^n(z)=e^{in\theta}z+e^{i(n-1)\theta}\left(\frac{c_{3}(0)}6z^3\frac{1-e^{2in\theta}}{1-e^{2i\theta}}+n\frac{c_{2}(0)}2z^2\bar z+\frac{c_{1}(0)}2z\bar z^2\frac{1-e^{-2in\theta}}{1-e^{-2i\theta}}
+\frac{c_{0}(0)}6\bar z^3\frac{1-e^{-4in\theta}}{1-e^{-4i\theta}}\right)+r_4(z),
 \]
From this expression, we get 
\begin{align*}
|Z^{n}(z)|^2=&|z|^2+n\mathrm{Re}(e^{-i\theta}c_2(0))|z|^4\\
&+2\mathrm{Re}\left(e^{-i\theta}\left(\frac{c_{3}(0)}6|z|^2z^2\frac{1-e^{2in\theta}}{1-e^{2i\theta}}+\frac{c_{1}(0)}2|z|^2\bar z^2\frac{1-e^{-2in\theta}}{1-e^{-2i\theta}}
+\frac{c_{0}(0)}6\bar z^4\frac{1-e^{-4in\theta}}{1-e^{-4i\theta}}\right)\right)+r_5(z)
\end{align*}
We have 
\[
\left|\mathrm{Re}\left(e^{-i\theta}\left(\frac{c_{3}(0)}6|z|^2z^2\frac{1-e^{2in\theta}}{1-e^{2i\theta}}+\frac{c_{1}(0)}2|z|^2\bar z^2\frac{1-e^{-2in\theta}}{1-e^{-2i\theta}}
+\frac{c_{0}(0)}6\bar z^4\frac{1-e^{-4in\theta}}{1-e^{-4i\theta}}\right)\right)\right|\leq C |1-e^{2in\theta}||z|^4
\]
for some constant $C\in\R_\ast^+$. Moreover, every $\theta\in\R$ can be approximated arbitrarily close by a rational multiple of $\pi$. That is to say, one can choose $n$ sufficiently large in order to make $|1-e^{2in\theta}|$ arbitrarily small.\footnote{In particular, if $\theta=\frac{p}{q}\pi$ for some $p,q\in \Z$, then we obviously have $e^{2iq\theta}=1$. When $\theta$ is irrational multiple of $\pi$, then the continued fraction algorithm provides such $n$.}
Therefore, if 
\begin{equation}
\mathrm{Re}(e^{-i\theta}c_2(0))<0,
\label{STABCOND}
\end{equation}
then, provided that $n$ is suitably chosen sufficiently large there exists $a,\epsilon>0$ such that 
\[
|Z^{n}(\frac12+\delta,z)|^2\leq |z|^2(1-a|z|^2),\ \forall |\delta|,|z|<\epsilon.
\]
Independently, let $(\delta,z)$ be sufficiently close to $(0,0)$ so that 
\[
(e^{\overline{\epsilon}},\mu+\Delta),(e^{\overline{\epsilon}},1-\mu+\Delta)\in U
\]
Using the symmetry (Hf3) as in the proof of Lemma \ref{CONSTRAINTS-X-N2}, we obtain
\[
|2 \hat\mu(\frac12+\delta,z)-1|\leq 2|\delta|\sup_{(\rho,x)\in U}|f'_x(\rho,x)|.
\] 
The assumption ${\displaystyle \sup_{(\rho,x)\in U}}|f'_x(\rho,x)|\leq 1$ ensures that $|2 \hat\mu(\frac12+\delta,z)-1|<2\epsilon$. Hence, the inequality above on $|Z^{n}|^2$ holds for the iterated pair $(\hat\mu(\frac12+\delta,z),Z(\delta,z))$, and then, by repeating the argument, for all subsequent $nt$-iterates, implying the limit \eqref{LIMITZ}.

\paragraph{Analysis of the stability condition.}
In order to complete the proof of the Theorem, it remains to express the stability condition \eqref{STABCOND} in terms of the components of the original map $F_\mathrm{skew}$.

To that goal, we compute the coefficient $c_2(0)$ from the expansions of the $3^\mathrm{rd}$-order expansions of the functions $X(\mu+\delta,x,y)$ and $Y(\mu+\delta,x,x)$ at $(\frac12,0,0)$. The resulting expression is 
\begin{align*}
&c_2(0)=\frac18 \left(X'''_{x^3}(0) + X'''_{xy^2}(0)  + Y'''_{x^2y}(0) +  Y'''_{y^3}(0) +i(Y'''_{x^3}(0)-X'''_{x^2y}(0) +  Y'''_{xy^2}(0)- X'''_{y^3}(0) )   \right)
\end{align*}
Accordingly, we need to compute the $3^\mathrm{rd}$-order derivatives at $(\frac12,0,0)$ of the functions $X(\mu,x,y)$ and $Y(\mu,x,x)$. The expressions are given in Appendix \ref{A-DERIV}. Given these expressions, we obtain the following expressions for the real part of the coefficient $c_2(0)$, up to the factor $\frac18$
\begin{align*}
(1-\alpha ) \left(4  f_p g_p+ \frac{  f_p (1-\cos \theta)}{g_p^2}K_g+12g_p f''_{\rho^2}(1,\frac12)+4  g_p f'''_{\rho^3}(1,\frac12 )+ (3-2 \cos \theta) f'''_{\rho^2 x}(1,\frac12 ) \right)
\end{align*}
Similarly, for the imaginary part of $c_2(0)$, we have, up to the factor $\frac18$, using again the definition \eqref{DEFCOS} appropriately
\begin{align*}
&-(1-\alpha )\left(\frac{4 f_p g_p \cos \theta}{\sin \theta}+\frac{ f_p (1-\cos \theta)\cos \theta}{g_p^2 \sin \theta}K_g+\frac{12  g_p \cos \theta }{\sin \theta}f''_{\rho^2}(1,\frac12 )+\frac{4  g_p \cos \theta }{\sin \theta}f'''_{\rho^3}(1,\frac12 )\right.\\
&\left. +\frac{(3-2\cos\theta)\cos\theta-1}{\sin \theta}f'''_{\rho^2 x}(1,\frac12 )\right)
\end{align*}

Altogether, the real part 
\[
8\mathrm{Re}(e^{-i\theta}c_2(0))=(X'''_{x^3}(0) + X'''_{xy^2}(0)  + Y'''_{x^2y}(0) +  Y'''_{y^3}(0))\cos\theta+(Y'''_{x^3}(0)-X'''_{x^2y}(0) +  Y'''_{xy^2}(0)- X'''_{y^3}(0))\sin\theta, 
\]
writes $(1-\alpha )f'''_{\rho^2 x}(1,\frac12)$, 
from where the last inequality in equation  \eqref{CONDERIV} in the statement of the Theorem immediately implies the stability condition \eqref{STABCOND}.  

\section{Conclusion}
In this paper, a behavioural model for the population dynamics in OTC wholesale fresh product markets, that aims to address the limitations of previous modelling, has been introduced and mathematically analyzed. The evolution rules, which  involve a negative regulatory feedback scheme,  are based on elementary buyer-seller interactions that are inspired from empirical observations and accounts by the various actors. The analysis proves that the dynamics self-regulates and feedback-induced oscillations of limited amplitude must prevail forever, as far as clientele fractions and prices ratio are concerned. 

One particularly important result is that the oscillations must be asymptotically damped unless the prices themselves vanish in the long term. In the case of $N=2$ and under the symmetry assumption, we have proved that, provided that the fraction of loyal buyers is high enough or the rate at which sellers react to clientele change is sufficiently small, the fixed point $(\frac12,\frac12,0)$ is elliptic and under additional technical conditions, the oscillations are indeed damped (at least when starting in a sufficiently small neighbourhood of this fixed point), implying  long-term equilibration of prices and clientele. Altogether, our results indicate that, despite the limited number of ingredients and their simplicity, the model exhibits a number of realistic functioning modes, including long-term equilibration.  

Still in the $N=2$ case, the fact that equilibria are unstable (hyperbolic) when loyalty is small and sellers are highly reactive suggests that the orbits are unlikely to converge to a stationary point. In fact, \cite{E25} proves that 2-periodic orbits must exist in this case (provided that $f$ is piecewise affine as in Fig.\ \ref{GRAPHF-AFFINE}), and proves also the existence of family of maps with stable 2-periodic orbits. Hence, there are examples of maps and orbits that converge to a (non-stationary) periodic orbit, implying vanishing prices in this case. Such behaviour raises interesting questions about the relevance of modelling in this case. Indeed, is asymptotically vanishing prices compatible with perennial business? If not, does this property implies that little loyalty and high reactivity are unrealistic features of such markets? 

Another question of interest is the extension of the nonlinear stability result to non-symmetric map $f$ when $N=2$ and also to markets with more than 2 sellers. On the broader horizon of improving modelling, to introduce some heterogeneity in the buyers population is another challenge that will be addressed in future studies. 


\paragraph{Acknowledgements.} 
We are grateful to N.\ Raffin, A.\ Vignes and J.\ Wenzelburger for fruitful discussions and relevant comments. The research in this article has been accomplished in the framework of the project ANR LabCom LOPF, ref ANR-20-LCV1-0005.

\appendix

\section{Emerging bounds on clientele volumes in the model with price ratios $\frac{p_i}{\langle \mathbf p\rangle}$}\label{A-ALTERNATIVE}
As explained in the definition of the iterations \eqref{DEFDYNAM}, the reason for considering the ratios $\frac{p_i}{\langle \mathbf p\rangle_i^c}$ instead of $\frac{p_i}{\langle \mathbf p\rangle}$, is to avoid any bias in the dynamics that would result from the breakdown of the symmetry $\rho\to \frac1{\rho}$. In this appendix, we emphasize on the difference between the two models by exhibiting an asymmetric feature of the constraints on clientele volumes when considering $\frac{p_i}{\langle \mathbf p\rangle}$, even though the family $\{f_\rho\}$ is invariant under such symmetry.

To be more specific, we assume that $N=2$ and consider the following iterations
\begin{equation}
\left\{\begin{array}{l}
x_i^{t+1}=f_{\alpha,\frac{2p_i^{t+1}}{p_1^{t+1}+p_2^{t+1}}}(x_i^{t})\\
p_i^{t+1}=p_i^{t}\left(1+g(x_i^{t}-\langle \mathbf x^t\rangle_i^c)\right)
\end{array}\right.\ \text{for}\ i\in [1,2],
\label{ALTERNATIVE}
\end{equation}
where the maps $f_\rho$ are given by \eqref{LINEARMAP} with $c(\rho)=e^{-|\ln\rho|}$. We have the following result whose second statement contrasts the statement in Lemma \ref{CONSTRAINTS-X-N2}.
\begin{Claim}
(i) The following property holds for every orbit of the system \eqref{ALTERNATIVE}
\[
\langle \mathbf x^{t}\rangle>\frac12\quad\Longrightarrow \quad \langle \mathbf x^{t+1}\rangle>\frac12.
\]

\noindent
(ii) There are orbits of the system \eqref{ALTERNATIVE} for which there exists $t\in\R$ such that $\langle \mathbf x^{t}\rangle<\frac12$ and $\langle \mathbf x^{t+1}\rangle>\frac12$. 
\end{Claim}
\begin{proof}
As usual, it suffices to prove the result for $\alpha=0$. Removing the superscripts $t$ for simplicity, we have
\[
2\langle \mathbf x^{t+1}\rangle=f_{\frac{2\rho_1}{1+\rho_1}}(x_1)+f_{\frac{2}{1+\rho_1}}(x_2)
\]
In particular, when $\rho_1\geq 1$, we have 
\[
2\langle \mathbf x^{t+1}\rangle=\frac{1+\rho_1}{2\rho_1}x_1+1-\frac2{1+\rho_1}(1-x_2)=1+\left(\frac{1+\rho_1}{2\rho_1}-\frac2{1+\rho_1}\right)x_1-\frac2{1+\rho_1}(1-2\langle \mathbf x^{t}\rangle)
\]
Therefore, we have $\langle \mathbf x^{t+1}\rangle>\frac12$ when 
\[
\left(\frac{1+\rho_1}{2\rho_1}-\frac2{1+\rho_1}\right)x_1>\frac2{1+\rho_1}(1-2\langle \mathbf x^{t}\rangle),
\]
from where both statements {\em (i)} and {\em (ii)} easily follow in the case $\rho_1\geq 1$. The proof for $\rho_1\leq 1$ is similar.
\end{proof}


\section{Emerging bounds on clientele volumes for $N=2$ in absence of symmetry}\label{A-NONSYM}
In the case $N=2$, Lemma  \ref{CONSTRAINTS-X-N2}, which provides some control on the convergence of $\langle \mathbf x^t\rangle$ to $\frac12$, relies on the symmetry assumption (Hf3). Yet, some constraints on $\langle \mathbf x^t\rangle$ prevail in absence of the full symmetry, when only the derivatives at the two fixed points are equal and provided that the deviations from the linearizations at these fixed points are under control. The accurate result is given in the following statement.
\begin{Claim}
Let $N=2$ and assume that $g$ satisfies {\rm (Hg1)}, $f$ satisfies {\rm (Hf1-2)}, $f'_x(\rho,0)=f'_x(\frac1{\rho},1)$ for all $\rho\geq 1$ and there exists $\gamma\in (0,1)$ such that we have for $\rho>1$
\[
f'_x(\rho,0)x\leq f(\rho,x)\leq f'_x(\rho,0)x+\gamma (1-f'_x(\rho,0)),\ \forall x\in [0,1],
\]
and for $\rho<1$
\[
1-f'_x(\rho,1)(1-x)-\gamma (1-f'_x(\rho,1))\leq f(\rho,x)\leq 1- f'_x(\rho,1)(1-x),\ \forall x\in [0,1]
\]
Then, for every orbit $\{(\mathbf x^t,\rho_1^t)\}_{t\in\N}$ of \eqref{BASIS}, we have when $\rho_1^{t+1} >1$
\[
2\langle \mathbf x^{t+1}\rangle-1-\gamma \leq f'_x(\rho_1^{t+1},0)(2\langle \mathbf x^{t}\rangle-1-\gamma)
\]
and 
\[
2\langle \mathbf x^{t+1}\rangle-1+\gamma \geq f'_x(\rho_1^{t+1},1)(2\langle \mathbf x^{t}\rangle-1+\gamma).
\]
Moreover, similar inequalities hold with $f'_x(\rho_1^{t+1},0)$ and $f'_x(\rho_1^{t+1},1)$ exchanged, when $\rho_1^{t+1}<1$.
\end{Claim}
As a consequence the interval $[\frac{1-\gamma}2,\frac{1+\gamma}2]$ is an invariant set of the variable $\langle \mathbf x^{t}\rangle$, to which $\langle \mathbf x^{t}\rangle$ must approach in every orbit for which $\rho_1^t$ does not approach 1 too fast.\footnote{Indeed, $2\langle \mathbf x^{t}\rangle-1-\gamma$ must decrease when positive (and $\rho_1^{t+1}\neq 1$). Moreover, the same inequality implies that $2\langle \mathbf x^{t+1}\rangle-1-\gamma\leq 0$ when $2\langle \mathbf x^{t}\rangle-1+\gamma\leq 0$, and similarly when $2\langle \mathbf x^{t}\rangle-1-\gamma\geq 0$. This means that $\langle \mathbf x^{t}\rangle$ cannot jump over the interval. Hence, we are sure that $\langle \mathbf x^{t}\rangle$ must approach the interval when outside.} 
\begin{proof}
Assume that $\rho_1^{t+1}>1$. Then, from the right inequalities in the conditions on $f(\rho,x)$ and using that $f'_x(\frac1{\rho_1^{t+1}},1)=f'_x(\rho_1^{t+1},0)$, we obtain
\[
2\langle \mathbf x^{t+1}\rangle \leq 1+ f'_x(\rho_1^{t+1},0)(2 \langle \mathbf x^{t}\rangle -1)+\gamma (1-f'_x(\rho_1^{t+1},0)),
\]
from where the first inequality follows. The second one is obtained similarly by using the left inequalities in the conditions on $f(\rho,x)$. 
The proof for $\rho_1^{t+1}<1$ is similar. 
\end{proof}

\section{Expressions of the derivatives of the functions $X$ and $Y$}\label{A-DERIV}
This appendix reports the results of the computations of the $3^\mathrm{rd}$-order derivatives at $(\frac12,0,0)$ of the functions $X(\mu,x,y)$ and $Y(\mu,x,x)$, whose expressions compose the coefficient $c_2(0)$ involved in the stability condition \eqref{STABCOND}. Throughout, we omit the explicit dependence on $\mu$ because it is superfluous. 

\paragraph{Expressions of the $3^\mathrm{rd}$-order derivatives of $X$.}
Writing 
\[
X(x,y)=x+\log u(\frac{\cos\theta-1}{2g_p} x-\frac{\sin\theta}{2g_p}y)\quad \text{where}\quad\ u(x)=\frac{1+g(x)}{1+g(-x)},
\]
and using the equalities
\[
u(0)=1,\quad u'(0)=2g_p,\quad u''(0)=4g_p^2,\quad\text{and}\quad u'''(0)=2  (6  gp^3 - 3  gp  g''(0) +g'''(0)),
\]
we get 
\begin{align*}
&X'''_{x^3}(0)=\left(\frac{\cos\theta-1}{2g_p}\right)^3K_g\\
&X'''_{x^2y}(0)=-\left(\frac{\cos\theta-1}{2g_p}\right)^2\frac{\sin\theta}{2g_p}K_g\\
&X'''_{xy^2}(0)=\left(\frac{\sin\theta}{2g_p}\right)^2\frac{\cos\theta-1}{2g_p}K_g\\
&X'''_{y^3}(0)=-\left(\frac{\sin\theta}{2g_p}\right)^3K_g
\end{align*}
where $K_g=4g_p^3-6g_pg''(0)+2g'''(0)$. 

\paragraph{Expressions of the $3^\mathrm{rd}$-order derivatives of $Y$.}
Using the expression 
\[
Y(x,y)=\frac{\cos\theta-1}{\sin\theta}X(x,y)-\frac{4g_p}{\sin\theta}\tilde{\Delta}(x,y),
\]
and the equalities
\[
X(0)=0,\quad X'_x(0)=\cos\theta,\quad X'_y(0)=-\sin\theta,\quad X''_{\omega_1\omega_2}(0)=0,\ \text{for}\  \omega_1\omega_2\in \{x^2,xy,y^2\},
\]
and
\[
f''_{\rho x}(1,\frac12 )=f'''_{x^3}(1,x)=0,\ \forall x\in [0,1],
\]
and replacing $\cos\theta -1$ by $2(1-\alpha)f_pg_p$ where appropriate, we get the following expressions of the derivatives of $Y$, by computing the $3^\mathrm{rd}$-order derivatives of $\overline{\Delta}$, and after substantially algebraic manipulation
\begin{align*}
&Y'''_{x^3}(0)=-(1-\alpha )\left(\frac{4 f_p g_p \cos ^3\theta}{ \sin\theta}+\frac{2  f_p g_p}{\sin\theta}X'''_{x^3}(0) +\frac{12  g_p \cos ^3\theta}{\sin \theta} f''_{\rho^2}(1,\frac12 )+\frac{4 g_p \cos ^3\theta}{\sin \theta} f'''_{\rho^3}(1,\frac12 )\right.\\
&\left. -\frac{3  (1-\cos \theta) \cos^2 \theta}{\sin \theta} f'''_{\rho^2 x}(1,\frac12 )\right)\\
&Y'''_{x^2y}(0)= (1-\alpha)\left(4  f_p g_p \cos ^2\theta-\frac{2   f_p g_p}{\sin\theta} X'''_{x^2y}(0)+ 12 g_p \cos ^2\theta f''_{\rho^2}(1,\frac12 )+ 4g_p \cos ^2\theta f'''_{\rho^3}(1,\frac12 )\right.\\
&\left. +(3 \cos \theta-2) \cos \theta f'''_{\rho^2 x}(1,\frac12 )\right)\\
&Y'''_{xy^2}(0)=-(1-\alpha ) \left(4 f_pg_p  \sin \theta\cos \theta+\frac{2  f_pg_p}{\sin\theta} X'''_{\rho x^2}(0) +\sin \theta \left(12 g_p \cos \theta f''_{\rho^2}(1, \frac12 ) +4 g_p \cos \theta f'''_{\rho^3}(1, \frac12 )\right.\right.\\
&\left. \left. -(1-3 \cos \theta) f'''_{\rho^2 x}(1, \frac12 )\right)\right)\\
&Y'''_{y^3}(0)=(1-\alpha )\left(4  f_p g_p \sin^2\theta- \frac{2 f_p g_p}{\sin\theta}X'''_{y^3}(0)+\sin^2\theta\left(12 g_p f''_{\rho^2}(1,\frac12)+4 g_p  f'''_{\rho^3}(1,\frac12 )\right)\right)
\end{align*}

\end{document}